\documentclass[final,3p,times]{elsarticle}
\journal{Nuclear Physics B}
\pdfoutput=1
\usepackage{bm}
\usepackage{multirow}
\usepackage{cancel}
\usepackage{amsmath}
\usepackage{amssymb}
\usepackage{hyperref}
\biboptions{sort&compress}
\newcommand{\beq}{\begin{equation}}
\newcommand{\eeq}{\end{equation}}
\newcommand{\bea}{\begin{eqnarray}}
\newcommand{\eea}{\end{eqnarray}}
\newcommand{\bpmat}{\begin{pmatrix}}
\newcommand{\epmat}{\end{pmatrix}}

\begin{document}
\begin{frontmatter}{}
\title{ Purely Triplet Seesaw and Leptogenesis within  Cosmological Bound, Dark Matter, and Vacuum Stability }
\author[1]{Mina Ketan Parida\corref{cor1}}%
\ead{minaparida@soa.ac.in}
\author[2]{Mainak Chakraborty}
\ead{mainak.chakraborty2@gmail.com}
\author[1]{Swaraj Kumar Nanda}
\ead{swarajnanda.phy@gmail.com}
\author[1]{Riyanka Samantaray}
\ead{riyankasamantaray59@gmail.com}
\address[1]{Centre of Excellence in Theoretical and Mathematical Sciences,
Siksha `O' Anusandhan, Deemed to be University, Khandagiri Square, Bhubaneswar 751030, India}
\address[2]{Department of Physics, University of Calcutta, 92 Acharya Prafulla Chandra Road, Kolkata
700009, India}
\cortext[cor1]{Corresponding author}
\begin{abstract}
In a novel standard model extension it has been suggested that, even in the absence of right-handed neutrinos and
type-I seesaw, purely triplet leptogenesis leading to baryon asymmetry of the universe can be realized  by two heavy 
Higgs triplets which  also provide type-II seesaw ansatz for neutrino masses. In this work we discuss this model
predictions for hierarchical neutrino masses in concordance with recently determined cosmological bounds and oscillation
data including $\theta_{23}$ in the second octant and large Dirac CP phases. We find that for both normal and inverted 
orderings, the model fits the oscillation data with the  sum of the three neutrino masses consistent with current cosmological bounds determined from Planck satellite data. In addition, using this model ansatz for CP-asymmetry and solutions of Boltzmann equations, we also show how successful predictions of baryon asymmetry  emerges in the  cases of both  unflavoured and two-flavoured leptogeneses. With additional $Z_2$ discrete symmetry, a minimal extension of this model is further shown to predict a scalar singlet WIMP dark matter in agreement with direct and indirect observations which also resolves the issue of vacuum instability persisting in the original model. Although the combined constraints due to relic density and direct detection cross section allow this scalar singlet dark matter mass to be  $m_{\xi}=750$ GeV, the additional vacuum stability constraint pushes this limiting value to $m_{\xi}=1.3$ TeV which is verifiable by ongoing experiments. We also discuss  constraint on the  model parameters for the radiative stability of the standard Higgs mass.
\end{abstract}
\end{frontmatter}{}
\section{Introduction}\label{sec.1}
Neutrino oscillation \cite{nudata,Forero:2014,Esteban:2018}, baryon asymmetry of the universe
\cite{BAUexpt,Planck15} and  dark matter \cite{DMexpt} are
the three most prominent physics issues which can not be explained
within the purview of the 
standard model (SM). However, seesaw mechanisms have been widely recognized as possible origins 
of tiny neutrino masses where  leptogenesis  caused by the decay  of
mediating heavy particles are believed to be the underlying sources
of baryon asymmetry through sphaleron interactions \cite{sphaleron}. 
Large number of leptogenesis models using right-handed neutrino (RHN) mediated type-I seesaw  \cite{type-I,Valle:1980,Fuku-Yana:1986}, or other seesaw mechanisms, have been proposed for successful baryon asymmetry generation and  a partial list of such extensive investigations is given in \cite{Nir:2008,leptogenesis,ASJWR,Adhikary:2014qba}.  Since  SM itself does not have RHNs, it has to be extended for the implementation of type-I seesaw.   
 But, in a novel
 interesting proposal against the conventional lore and without using any RHNs or  supersymmetry, realisation
of neutrino masses and baryon asymmetry have been also shown to be
possible \cite{Ma-Us:1998} through the SM extension  by two heavy scalar triplets, $\Delta_1$ and $\Delta_2$, each of which generates neutrino mass by another
popular mechanism, called type-II seesaw \cite{type-II}. 
The tree level dilepton decay of any one of these
triplets combined with loop contribution generated by their
collaboration predicts the desired CP-asymmetry
formula for leptogenesis leading to observed baryon asymmetry of the Universe (BAU).
In an  implementation of this leptogenesis idea \cite{Ma-Us:1998} in non-supersymmetric (non-SUSY) SM extension in the scalar sector, quasi-degenerate (QD) neutrino masses of order $\sim 1$ eV and solution to simplified Boltzmann equations have been used to predict the baryon asymmetry of the Universe
$Y_{B} \sim 10^{-11}$. 
Leptogenesis with or without RHNs  has  been also
 implemented including or excluding supersymmetry \cite{leptogenesis,Ham-Ma-Us:2001,Ma-Ss-Us:1999,Ham-gs:2003,Ham:2012,Ham-Ma:2006,Ham-Strum:2006,Ham:2005,Gu:2016}.
The QD neutrino mass hypothesis used in \cite{Ma-Us:1998} has also a very interesting
outcome of predicting neutrinoless double beta decay rates \cite{bbexpt}, radiative magnification of neutrino mixings \cite{Balaji:2000}, and unification of  quark and neutrino
mixings at high scales \cite{mpr:2004} . 
On the other hand recent Planck satellite data \cite{Planck15}   have
set  the 
 cosmological upper bound on the sum of three light neutrino masses 
\beq
  \sum m_i\le 0.23 \,\,{\rm  eV} \equiv \Sigma_{Planck} .      \label{eq:cb}
\eeq
which is consistent with the standard $\Lambda$CDM big bang cosmology of the Universe \cite{Planck15}.  
Although Planck satellite data \cite{Planck15} also permits a much larger value  $\Sigma_C \simeq 0.71$ eV, this latter type of solution has been shown to be possible only in the absence of the cosmological constant ($\Lambda$). Compared to Planck satellite data \cite{Planck15}, somewhat lower value of the cosmological bound $\Sigma_{new}=0.12$ eV in the $\Lambda$CDM model has been also noted \cite{Sunny:2018}
\beq
 \sum m_i\le 0.12\, {\rm eV} \equiv \Sigma_{new}. \label{eq:cbnew}
\eeq
 Sensitivity of non-standard interactions to neutrino masses has been  investigated  \cite{Ohlsson:2019}.
 As against such cosmological bounds of eq.(\ref{eq:cb}) and 
eq.(\ref{eq:cbnew}),
KATRIN Collaboration \cite{KATRIN} has recently set the upper limit on the
neutrino mass scale $m_0\le 1$ eV which  predicts for QD neutrinos  
\beq
\sum m_i \le 3\,\, {\rm eV}\equiv \Sigma_{\rm KATRIN}. \,\,\label{eq:katrin} 
\eeq
 For a  QD neutrino mass scale as low as $m_0\simeq
 0.2$ eV or heavier, the neutrino  would manifest
in the direct experimental detection of neutrinoless double beta decay \cite{bbexpt} establishing its Majorana nature which has remained elusive so far. In any case it is quite important
to investigate the impact of the cosmological bounds \cite{Planck15,Sunny:2018} and the recently measured neutrino
oscillation data \cite{nudata,Forero:2014,Esteban:2018}  on the two-Higgs triplet seesaw and purely triplet leptogenesis \cite{Ma-Us:1998} in the absence of RHNs.

 Quite recently certain new features have been revealed in the neutrino oscillation
 data \cite{nudata,Forero:2014,Esteban:2018} which have to be
 explained in any theoretical model. The new data reveal the  values on atmospheric
 neutrino mixing angle to be in the second octant with $\theta_{23}\simeq 49.6^{\circ}$ and the
 Dirac CP phase to be large, $\delta  \sim 214^{\circ}$.  The impact of  new cosmological bounds \cite{Planck15,Sunny:2018} or the new oscillation data \cite{nudata,Forero:2014,Esteban:2018} have not been examined on the triplet leptogenesis model \cite{Ma-Us:1998,Sierra:2014tqa,Sierra:2011ab}. On the other hand, 
type-II seesaw dominance in SO(10) with scalar dark matter and vacuum stability
has been shown to be capable of providing excellent representation of the neutrino data \cite{cps:2019} where RHN loop mediated triplet leptogenesis
explains the baryon asymmetry of the universe. Unlike type-I seesaw, the type-II seesaw dominant mass matrix elements have one-to-one correspondence with
the mass matrix constructed using the oscillation data, a fact which underlines the importance of type-II over type-I. 
The model under discussion \cite{Ma-Us:1998} has the property of predicting two different type-II seesaw mass matrices mediated by the respective heavy triplets and has the ability that one of them can dominate over the other. In fact this type-II seesaw dominance property has been utilized in the original model
\cite{Ma-Us:1998} with quasi-degenerate neutrino masses.
It is, therefore, quite pertinent to examine whether the  seesaw model \cite{Ma-Us:1998} can
 fit the current neutrino data \cite{nudata,Forero:2014,Esteban:2018} for hierarchical neutrino masses satisfying the cosmological  bounds \cite{Planck15,Sunny:2018} while
  successfully predicting the observed baryon asymmetry of the
 universe. 

Supersymmetric type-I seesaw leptogenesis  with RHN mass scale $M_N \ge 10^9$ GeV \cite{DI:2002} is known to predict over-production of gravitinos in the early universe affecting relic abundance of light elements. Resolution of gravitino problem \cite{Khlopov} in the supersymmetric triplet seesaw model  has been discussed  \cite{Ham-Ma-Us:2001}. A clear advantage of  non-supersymmetric (non-SUSY) leptogenesis models including \cite{Ma-Us:1998} is the absence of the gravitino problem ensuring  cosmologically safe relic abundance \cite{Khlopov}. 
On the other hand, the SM Higgs mass ($m_{\phi}$)  in \cite{Ma-Us:1998} is not protected against radiative correction  which is likely to give $\delta m_{\phi} \simeq 10^{13}$ GeV $=$ the heaviest triplet mass in the model \cite{Drees:1996} tending to destabilise the electroweak gauge hierarchy. This calls for exploring fine-tuned naturalness (or stability) constraint 
on the model parameters as derived in this work  which might restrict such 
correction \cite{Vissani:1998,Casas:2004} not to exceed the Higgs mass.

 As the purely triplet seesaw model \cite{Ma-Us:1998,Sierra:2014tqa} does not have dark matter prediction to explain observed relic density and mass bounds 
 determined by direct and indirect
detection experiments \cite{Akerib:2016,Aprile:2017,Aprile:2018,Cui:2017,wmap},
it would enhance the model capabilities if  the dark matter phenomena  can be accommodated in its simple minimal
extension as suggested in this work.

 Despite the presence of two
heavy Higgs triplets,  we note that the renormalisation group running renders the  Higgs quartic coupling in the model \cite{Ma-Us:1998,Sierra:2014tqa} to
 acquire negative values in the interval  $|\phi|=(5\times 10^{9}-10^{13})$ GeV leading to vacuum instability \cite{Espinosa,Lebedev}. Noting that it is a natural compulsion to guarantee  vacuum stability of the scalar potential in any of the model applications such as neutrino masses, baryon asymmetry and dark matter, we have shown in this work how the
 minimally extended model with scalar singlet dark matter ensures such a stability.

Compared to earlier works on purely triplet seesaw ansatz \cite{Ma-Us:1998,Sierra:2014tqa}, 
in this work
 we have fitted the most recent neutrino data \cite{nudata,Forero:2014,Esteban:2018}  including $\theta_{23}$ in the second octant
and large Dirac CP-phase ($\delta\simeq 214^{\circ}$). We have further exposed the success of the model potential to be compatible with recent cosmological bounds determined from Planck satellite data \cite{Planck15,Sunny:2018}. In  our approach, the dominant of the two matrices being  completely determined from neutrino data, provides known values of lepton flavour and number violating couplings occuring in the CP-asymmetry parameters. In addition, optimal or randomised  
phase differences between the elements of the two matrices provide a rich structure for CP- asymmetry parameters for unflavoured and flavoured leptogenesis. 
Using  our model CP-asymmetry inputs, we find solutions of  Boltzmann equations successfully predicting baryon asymmetry of the Universe in the unflavoured as well as two-flavoured regimes.

Highlights of the present work  are 
\begin{itemize}
\item{The two-Higgs triplet seesaw model \cite{Ma-Us:1998} is found to fit the most recent neutrino data
  including $\theta_{23}$ in the second octant and large Dirac
  CP-phases  for both normal ordering (NO) and
  inverted ordering (IO) of neutrino masses in concordance with cosmological bounds determined from Planck satellite measurements \cite{Planck15,Sunny:2018}}.
  
\item{ The model ansatz for CP-asymmetry and solutions of Boltzmann equations
are found to predict the observed value of baryon asymmetry of the Universe  
in the case of unflavoured (two-flavoured) leptogenesis  for the lighter triplet mass values $M_{\Delta_2} \simeq
{\cal O} (10^{12})$ GeV ($M_{\Delta_2}\simeq {\cal O}(10^{11})$ GeV )  with corresponding values of lepton number violating coupling $\mu_{\Delta_2}$ in each case.} 
\item{ Whereas the original model \cite{Ma-Us:1998} does not have dark matter (DM), a  simple extension of the model is found to predict  a real scalar singlet WIMP \cite{WIMP} dark matter \cite{GAMBIT} in agreement with observed relic density and direct detection measurements which set the lower bound $m_{\xi}=750$ GeV.} 
\item{This real scalar DM is  also found to remove the vacuum instability
of the scalar potential existing in
the original model.}
\item{When the vacuum stability constraint is combined with those due to relic density and direct detection measurements, this real scalar singlet mass limit is pushed from $m_{\xi}=750$ GeV to $m_{\xi}=1.3$ GeV which is verifiable by ongoing experiments.}
\item{Despite the two heavy triplet scalar masses, the model parameters are noted to satisfy fine-tuned conditions necessary for the radiative stability of the
standard Higgs mass.} 
 \item{Using the two-Higgs triplet model \cite{Ma-Us:1998} and its further simple
   extension,  we have thus successfully addressed four important issues
   confronting the standard model: neutrino masses and mixings within cosmological
   bound, baryon asymmetry of the universe, dark matter, and vacuum stability of the scalar potential. In addition, we have derived necessary constraint on the model parameters for  the radiative stability of standard Higgs mass.}
\end{itemize}

This paper is organised in the following manner.
In Sec.\ref{sec:model} we discuss the triplet leptogenesis model \cite{Ma-Us:1998}. In
Sec.\ref{sec:nufit} we discuss how the current neutrino data is fitted by
two-triplet generated type-II seesaw formula with  NO or IO
masses consistent with cosmological bound where we also
  derive possible values
of  the scalar triplet masses and trilinear couplings for leptogenesis. Prediction of baryon
asymmetry with detailed numerical analyses supported by proper graphical representation is presented in Sec.\ref{sec:bauest}.
Its subsections deal with different regimes of leptogenesis taking into account various schemes to choose the data set for 
leptogenesis calculation. 
Extension of the model to accommodate  dark matter
is discussed in Sec.\ref{sec:dmvac}, Sec.\ref{sec:redm}, and Sec.\ref{sec:relic}. In Sec.\ref{sec:vstab}  we discuss the issue of 
vacuum instability of the scalar potential and its resolution with a summary on DM investigation in Sec.\ref{sec:sumDM}. 
In Sec.\ref{sec:nat}
we discuss Higgs mass stability constraint on the model parameters.
The work is summarised  in Sec.\ref{sec:sum}. Explanation and definition of different functions and parameters associated with Boltzmann equations are given in Sec.\ref{sec:ap1}, Sec.\ref{sec:ap2} and Sec.\ref{sec:ap3}  while renormalisation group equations for gauge, scalar and top-quark Yukawa couplings have been discussed in Sec.\ref{sec:rge} of the Appendix.
\section{The two-triplet model}\label{sec:model}
Whereas in majority of models the RHNs have been found instrumental in
theories of neutrino masses and leptogenesis,  along with type-II dominance seesaw ansatz for
neutrino mass  possible realisation of leptogenesis in
two-triplet extensions of standard model has
been proposed in \cite{Ma-Us:1998}  without any RHN.
 
In other triplet seesaw and leptogenesis models \cite{Ham-gs:2003,Sierra:2014tqa,cps:2019}, heavy RHNs are needed for loop mediation even though the triplet in collaboration with SM Higgs doublet is capable of explaining neutrino masses. But this model \cite{Ma-Us:1998} does not need any RHN to implement both the phenomena: neutrino mass and leptogenesis for successful prediction of baryon asymmetry of the Universe.  
The resulting scalar potential in this model has different terms depending upon the SM Higgs field values $\mu=|\phi|$ as discussed in Sec.\ref{sec:dmvac}.
The charges of fermions and scalars have been also defined in Sec.\ref{sec:dmvac} in Table \ref{tab:smext}.
 Thus, in addition to the usual SM interactions and their modifications, the nonstandard 
part of the Lagrangian that contributes to type-II seesaw and leptogenesis is
\beq
-{\cal L}_{ext} =\sum_{\alpha=1}^2\left ((D_{\mu}{\vec {\Delta}}_{\alpha})^{\dagger}.(D^{\mu}{\vec {\Delta}}_{\alpha})- M^2_{\Delta_\alpha}
Tr(\Delta_\alpha^\dagger \Delta_\alpha) +  [\frac{1}{2}y^{(\alpha)}_{ij} L^T_i Ci\tau_2 \Delta_\alpha L_j 
 - \mu_{\Delta_\alpha} \phi^T i\tau_2 \Delta_\alpha \phi
+h.c.]\right). \label{Yukhiggs}
\eeq
Here $i,j=1,2,3$ denote the three lepton flavors represented by the lepton
doublets $L_i$ but $\alpha=1,2$ denote  the two scalar triplets. $M_{\Delta_\alpha}=$ mass of the triplet $\Delta_\alpha$, $y^{(\alpha)}_{ij}=$ 
Majorana coupling of $\Delta_\alpha$ with $L_i$ and $L_j$ and $\mu_{\Delta_\alpha}=$  lepton-number violating trilinear coupling of $\Delta_\alpha$ with
standard Higgs doublet $\phi$.

Defining the induced triplet VEVs $V_{L_{\alpha}} (\alpha=1,2)$
\beq
V_{L_{\alpha}}= \frac{\mu_{\Delta_\alpha} v^2}{2 M_{\Delta_\alpha}^2}, \label{vlk} 
\eeq
the formula for the light neutrino mass matrix $m_{\nu}$ is
\bea
m_{\nu} &=& 2 y^{(1)} V_{L_{1}}+2 y^{(2)} V_{L_{2}}, \nonumber\\
        &=& y^{(1)}\frac{\mu_{\Delta_1} v^2}{ M_{\Delta_1}^2}  +  y^{(2)} \frac{\mu_{\Delta_2} v^2}{ M_{\Delta_2}^2} \nonumber\\
&\equiv& m^{(1)}_{\nu}+m^{(2)}_{\nu}. \label{mnu_t2}
\eea    
Here $v=246$ GeV, the standard Higgs vacuum expectation value (VEV).

\subsection{CP asymmetry}\label{cpasy}
\begin{figure}
\begin{center}
 \includegraphics[width=4cm,height=3cm]{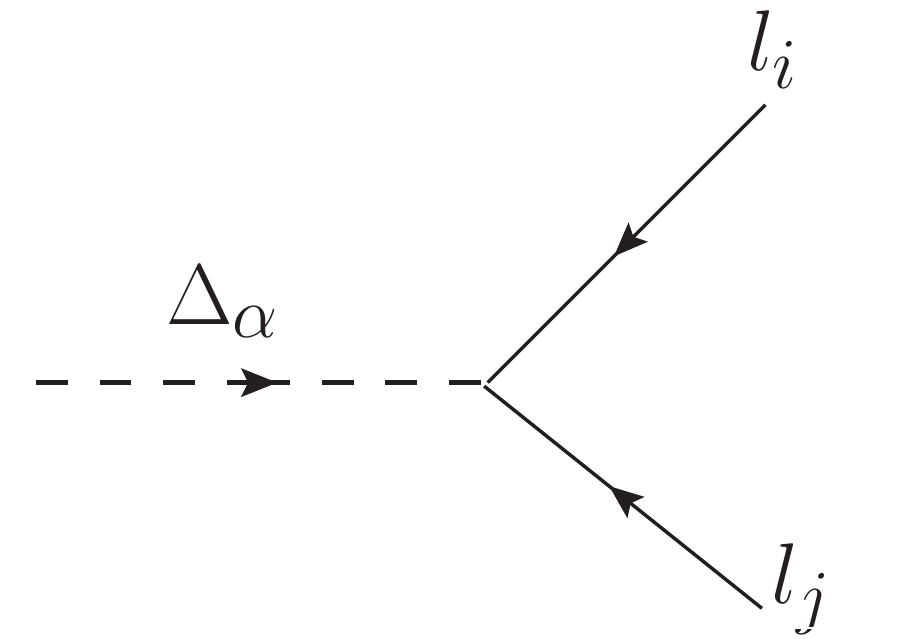}
 \includegraphics[width=5cm,height=3cm]{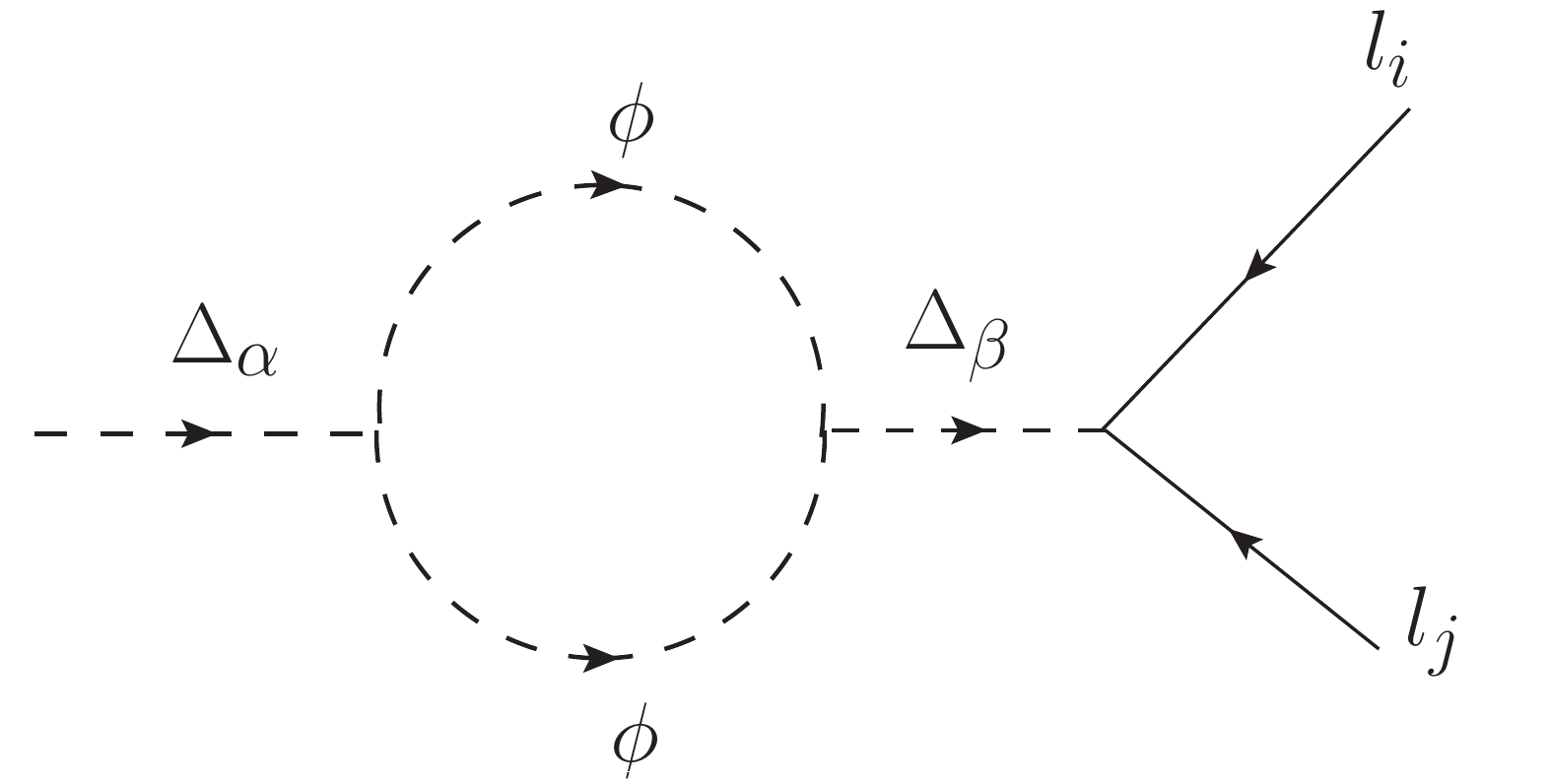}
 \includegraphics[width=5cm,height=3cm]{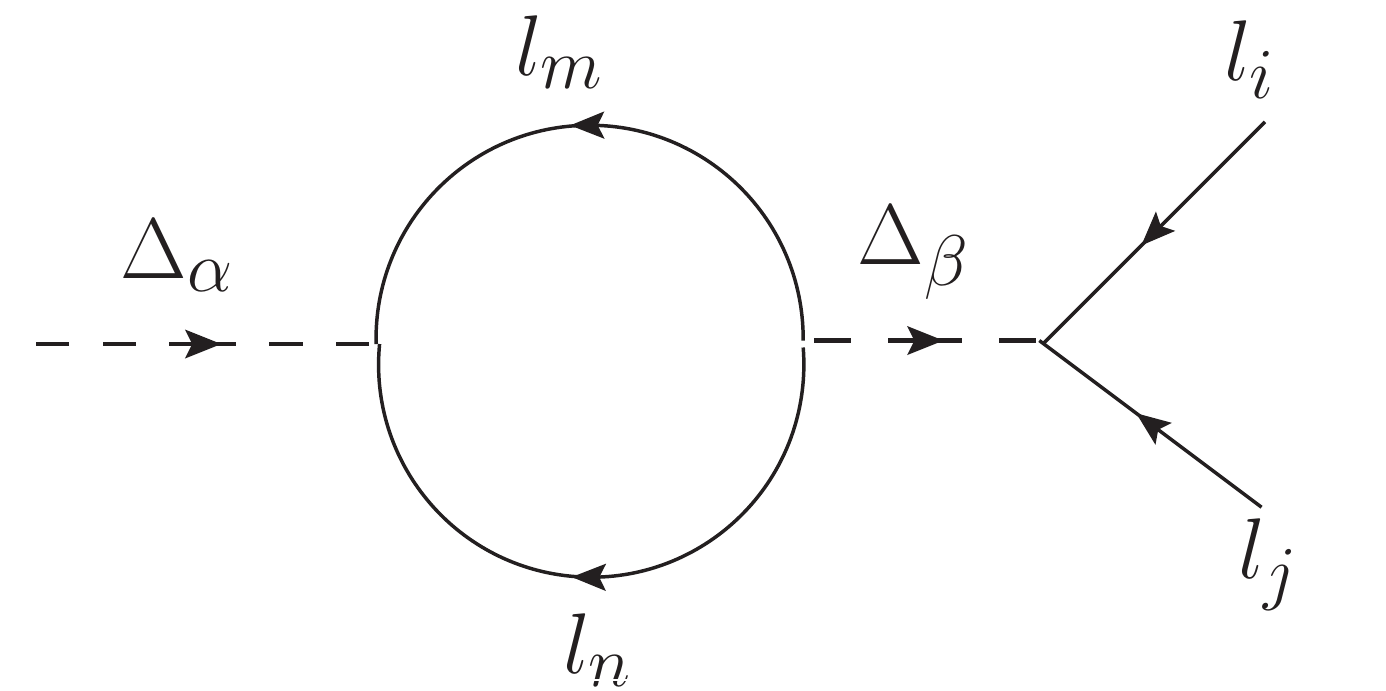}
 \caption{Tree level and one loop Feynman diagrams for a triplet decaying to bi-leptons in case of flavoured leptogenesis}
 \label{feyn-flav}
 \end{center}
\end{figure}

Now we discuss about the source of lepton number violation and generation of CP asymmetry. It is clear from the interaction 
lagrangian (eq.\ref{Yukhiggs}) that lepton number violation (LNV)
is possible due to the coexistence of the Higgs triplet-bilepton Yukawa matrix $y$ along with the trilinear coupling 
$\mu_{\Delta_\alpha}(\alpha=1,2)$. Since heavy RHNs are absent in the theory, the entire CP asymmetry is created due to the 
decay of the heavy scalar triplets. In tree level the scalar triplets can decay to bi-leptons as well as two SM Higgs. 
The corresponding branching ratios of $\Delta_\alpha$ decay to leptons and SM Higgs are respectively
\begin{eqnarray}
&& B^\alpha_l=\sum_{i=e,\mu,\tau} B^\alpha_{l_{i}} = \sum_{i,j=e,\mu,\tau} B^\alpha_{l_{ij}} =\sum_{i,j=e,\mu,\tau} 
\frac{M_{\Delta_\alpha}}{8\pi \Gamma^{tot}_{\Delta_\alpha}} |y^{(\alpha)}_{ij}|^2 ~{\rm and}~\\
&& B^\alpha_\phi =\frac{|\mu_{\Delta_\alpha}|^2}{8\pi M_\Delta \Gamma^{tot}_{\Delta_\alpha}}, 
\end{eqnarray}
which obviously satisfy $B^\alpha_l+B^\alpha_\phi=1$,
where $\Gamma^{tot}_{\Delta_\alpha}$ is the total decay width of $\Delta_\alpha$, given by
\begin{equation}
 \Gamma^{tot}_{\Delta_\alpha}=\frac{M_{\Delta_\alpha}}{8\pi}\Big( \sum_{i,j} |y^{(\alpha)}_{ij}|^2 +
 \frac{|\mu_{\Delta_\alpha}|^2}{M^2_{\Delta_\alpha}} \Big)~.
\end{equation}
The decay to 
bi-leptons can also occur due to one loop process where the loop is mediated by either SM Higgs or leptons as shown in 
Fig.\ref{feyn-flav}\footnote{It is to be noted that we wish to study the leptogenesis phenomena in both the regimes above and below $10^{12}$ GeV. Lepton flavours become distinguishable below $10^{12}$ GeV. The resulting leptogenesis is termed as flavoured leptogenesis. Thus in the Feynman diagram we have denoted the lepton flavors with different indices. }.
The CP asymmetry arises due to the interference of the tree level contribution with that of the one loop wave function diagrams as shown in Fig.\ref{feyn-flav}. Again the total flavoured CP asymmetry consists of two pieces, the scalar loop gives gives rise to a part which violates both lepton number and flavour  whereas the one with lepton loop gives rise to only flavour violating asymmetry. So the total flavoured asymmetry (taking into account both lepton number + flavour violations, denoted by ($\not L,\not F$),   and only lepton flavour violation denoted by $\not F$) is given by
\begin{equation}
\epsilon^{l_i}_{\Delta_\alpha}= \epsilon^{l_i(\not L,\not F)}_{\Delta_\alpha}+ \epsilon^{l_i(\not F)}_{\Delta_\alpha}~.
\label{flav_ep}
\end{equation}
In the present work we have considered only two  heavy triplets corresponding to $\alpha=1,2$. It is well known that lepton asymmetry
will be produced mostly due to the decay of the lighter triplet which is $\Delta_2$ in our case while the asymmetry due to the 
heavier one will be washed out. Dynamical origin of neutrino masses is thus made consistent with the dominance of the $\Delta_2$ to neutrino mass matrix over $\Delta_1$.  Therefore, in all the future expressions of CP asymmetries, branching ratios, decay widths we will omit the index on the scalar triplet.
It is to be understood that $\epsilon_\Delta \equiv \epsilon_{\Delta_2},B_l \equiv B^2_l, B_\phi \equiv B^2_\phi$. The above mentioned two pieces of CP asymmetry
in eq.(\ref{flav_ep}) arising due to
$\Delta_2$ decay are given by
\begin{eqnarray}
&& \epsilon^{l_i(\not L,\not F)}_{\Delta} = \frac{1}{2 \pi} \frac{Im \left\{ \sum\limits_{n}^{} \left( y^{(2)} \right)^\ast_{ni} 
 \left({y^{(1)}}\right)_{ni} \mu^\ast_{\Delta_2} \mu_{\Delta_1} \right \}}{ M^2_{\Delta_2} Tr (  y^{(2)}  {y^(2)}^\dagger ) +|\mu_{\Delta_2}|^2 }g (x_{12}), \label{flav_ep1}\\
&& \epsilon^{l_i(\not F)}_{\Delta} = \frac{1}{2 \pi} \frac{Im \left\{ ( {y^{(2)}}^\dagger {y^{(1)}} )_{ii} Tr(y^{(2)} {y^{(1)}}^\dagger)  \right \}}{ M^2_{\Delta_2} Tr (  y^{(2)}  {y^(2)}^\dagger ) +|\mu_{\Delta_2}|^2 }g (x_{12}),\label{flav_ep2}
\end{eqnarray}
where  \\
\bea
 g(x_{\alpha\beta})&=&\frac{x_{\alpha\beta}(1-x_{\alpha\beta})}{(1-x_{\alpha\beta})^2 + x_{\alpha\beta}y }, \nonumber\\ 
x_{\alpha\beta}&=&\frac{M^2_{\Delta_\alpha}}{M^2_{\Delta_\beta}}, \nonumber\\
 y&=&\left( \frac{\Gamma^{tot}_{\Delta_\beta}}{M_{\Delta_\beta}} \right)^2.\label{eqg}
\eea
The indices $n,i$ in the above expressions of CP asymmetries stand for the lepton flavour indices $(e,\mu,\tau)$. It can be easily understood 
that if we sum over the flavour indices, the second piece of the CP asymmetry which is solely due to  flavour violation vanishes identically, i.e
\begin{equation}
 \sum_{i=e,\mu,\tau} \epsilon^{l_i(\not F)}_{\Delta} =0~.
\end{equation}
This CP asymmetry parameter survives only in the case of flavoured leptogenesis. Since this asymmetry does not involve any lepton number
violation, some times it is called purely flavoured asymmetry and the corresponding leptogenesis scenario which is dominated by the 
flavour violating CP asymmetry $(\epsilon^{l_i(\not F)}_{\Delta} \gg \epsilon^{l_i(\not L,\not F)}_{\Delta})$ is referred to as 
purely flavoured (PFL) leptogenesis. The condition to get PFL leptogenesis can be shown to be \cite{Sierra:2014tqa}
\begin{equation}
\mu^\ast_{\Delta_2} \mu_{\Delta_1} \ll M^2_{\Delta_2} Tr(y^{(2)} {y^{(1)}}^\dagger)~.
\end{equation}
Therefore, consistent with dominance of $m_{\nu}^{(2)}$ over  $m_{\nu}^{(1)}$ and the condition that $M_{\Delta_2} < M_{\Delta_1}\simeq \mu_{\Delta_1}$,  PFL is not always guaranteed in flavoured leptogenesis regime. During our actual numerical analysis of flavoured leptogenesis we will examine whether PFL is achievable in our case.  \\\\
The following set of parameters have been used for the model
predictions in \cite{Ma-Us:1998} 
\begin{eqnarray}
&& y^{(1)}=1,y^{(2)}=0.1, |\mu_{\Delta_1}|=10^{13} {\rm GeV},|\mu_{\Delta_2}|=2\times 10^{12} {\rm GeV},\nonumber\\
&& M_{\Delta_1}=3\times 10^{13}{\rm  GeV},  M_{\Delta_1}= 10^{13}
 {\rm GeV}, \label{parmaus}
\end{eqnarray}
consistent with QD neutrino mass eigen values 
\beq
m_1\sim m_2 \sim m_3\sim 1.2 eV \label{mimaus} 
\eeq
This choice leads to an interesting prediction for observable neutrinoless double beta
decay close to the current experimental limit
\cite{bbexpt}.  KATRIN \cite{KATRIN} experimental
search program has recently set the neutrino mass limit to
$m_0 \le 1$ eV.  

However, recent estimations derived from Planck satellite data
appear to constrain the QD spectrum considerably
\cite{Planck15,Sunny:2018} as stated through eq.(\ref{eq:cb}) and eq.(\ref{eq:cbnew}). In addition  the recent neutrino data has revealed
certain significant interesting changes over previous results with the
atmospheric neutrino mixing angle in the second octant $\theta_{23}\ge
45^{\circ}$ and the Dirac CP phase  $\delta \simeq
214^{\circ}$. Success of a class of type-II
seesaw dominant models  but with RHN loop mediated triplet
leptogenesis  has been investigated along with
predictions of new CP-asymmetry formulas in concordance with the recent oscillation data \cite{cps:2019} and cosmological bound \cite{Planck15}.      
It is thus  quite important to  examine whether purely triplet seesaw
and leptogenesis predictions without RHNs could be compatible with the recent
data, cosmological bound \cite{Planck15},
baryon asymmetry \cite{Planck15,Sunny:2018} and
dark matter while ensuring vacuum stability of the scalar potential.   \\ 
In order to fit the recent neutrino data by the present formulation, 
we use the approximation that the type-II seesaw formula generated by
lighter of the two triplets (with $M_{\Delta_2}\ll M_{\Delta_1}$) has
dominant contribution, i.e  
\beq
m_{\nu}\simeq m^{(2)}_{\nu}=m_{\nu}^{(DATA)}.  \label{eq:apmnu}
\eeq
$m_\nu^{(1)}$ and $m_\nu^{(2)}$ matrices can always be represented as
\begin{equation}
(m_\nu^{(1)})_{ij}=|{m^{(1)}_\nu}_{ij} | e^{i\phi^{(1)}_{ij}} ,~ (m_\nu^{(2)})_{ij}=|{m^{(2)}_\nu}_{ij} | e^{i\phi^{(2)}_{ij}},
\end{equation}
which in turn gives
\begin{equation}
\frac{(m_\nu^{(1)})_{ij}}{(m_\nu^{(2)})_{ij}}= \frac{|{m^{(1)}_\nu}_{ij} |}{|{m^{(1)}_\nu}_{ij} |}   e^{i(\phi^{(1)}_{ij}-\phi^{(2)}_{ij})} ~.
\end{equation}
Under this assumption  each element of $m_\nu^{(1)}$ is connected to the corresponding element of $m_\nu^{(2)}$ through a  multiplicative factor
which is a complex number in general and can be represented as
\begin{equation}
(m_\nu^{(1)})_{ij} =F_{ij} e^{i(\phi^{(1)}_{ij}-\phi^{(2)}_{ij})} (m_\nu^{(2)})_{ij} \label{mnu1_2}
\end{equation}
where $F_{ij}=|{m^{(1)}_\nu}_{ij} |/|{m^{(2)}_\nu}_{ij} |$ which is a real ratio.
In order to ensure the dominance of $m_\nu^{(2)}$ over $m_\nu^{(1)}$ we assume $F_{ij} \le 0.1$ for all $i,j$.
In general $F_{ij}$s and $(\phi^{(1)}_{ij}-\phi^{(2)}_{ij})$ can have different values for different combinations of $i$ and $j$.
Using eq.(\ref{mnu_t2}) and very near equality of $m^{(2)}_{\nu}$ with $m_{\nu}^{(DATA)}$, the numerical value of the Yukawa coupling matrix $y^{(2)}$ 
can be easily found for a known set of $(M_{\Delta_2},\mu_{\Delta_2})$. Again for any random value of the ratio $F_{ij}$ and the phase difference
$(\phi^{(1)}_{ij}-\phi^{(2)}_{ij})$, the other Yukawa coupling  $ y^{(1)}$ can be computed from eq.(\ref{mnu1_2}) provided the corresponding
trilinear coupling $(\mu_{\Delta_1})$ and the triplet mass $(M_{\Delta_1})$ are already known. We may assume some numerical values 
$(M_{\Delta_2},\mu_{\Delta_2})$ depending upon the regime of leptogenesis(flavoured/unflavoured) and then $(M_{\Delta_1},\mu_{\Delta_1})$
can be accordingly chosen to keep $m^{(1)}_\nu$ sub-dominant, which in turn requires $M_{\Delta_1} \gg M_{\Delta_2}$. This also ensures the 
contribution of $\Delta_1$ towards leptogenesis to be negligible. Knowledge of all these parameters along with some random value 
of the phase difference and the ratio enables us to calculate the flavoured CP asymmetry(eq.(\ref{flav_ep1},\ref{flav_ep2})) parameters.
However in the unflavoured regime the purely flavoured CP asymmetry part vanishes and we are left with (lepton number + flavour ) violating
part which can be represented in terms of the experimental value of light neutrino mass matrix (denoted as $m_\nu(=m^{\rm DATA}_\nu)$)
as
\begin{eqnarray}
\epsilon^l_\Delta &=& \sum_i  \epsilon^{l_i(\not L,\not F)}_{\Delta} \nonumber\\
                &=& \frac{ M^2_{\Delta_1} M^2_{\Delta_2} }{2\pi v^4} 
                    \frac{\sum\limits_{ij} F_{ij} |(m_\nu)_{ij}|^2 \sin (\phi^{(1)}_{ij}-\phi^{(2)}_{ij}) }
                    {M^2_{\Delta_2} Tr (  y^{(2)}  {y^{(2)}}^\dagger ) +|\mu_{\Delta_2}|^2 }g (x_{12})  \label{unflav-asy1} \\                    
                 &\simeq& \frac{ M^2_{\Delta_1} M^2_{\Delta_2} }{16 \pi^2 v^4}    
                     \frac{\sum\limits_{ij} F_{ij} |(m_\nu)_{ij}|^2 \sin (\phi^{(1)}_{ij}-\phi^{(2)}_{ij}) }
                     { (M^2_{\Delta_1} -M^2_{\Delta_2}) } \left( \frac{M_{\Delta_2}}{\Gamma^{tot}_{\Delta_2}} \right) ~.\label{unflav-asy2} 
\end{eqnarray}
Then for  $M_{\Delta_1}\gg M_{\Delta_2}$, the CP-asymmetry is 
\beq
\epsilon^l_\Delta=\frac{ M^2_{\Delta_2} }{16 \pi^2 v^4}    
                     \sum\limits_{ij} F_{ij} |(m_\nu)_{ij}|^2 \sin (\phi^{(1)}_{ij}-\phi^{(2)}_{ij})  \left( \frac{M_{\Delta_2}}{\Gamma^{tot}_{\Delta_2}} \right) ~.\label{unflav-asy3} 
\eeq

\subsection{Boltzmann equations for leptogenesis}
Boltzmann equations are used to track the evolution of the particle
asymmetries in the early universe where the hot plasma is composed of large number of particle 
species resulting in numerous reactions. However there is no need take into
account all of them. Only those reactions are important whose rates at that temperature
are comparable to the Hubble rate (i.e $\Gamma (T) \sim H(T)$).

Lepton number violation is embedded in the interaction lagrangian (eq.(\ref{Yukhiggs})) through the Majorana type coupling of the 
triplet Higgs with bi-leptons. Lepton number is violated by two units whenever $\Delta$ decays to $(l_i,l_j)$. As it is a baryon
number conserving process, $(B-L)$ is also violated by two units. So our aim is to find out the evolution of abundance 
of $(B-L)$ which at later stage gets converted into baryon number through sphaleron transition process. It is worthwhile to 
mention that during the sphaleron process the quantity $(B-L)$ $(B/3-L_i)$ is conserved in case of unflavoured (flavoured) leptogenesis.
Accordingly the asymmetry parameter whose evolution with temperature has to be traced is $(B-L)$(($B/3-L_i$)) for unflavoured (flavoured) leptogenesis scenario. 
It is not possible to compute the evolution of $(B-L)$ or $(B/3-L_i)$ independently as
it includes other parameters which also evolve with temperature. In
fact the Boltzmann equations consist of a set of coupled  differential equations which have to be solved 
simultaneously to find solution for any of the variables. In this purely triplet leptogenesis model the asymmetry can only be
generated by the decay of heavy scalar triplet. Therefore, along with the first order differential of $(B-L)$ or $(B/3-L_i)$,  the 
Boltzmann equations contain first order differentials of scalar triplet density and scalar triplet asymmetry.
This scalar triplet asymmetry arises due to the fact that $\Delta_2$ and $\Delta_2^{\dagger}$ are not self-conjugate.
The right hand side of relevant Boltzmann equations contains interaction terms that tend to change the density of the corresponding 
variable. Considering all such interactions, the network of lepton flavour dependent  coupled Boltzmann 
equations are \cite{Sierra:2011ab,Sierra:2014tqa,cps:2019}

\begin{eqnarray}
&& \dot{Y}_\Sigma=-\Big(\frac{Y_\Sigma}{Y_\Sigma^{eq}}-1 \Big )\gamma_D -2\Big[\Big(\frac{Y_\Sigma}{Y_\Sigma^{eq}}\Big )^2-1 \Big ]\gamma_A \label{boltz_big2},\\
&& \dot{Y}_{\Delta_\Delta}=-\Big[\frac{Y_{\Delta_\Delta}}{Y_\Sigma^{eq}}-\sum_{k} \Big(\sum_i B_{l_i}C_{ik}^l -B_\phi C_k^\phi \Big)\frac{Y_{\Delta_k}}{Y_l^{eq}}\Big ]\gamma_D\label{boltz_big3},\\
&& \dot{Y}_{\Delta_{B/3-L_i}}= 
- \Big[ \Big( \frac{Y_\Sigma}{Y_\Sigma^{eq}} -1 \big) \epsilon^{l_i}_\Delta -2 \sum_j \Big( \frac{Y_{\Delta_\Delta}}{Y_\Sigma^{eq}} -
\frac{1}{2} \sum_k C^l_{ijk} \frac{Y_{\Delta_k}}{Y_l^{eq}}\Big)B_{l_{ij}} \Big]\gamma_D \nonumber\\
&&\hspace{1.4cm}-2 \sum_{j,k}\Big(C^\phi_k +\frac{1}{2}C^l_{ijk}\Big)\frac{Y_{\Delta_k}}{Y^{eq}_l}\Big( \gamma^{\prime \phi \phi}_{l_i l_j}
+ \gamma^{\phi l_j}_{\phi l_i} \Big)
 -\sum_{j,m,n,k} C^l_{ijmnk} \frac{Y_{\Delta_k}}{Y^{eq}_l}\Big( \gamma^{\prime l_n l_m}_{l_il_j} 
+\gamma^{l_m l_j}_{l_il_n} \Big)~~.\nonumber\\
\label{boltz_big4}
\end{eqnarray}
Notational conventions adopted here are as follows: $Y_{\Delta_X}$ stands for the ratio of number density (or difference of number density) 
to the entropy density, i.e $Y_{\Delta_X}=\frac{n_X-n_{\bar{X}}}{s}$, where $n_X~(n_{\bar{X}})$ is
the $X~({\bar X})$ number density. Standard mathematical forms of equilibrium number densities of different particle species $X~(\bar{X})$ are given in 
 Sec.\ref{sec:ap1} of Appendix. It is implied that the variables of the differential equations $(Y_{\Delta_\Delta},Y_\Sigma,Y_{\Delta_{ B/3-L_i}})$ are function of $z=M_\Delta/T$.
Here $\dot{Y}_X$ denotes $\dot{Y}_X\equiv\dot{Y}_X(z)=s(z)H(z)\frac{dY_X(z)}{dz}$. The scalar triplet density and asymmetry are denoted as 
$\Sigma=\Delta+\Delta^\dagger$ and $\Delta_\Delta=\Delta-\Delta^\dagger$, respectively. Superscript \textquoteleft$eq$\textquoteright~ denotes the equilibrium value
of the corresponding quantity. Functional forms of all such equilibrium densities are presented in Sec.\ref{sec:ap1} of Appendix.
The total reaction density of the triplet including its decay and inverse decay to lepton pair or scalars  is represented as $\gamma_D$.
The gauge induced $2\leftrightarrow2$ scattering of triplets to fermions, scalars and gauge bosons is denoted by $\gamma_A$.
Lepton flavour and number ($(\Delta L=2)$) violating Yukawa scalar induced $s$ channel $(\phi \phi \leftrightarrow \bar{l_i} \bar{l_j})$ and $t$ channel
$(\phi l_j \leftrightarrow \bar{\phi}\bar{l_i} )$ scattering related reaction densities are denoted as 
$\gamma^{\phi \phi}_{l_i l_j}$ and $\gamma^{\phi l_j}_{\phi l_i}$, respectively. Similarly reaction densities related to
Yukawa induced triplet mediated lepton flavour violating $2\leftrightarrow2$ $s$ channel and $t$ channel processes are 
denoted by $\gamma^{l_nl_m}_{l_i l_j}$ and $\gamma^{l_j l_m}_{l_il_n}$. The primed $s$ channel reaction densities
are given by $\gamma^\prime=\gamma-\gamma^{\rm on~ shell}$.
We present the explicit expressions of these reaction densities in  Sec.\ref{sec:ap2} of Appendix.
The asymmetry coupling matrices $C^l_{ijk}$ and $C^l_{ijmnk}$ are defined as \cite{Sierra:2014tqa}
\begin{eqnarray}
&&C^l_{ijk}=C^l_{ik} + C^l_{jk},\nonumber\\
&&C^l_{ijmnk}=C^l_{ik} + C^l_{jk}-C^l_{mk}-C^l_{nk},
\end{eqnarray}
where $C^l$ matrix connects the asymmetry of lepton doublets with that of $B/3-L_i$ whereas $C^\phi$ establishes a relation between the 
asymmetry of scalar triplet and $B/3-L_i$, i.e,
\begin{eqnarray}
&& Y_{\Delta_{l_i}}=-\sum_k C^l_{ik} Y_{\Delta_k}\nonumber\\
&& Y_{\Delta_\phi}=-\sum_k C^\phi _k Y_{\Delta_k}
\end{eqnarray}
where $Y_{\Delta_k}$ represents the components of the asymmetry vector $\vec{Y}_{\Delta}$,
\begin{equation}
\vec{Y}_{\Delta} \equiv  (Y_{\Delta_\Delta}, Y_\Delta{_{B/3-L_k}} )^T .
\end{equation}
The generation index $k$ in the above equation runs from $1$ to $3$ for fully (three) flavoured leptogenesis
whereas it takes values $1,~2$ for two flavoured leptogenesis which dictates the corresponding $\vec{Y}_{\Delta}$ will be a column matrix
with four or three entries, respectively. $C^l$ and $C^\phi$ matrices are determined from chemical equilibrium conditions. 
Their detailed structure and dimensionality in different temperature regimes are given in Sec.\ref{sec:ap3} of Appendix. The flavoured Boltzmann equations
presented in eqs.(\ref{boltz_big2},\ref{boltz_big3},\ref{boltz_big4}) have to be solved simultaneously upto a large value of $z$ (where the asymmetry gets frozen).
Then the final value of Baryon asymmetry parameter is computed to be 
\begin{equation}
Y_B \equiv Y_{\Delta_B}=3 \times \frac{12}{37}\sum_i  Y_{\Delta_{B/3-L_i}} \label{yb}
\end{equation}
where the factor $3$ takes care of different $SU(2)$ degrees of freedom of the scalar triplet. When the mass of the decaying heavy 
particle (or equivalently the temperature for asymmetry generation) exceeds $10^{12}$ GeV, the charged lepton Yukawa interactions 
go out of equilibrium and, as a result, the lepton flavours lose their distinguishability. Thus we need not treat the flavours
separately and as a result corresponding Boltzmann equations are free of lepton flavour index. 
This variant of leptogenesis is referred to as the unflavoured leptogenesis and the set of Boltzmann equations applicable to this case
are obtained through modifications of eq.(\ref{boltz_big2}-\ref{boltz_big4})\cite{Sierra:2014tqa} as
\begin{eqnarray}
&& \dot{Y}_\Sigma=-\Big(\frac{Y_\Sigma}{Y_\Sigma^{eq}}-1 \Big )\gamma_D -2\Big[\Big(\frac{Y_\Sigma}{Y_\Sigma^{eq}}\Big )^2-1 \Big]\gamma_A \label{boltz_uf1}\\ 
&& \dot{Y}_{\Delta_\Delta}=\Big[\frac{Y_{\Delta_\Delta}}{Y_\Sigma^{eq}}-\sum_{k} \Big( B_l C_k^l -B_\phi C_k^\phi \Big)\frac{Y_{\Delta_k}}{Y_l^{eq}}\big]\gamma_D,\label{boltz_uf2}\\ 
&& \dot{Y}_{\Delta_{B-L}}= - \Big[ \Big( \frac{Y_\Sigma}{Y_\Sigma^{eq}} -1 \big) \epsilon^{l}_\Delta -2 \Big( \frac{Y_{\Delta_\Delta}}{Y_\Sigma^{eq}} -
 \sum_k C^l_{k} \frac{Y_{\Delta_k}}{Y_l^{eq}}\Big)B_{l} \Big]\gamma_D 
-2 \sum_k \Big( C^\phi_k +C^l_k \Big ) \frac{Y_{\Delta_k}}{Y^{eq}_l} \Big (\gamma^{\prime \phi\phi}_{ll} +
\gamma^{\phi l}_{\phi l} \Big )
, \label{boltz_uf3}
\end{eqnarray}
where $\epsilon^{l}_\Delta~(=\sum_i \epsilon^{l_i}_\Delta)$ is the flavor summed or unflavoured CP asymmetry parameter and the asymmetry
vector $\vec{Y}_\Delta$ has now been reduced to a column vector with only two entries,$\vec{Y}_\Delta^T=(Y_{\Delta_\Delta},Y_{\Delta_{B-L}})$.
Thus, in this case too, the final baryon asymmetry is computed using the simple formula of eq.(\ref{yb}) where the whole quantity under the summation
should be replaced by a single asymmetry parameter $Y_{\Delta_{B-L}}$.

\subsection{Flavour decoherence and different regimes of leptogenesis} \label{fl-decoh}
 Whether the lepton flavours have to be treated separately at a certain temperature is decided
completely by the phenomenon of flavour decoherence \cite{Sierra:2014tqa}. It is a common 
practice to assume that flavour decoherence sets in as soon as the corresponding  charged lepton Yukawa interaction rate 
exceeds the Hubble rate at that very temperature. 
Along with this assumption, few intricate details of the underlying processes are also taken into account to deal with this flavour 
decoherence issue. In the present model under consideration (SM + two triplets $\Delta_1,\Delta_2$) with $\Delta_2\equiv \Delta$ and $M_{\Delta} < M_{\Delta_1}$, it is logical to assume survival of leptogenesis  caused by the $\Delta$ decay. Then   
the flavour decoherence  is dictated by the competition of two processes:
SM  charged lepton Yukawa interaction and inverse decay of leptons to triplet $\Delta$.
To clarify this statement let us assume that at some temperature $(T_h)$ during the 
evolution of  Universe , the charged lepton Yukawa interaction  is faster than the Hubble rate but slower compared to 
triplet inverse decay $(ll \rightarrow \bar{\Delta})$. 
As a result the charged leptons inverse decay before the triplet can undergo any charged lepton Yukawa interaction.
Even then it is still impossible to differentiate between the lepton flavours.
At some later stage of evolution when the temperature of the thermal bath becomes lower, the charged leptons inverse decay rate by virtue of being Boltzmann 
suppressed  gets reduced further.
Then, at a temperature $T=T_{decoh}$, when the lepton inverse decay rate to ${\bar \Delta}$ becomes less than the
lepton Yukawa interaction rate, the decoherence between the lepton flavours is achieved. 
Thus, between the temperature range $(T_h-T_{decoh})$, the flavour decoherence is not fully achieved, i.e within 
this intermediate temperature regime, it is not totally justified to use flavoured leptogenesis formalism. 
\paragraph{}
The decoherence temperature $(T_{decoh})$ is determined by the mass of the lighter of the two scalar triplets 
$(M_\Delta(=M_{\Delta_2}))$ and the effective decay parameter\cite{Sierra:2014tqa} 
\begin{equation}
\tilde{M}_{\Delta}^{eff}= \tilde{M}_\Delta \sqrt{\frac{1-B_\phi}{B_\phi}},\label{mdt_eff}
\end{equation}
where 
\begin{equation}
\tilde{M}_{\Delta}^2 =|\mu_{\Delta}|^2 \frac{v^4}{M_\Delta^4} Tr [Y
  Y^\dagger], \label{mdt_ex} 
\end{equation}
and $B_\phi$ is branching ratio of  $\Delta \to \phi\phi$.
\paragraph{}
Decoherence is fully achieved when our chosen parameter space satisfies the condition that, at a given temperature, lepton triplet inverse decay rate is slower than the SM charged lepton Yukawa interaction rate. 
Imposition of this condition will lead us to an upper limit on $M_\Delta$ as a function of $\tilde{M}_{\Delta}^{eff}$ which can be expressed as
\begin{equation}
\Gamma_{f_i} \geq B_l \Gamma_\Delta^{tot} \frac{Y_{\Sigma}^{eq}}{Y_l^{eq}} ~~~~~~({\rm with}~f_i=\tau,\mu).
\end{equation}
Here $B_l$ is the branching ratio of dilepton decay rate $\Delta_2\to ll$ that occurs due to Yukawa interaction.
This constraint relation can be translated into constraints over $M_\Delta$ and $\tilde{M}_{\Delta}^{eff}$ as\cite{Sierra:2014tqa}
\begin{eqnarray}
&& M_\Delta \leq 4 \times \Big ( \frac{10^{-3} {\rm eV}}{\tilde{M}_{\Delta}^{eff}} \Big ) \times10^{11} ~~{\rm GeV} ~~~~({\rm fully ~two ~flavoured}),\label{con_f2f}\\
&& M_\Delta \leq 1 \times \Big ( \frac{10^{-3} {\rm eV}}{\tilde{M}_{\Delta}^{eff}} \Big ) \times10^{9} ~~{\rm GeV} ~~~~({\rm fully ~three ~flavoured})~~. \label{con_f3f}
\end{eqnarray}
Following eq.(\ref{con_f2f}) and eq.(\ref{con_f3f}) we can say that, when the mass of the decaying triplet $M_\Delta > 4 \times 10^{11} $
GeV, all the lepton flavours act as a coherent superposition and the corresponding asymmetry generation proceeds through
unflavoured or single flavoured leptogenesis. When the temperature (or equivalently $M_\Delta$) drops below $4 \times 10^{11}$ GeV,
the $\tau$ flavour gets decoupled whereas $e+\mu$ still act indistinguishably. Thus, the coherent superposition is effectively
split into two flavours ($e+\mu$ and $\tau$) and the corresponding leptogenesis phenomena is termed as 2-flavoured (or $\tau$-flavoured) leptogenesis.
Below $10^{9}$ GeV, all the charged lepton Yukawa interactions reach equilibrium and flavour decoherence is fully attained
resulting in 3-flavoured (or fully flavoured) leptogenesis.

\section{Model fitting of neutrino data within cosmological bound}\label{sec:nufit} 
In this section at first we discuss the model capability to fit the most recent neutrino data satisfying the constraint imposed by cosmological bounds \cite{Planck15,Sunny:2018}.
Using the PDG convention \cite{Beringer:2012} we parameterize the
PMNS mixing matrix
\begin{equation}
 U_{\rm{PMNS}}= \left( \begin{array}{ccc} c_{12} c_{13}&
                      s_{12} c_{13}&
                      s_{13} e^{-i\delta}\cr
-s_{12} c_{23}-c_{12} s_{23} s_{13} e^{i\delta}& c_{12} c_{23}-
s_{12} s_{23} s_{13} e^{i\delta}&
s_{23} c_{13}\cr
s_{12} s_{23} -c_{12} c_{23} s_{13} e^{i\delta}&
-c_{12} s_{23} -s_{12} c_{23} s_{13} e^{i\delta}&
c_{23} c_{13}\cr
\end{array}\right) 
diag(e^{\frac{i \alpha_M}{2}},e^{\frac{i \beta_M}{2}},1)
\end{equation}
where $s_{ij}=\sin \theta_{ij}, c_{ij}=\cos \theta_{ij}$ with
$(i,j=1,2,3)$, $\delta$ is the Dirac CP phase and $(\alpha_M,\beta_M)$
are Majorana phases. We use the best fit values of the oscillation
data \cite{Forero:2014,Esteban:2018} as summarised below in
Table \ref{tab:nudata}.

\begin{table}[!h]
\caption{Input data from neutrino oscillation experiments \label{osc} \cite{Forero:2014,Esteban:2018}}
\label{tab:nudata}
\begin{center}
\begin{tabular}{|c|c|c|}
\hline
{ Quantity} & {best fit values} &{ $3\sigma$ ranges}\\
\hline
$\Delta m_{21}^2~[10^{-5}eV^2]$ & $7.39$ & $6.79-8.01$\\
$|\Delta m_{31}^2|~[10^{-3}eV^2](NO)$ & $2.52$ & $2.427-2.625$\\
$|\Delta m_{32}^2|~[10^{-3}eV^2](IO)$ & $2.51$ & $2.412-2.611$\\
$\theta_{12}/^\circ$ & $33.82$ & $31.61-36.27$\\
$\theta_{23}/^\circ (NO)$ & $49.6$ & $40.3-52.4$\\
$\theta_{23}/^\circ (IO)$ & $49.8$ & $40.6-52.5$\\
$\theta_{13}/^\circ (NO)$ & $8.61$ & $8.22-8.99$\\
$\theta_{13}/^\circ (IO)$ & $8.65$ & $8.27-9.03$\\
$\delta/^\circ (NO)$ & $215$ & $125-392$\\
$\delta/^\circ (IO)$ & $284$ & $196-360$ \\
\hline
\end{tabular}
\end{center}
\end{table}
Important among new interesting salient features of this set of data points 
are: (i) the best fit value of atmospheric  mixing angle $\theta_{23}$ is  in the second octant, (ii) large values of Dirac CP phases exceeding $\delta=200^{\circ}$.
The data Table \ref{tab:nudata} includes the reactor neutrino mixing  $\theta_{13}=8.6^\circ$ which was known earlier.\\

Using the  mass-squared differences from Table \ref{tab:nudata} and
choosing the lightest mass eigen value $m_1=0.001$ eV, we at first
determine the other two mass eigen values. Then  using the three 
 mass eigen values, mixing angles and phases given in Table
 \ref{tab:nudata} we derive  neutrino mass matrix consistent with best
 fit to the data through the standard  relation
  \begin{equation}
m_\nu = U_{PMNS}~diag(m_1, m_2, m_3) U_{PMNS}^T ~.\label{mnu}
\end{equation}
For NO and IO cases we get the following results:\\

\par\noindent{\bf Normal ordering (NO):}\\
\bea
&& m_1 = 0.001\, {\rm eV}, m_2=0.0086\, {\rm eV}, m_3=0.0502\, {\rm eV},\nonumber\\
&& \sum_i m_i = 0.0598\, {\rm eV}\ll \Sigma_{Planck},  {\rm or} \,\, \Sigma_{new}\label{eq:minonew}  
\eea
 where $\Sigma_{Planck}=0.23$ eV of eq.(\ref{eq:cb}) and $\Sigma_{new}=0.12$ eV of eq.(\ref{eq:cbnew}) are cosmological bounds derived in \cite{Planck15} and \cite{Sunny:2018}, respectively, using the Planck satellite data and the $\Lambda$CDM big-bang cosmological model of the Universe.
 We thus find that our best fit of the present neutrino data 
easily satisfies both the cosmological bounds in the NO case. We have noted 
\cite{Sahoo:2018wdt}
that within the $3\sigma$ uncertainty, the oscillation data can accommodate  lower (higher) values of $\sum_i m_i$ with $m_1 < 0.001$ eV ($ 0.001 {\rm eV} < m_1 \le 0.04$ eV) consistent with the cosmological bound $\Sigma_{Planck}$\cite{Planck15}. But when $m_1 > 0.04$ eV, the bound $\Sigma_{new}$ \cite{Sunny:2018} 
is violated. For best fit values of masses, the neutrino mass matrix constructed from neutrino data is

\beq
m_\nu^{NO}(\rm eV)=  \left( \begin{array}{ccc}
              0.00367-0.00105i  & -0.00205+0.00346i & -0.00634+0.00294i  \cr
               -0.00205+0.00346i & 0.03154+0.00034i & 0.02106-0.0001i \cr
                -0.00634+0.00294i & 0.02106-0.0001i & 0.02383-0.00027i\cr
                    \end{array}\right). \label{eq:bfp_no}
\eeq
This gives
\beq
\sum_{n,l} |m_{\nu, nl}^{\rm NO}|^2= 2.595\times 10^{-3} {\rm eV}^2~.\label{eq:modsqNO}
\eeq
Its close vicinity with $\Delta m_{31}^2$ value of Table
\ref{tab:nudata} in the NO case is noteworthy.   
\\

Although here we have presented the numerical values of $(m_{\nu})_{ij}$ only for the best fit of 
neutrino oscillation data, the analysis can be easily extended for $3\sigma$ range of the extant data. We have to start with a suitably 
chosen value of the lightest eigen value ($m_1$). Then the $3\sigma$ range of solar ($\Delta m_{21}^2$) and 
atmospheric ($\Delta m_{31}^2$) mass squared differences provides a range of values for $m_2(=\sqrt{m^2_1+\Delta m_{21}^2})$ and $m_3(=\sqrt{m^2_1+\Delta m_{31}^2})$ such that the set $(m_1,m_2,m_3)$ remains compatible with the $3\sigma$ limit. 
Plugging in these mass eigenvalues along with all possible combinations and the mixing angles within the $3\sigma$ uncertainty
of oscillation data as presented in the third column of Table.\ref{tab:nudata}) in eq.(\ref{mnu}), we can generate large
number of sets of $m_\nu$ matrix. It is easy to realise that elements of the resulting $m_\nu$ matrix are constrained to
vary within a range as dictated by the $3\sigma$ uncertainty of the oscillation data. Since we are dealing with purely triplet seesaw,
the Yukawa coupling matrix ( $=y^{(2)}$) has  one-to-one correspondence with $m_\nu$. Elements of $m_\nu$ and $y^{(2)}$ differ 
only by a scale factor ($\mu_{\Delta_2} v^2/ M_{\Delta_2}^2$). A detailed analysis on $3\sigma$ and $1\sigma$ fit of 
neutrino oscillation data (including both NO and IO) for Type-II seesaw has already been presented in our previous
work\cite{Sahoo:2018wdt}. Therefore we are not repeating the whole analysis here. \\

\par\noindent{\bf Inverted ordering (IO):}\\
\bea
&& m_1 = 0.04938\, {\rm eV}, m_2=0.0501\, {\rm eV}, m_3=0.001\, {\rm  eV},\nonumber\\
&& \sum_i m_i = 0.100\, {\rm eV} < \Sigma_{Planck}, {\rm or}\,\, \Sigma_{new},  \label{eq:miio}
\eea
 where $\Sigma_{Planck}=0.23$ eV  \cite{Planck15}  and $\Sigma_{new}=0.12$ eV \cite{Sunny:2018} given in eq.(\ref{eq:cb}) and eq.(\ref{eq:cbnew}), respectively. Both the bounds have been derived using Planck satellite data \cite{Sunny:2018}. It is clear that the best fit in the IO case also satisfies both the cosmological bounds.  

\beq
m_\nu^{IO}(\rm eV)=  \left( \begin{array}{ccc}
                  0.0484-0.00001i & -0.001122+0.0055i & -0.00137+0.00471i \cr
                   -0.001122+0.0055i  &  0.02075-0.00025i& -0.02459-0.00026i \cr
                    -0.00137+0.00471i & -0.02459-0.00026i &  0.02910-0.00026i\cr
                    \end{array}\right). \label{eq:bfp_io}
\eeq
The manifestly hierarchical nature of mass eigen values are evident
from eq.(\ref{eq:minonew}) and eq.(\ref{eq:miio}).
This gives
\beq
\sum_{n,l} |m_{\nu, nl}^{\rm IO}|^2=4.9\times 10^{-3}\,\,{\rm eV}^2 \label{eq:modsqIO}
\eeq
which is nearly $2$ times larger than the  $\Delta m_{32}^2$ value of
Table \ref{tab:nudata} in the IO case.
In both the NO and IO cases, the sum of the three neutrino masses are also consistent with the upper bound $\Sigma_{new}=0.12$ eV \cite{Sunny:2018}.

\section{Estimation of baryon asymmetry}\label{sec:bauest}
In general baryon asymmetry is expressed as excess of matter over anti-matter scaled by entropy density or photon density 
which, in practice, are expressed by two nearly equivalent quantities $Y_B$ or $\eta_B$ defined below, i.e
\begin{equation}
 Y_B=\frac{n_B-n_{\overline{B}}}{s}, \label{yb_def}
\end{equation}
where $n_B,n_{\overline{B}}$ are number densities of baryons and anti-baryons, respectively, and $s$ is the entropy density and 
\begin{equation}
\eta_B=\frac{n_B-n_{\overline{B}}}{n_\gamma}, \label{etab_def}
\end{equation}
where $n_\gamma=$  photon density.
Their values as observed by recent Planck satellite experiment\cite{Planck15} are
\begin{equation}
(\eta_B )_0= (6-6.6) \times10^{-10}, \label{etab_expt}
\end{equation}
 {or ~equivalently~} 
\begin{equation}
 (Y_B )_0=  (8.55-9.37) \times10^{-11}, \label{yb_expt}
\end{equation}
where subscript zero indicates that the value of the corresponding asymmetry parameter is at the present epoch. In the present model
under consideration the asymmetry is at first generated in the leptonic sector
where lepton flavour and number violating 
decays of the heavy scalar triplet to bi-leptons gives rise to the lepton asymmetry  which, later,  gets converted into 
baryon asymmetry through sphaleron process. As we have clarified through an exhaustive discussion in Sec.\ref{fl-decoh} that, depending upon the mass of the decaying particle, the asymmetry production and its evolution down the temperature occur via 
unflavoured or flavoured leptogenesis. In the sections below we present a systematic and elaborate study of scalar triplet
leptogenesis in unflavoured and flavoured regimes through support of proper numerical data with graphical analysis.

\subsection{Unflavoured regime}
In this regime the mass $M_{\Delta_2}=M_{\Delta}$ of the decaying particle $\Delta_2(=\Delta)$, which is mainly responsible for asymmetry generation, is $M_{\Delta} \gtrsim 4 \times 10^{11}$ GeV. As discussed earlier, 
lepton flavours in this regime are indistinguishable and their coherent superposition acts as a single entity. Therefore, to calculate
the baryon asymmetry, at first we have to compute the flavour summed CP asymmetry parameter $\epsilon^l_\Delta$ given in 
eqs.(\ref{unflav-asy1},\ref{unflav-asy2}) which has to be plugged into the set of unflavoured Boltzmann 
equations (\ref{boltz_uf1}-\ref{boltz_uf3}). Simultaneous solution of those equations upto a large value of $z=M_{\Delta}/T$ (or equivalently 
low temperature) will provide us the freeze-in value of $(B-L)$ asymmetry. A fraction of this freeze-in value  will be converted into baryon
asymmetry $(Y_B)$ through sphaleron process which is shown in eq.(\ref{yb}). Throughout the analysis we use a fixed set of
values for the heavier triplet mass and its associated tri-linear coupling  $M_{\Delta_1}=3 \times 10^{13},\mu_{\Delta_1}=10^{13}$ (GeV).
For the phase differences between the corresponding elements of $m^{(1)}_\nu$ and $m^{(2)}_\nu$ we follow two conventions for our numerical computations:
(i) Fixed phase differences for all the elements, 
(ii) Random and different values of phase differences.
 Numerical results using both these conventions are discussed in the following two subsections.

\subsubsection{Identical mass ratio and fixed phase difference connecting $m^{(1)}_\nu$ and $m^{(2)}_\nu$}
\begin{figure}[h!]
\begin{center}
\includegraphics[width=10cm,height=9cm,angle=0]{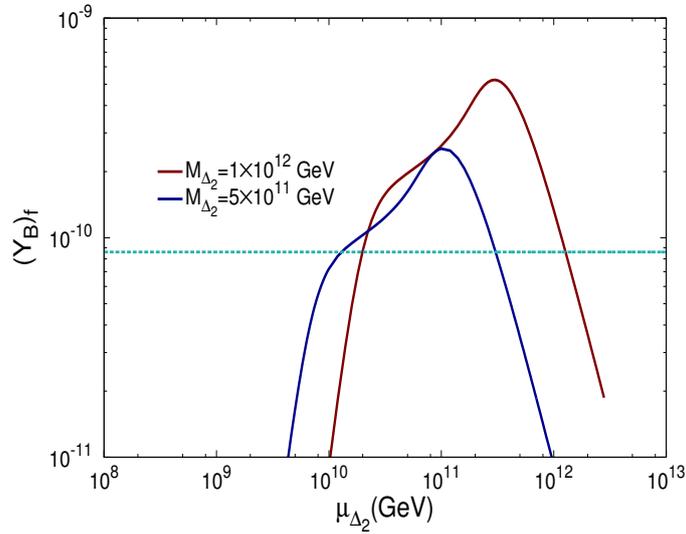}
\caption{Variation of final value of baryon asymmetry with trilinear coupling for two fixed benchmark values of lighter triplet mass using best fit values of NO type neutrino masses. 
The horizontal dashed line represents the experimental value of baryon asymmetry.}
\label{yb_mu_uf1}
\end{center}
\end{figure}

\begin{figure}[h!]
\begin{center}
\includegraphics[width=7.5cm,height=7cm,angle=0]{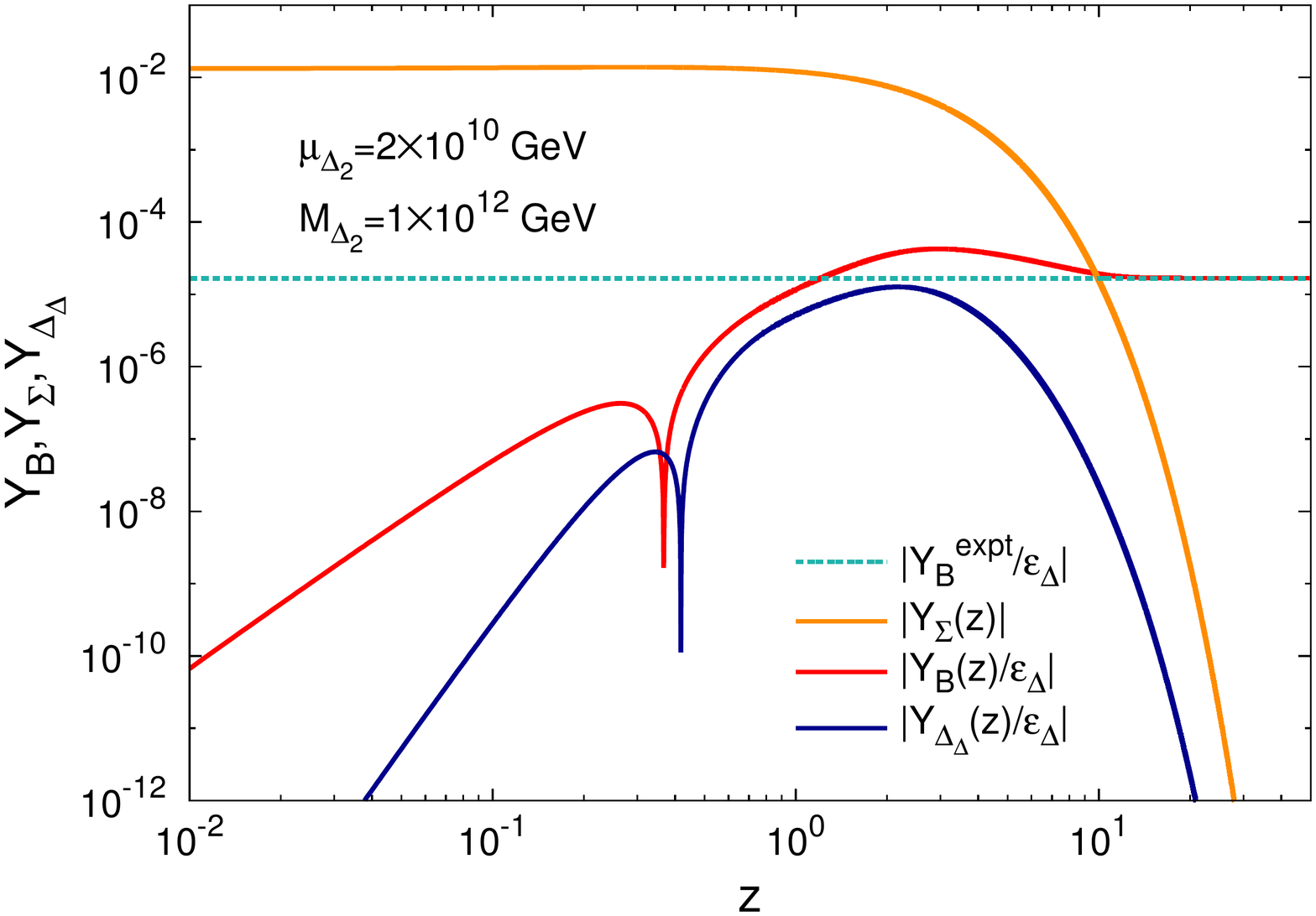}
\includegraphics[width=7.5cm,height=7cm,angle=0]{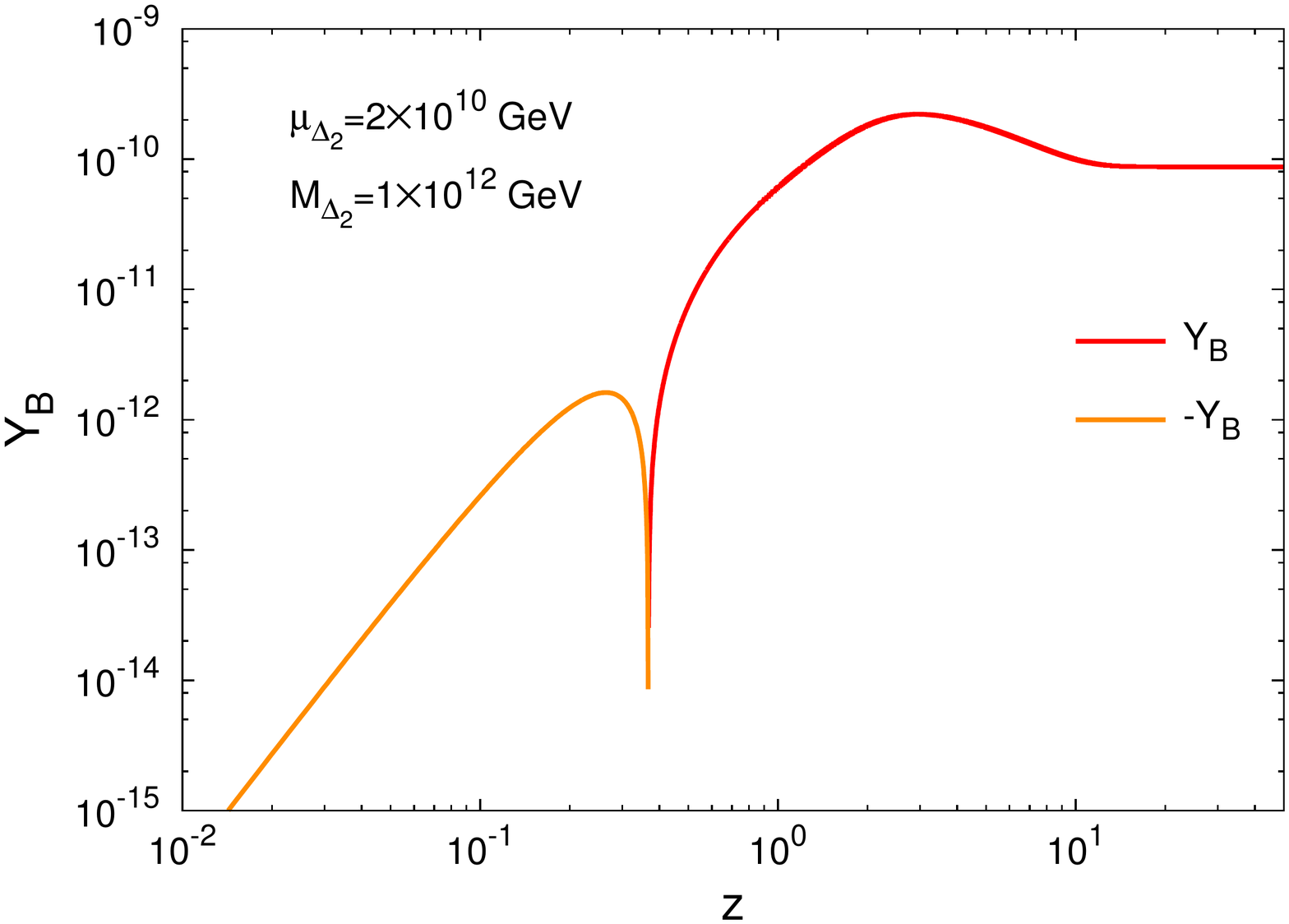}
\caption{left panel: Evolution of different variables of the Boltzmann equation with $z=M_{\Delta_2}/T$ for the fixed set 
($M_{\Delta_2}=10^{12},\mu_{\Delta_2}=2\times10^{10}$ GeV) and best fit to oscillation data with NO type neutrino masses. 
The horizontal dashed line represents the experimental value of baryon asymmetry.
$Y_B$ and $Y_{\Delta_\Delta}$ are scaled by the modulus value of unflavoured CP asymmetry parameter denoted by $\varepsilon_\Delta$
in the plot;
Right panel: variation of $Y_B$ with $z$ for the same fixed set of $(M_{\Delta_2},\mu_{\Delta_2})$.}
\label{yb_z_uf1}
\end{center}
\end{figure}

As we can see from the expression of the unflavoured CP asymmetry parameter (eq.(\ref{unflav-asy2})) two very important 
ingredients in its calculation are the modulus ratio $(F_{ij})$  and phase difference $(\phi^{(1)}_{ij}-\phi^{(2)}_{ij})$ 
between the corresponding elements of $m^{(1)}_\nu$ and $m^{(2)}_\nu$. In this section our numerical results will be limited to the first convention with 
a fixed choice of ratio  and phase differences which are again identical for all the elements, i.e we use 
$F_{ij}=0.1$  and $(\phi^{(1)}_{ij}-\phi^{(2)}_{ij})=-\pi/2$ for all $i,j$. For numerical value of $(m_\nu)_{ij}$ we 
use best fit values of the neutrino oscillation observables in normal mass ordering (NO) as presented in eq.(\ref{eq:bfp_no}).  
We proceed to calculate the baryon asymmetry
for two benchmark values of the lighter triplet mass ($M_{\Delta_2}=5\times10^{11},10^{12}$ GeV) in the unflavoured regime.
For each of the fixed benchmark value of $M_{\Delta_2}$ the corresponding trilinear coupling $\mu_{\Delta_2}$ is varied over 
a large range and for each combination of $(M_{\Delta_2},\mu_{\Delta_2})$ final value of baryon asymmetry $(Y_B)$ is evaluated.
The variation of final $Y_B$, denoted by ${(Y_B)}_f$ in the figure, with $\mu_{\Delta_2}$ for these two fixed values of $M_{\Delta_2}$
is shown graphically in Fig.\ref{yb_mu_uf1}.
The dashed line intersects the  ${(Y_B)}_f$ vs $\mu_{\Delta_2}$ curve in two places which signifies that for fixed value of $M_{\Delta_2}$
there are two $\mu_{\Delta_2}$ values which can produce baryon asymmetry within the experimental range. We now choose one such 
combination ($M_{\Delta_2}=10^{12},\mu_{\Delta_2}=2\times10^{10}$ GeV)\footnote{Although we are showing 
graphical representation of solution of Boltzmann equation only for this combination,  rigorous solution of the Boltzmann equation
has been carried out for each and every combinations of $(M_{\Delta_2},\mu_{\Delta_2})$ shown in Fig.\ref{yb_mu_uf1}.} 
from the  Fig.\ref{yb_mu_uf1} and show the evolution of different variables of the Boltzmann equation with $z$ in Fig.\ref{yb_z_uf1}. 
From the right panel of Fig.\ref{yb_z_uf1} it is clear that the final value of $Y_B$ indeed freezes to $\sim 8.6\times10^{-11}$
(which is well inside the range (eq.(\ref{yb_expt})) as observed by the Planck satellite experiment).\\

\begin{figure}[h!]
\begin{center}
\includegraphics[width=10cm,height=9cm,angle=0]{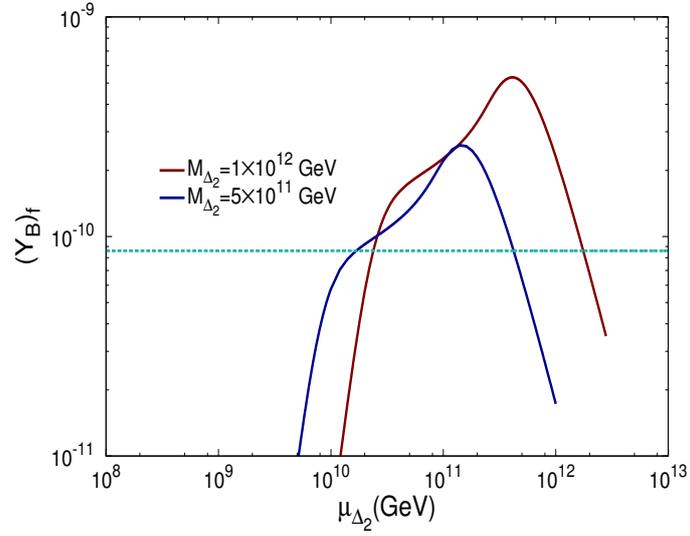}
\caption{Variation of final value of baryon asymmetry with trilinear coupling for two fixed benchmark values of lighter triplet mass $M_{\Delta_2}$ and IO type light neutrino masses.
The horizontal dashed line represents the experimental value of baryon asymmetry.}
\label{yb_mu_uf2}
\end{center}
\end{figure}

\begin{figure}[h!]
\begin{center}
\includegraphics[width=7.5cm,height=7cm,angle=0]{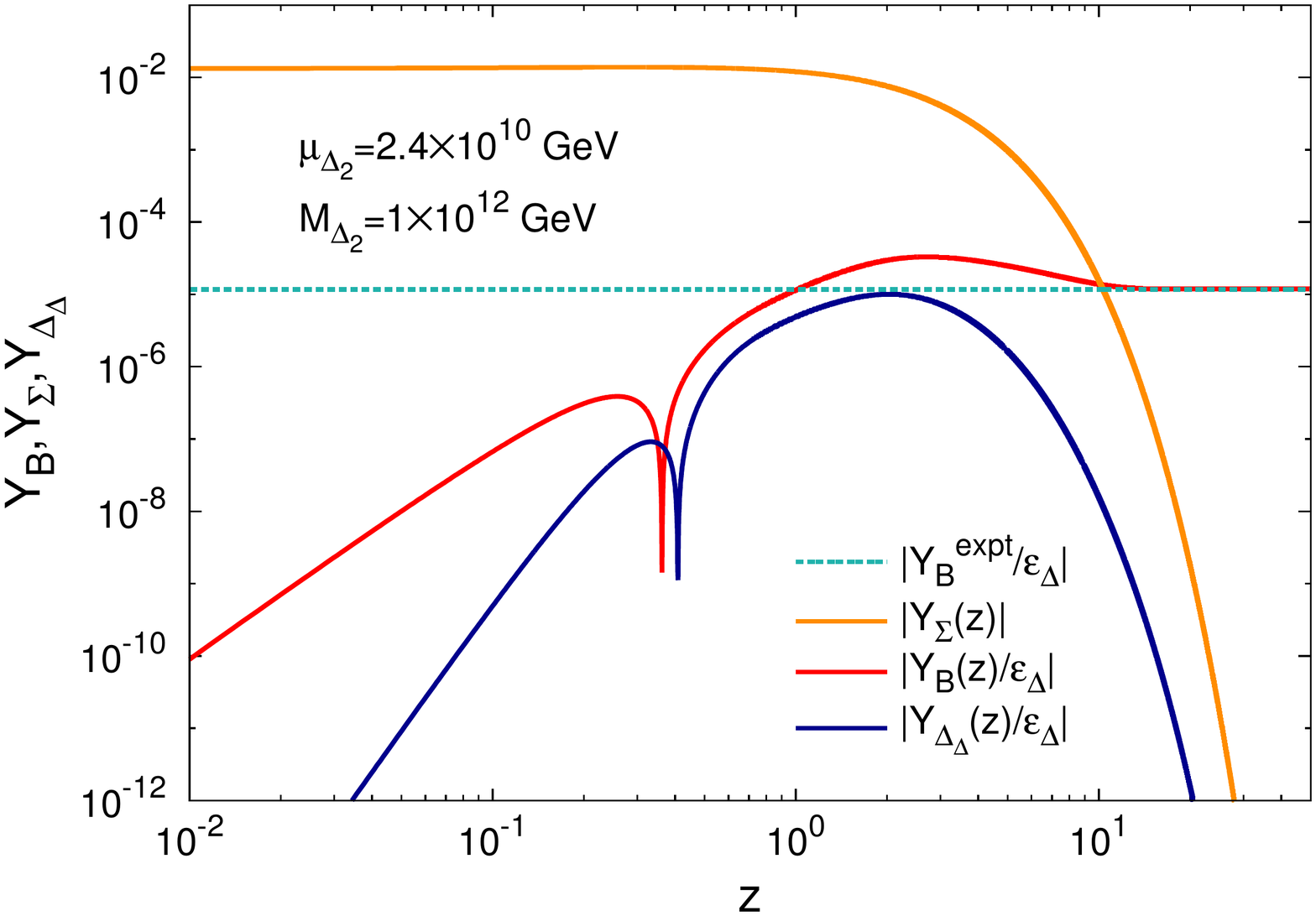}
\includegraphics[width=7.5cm,height=7cm,angle=0]{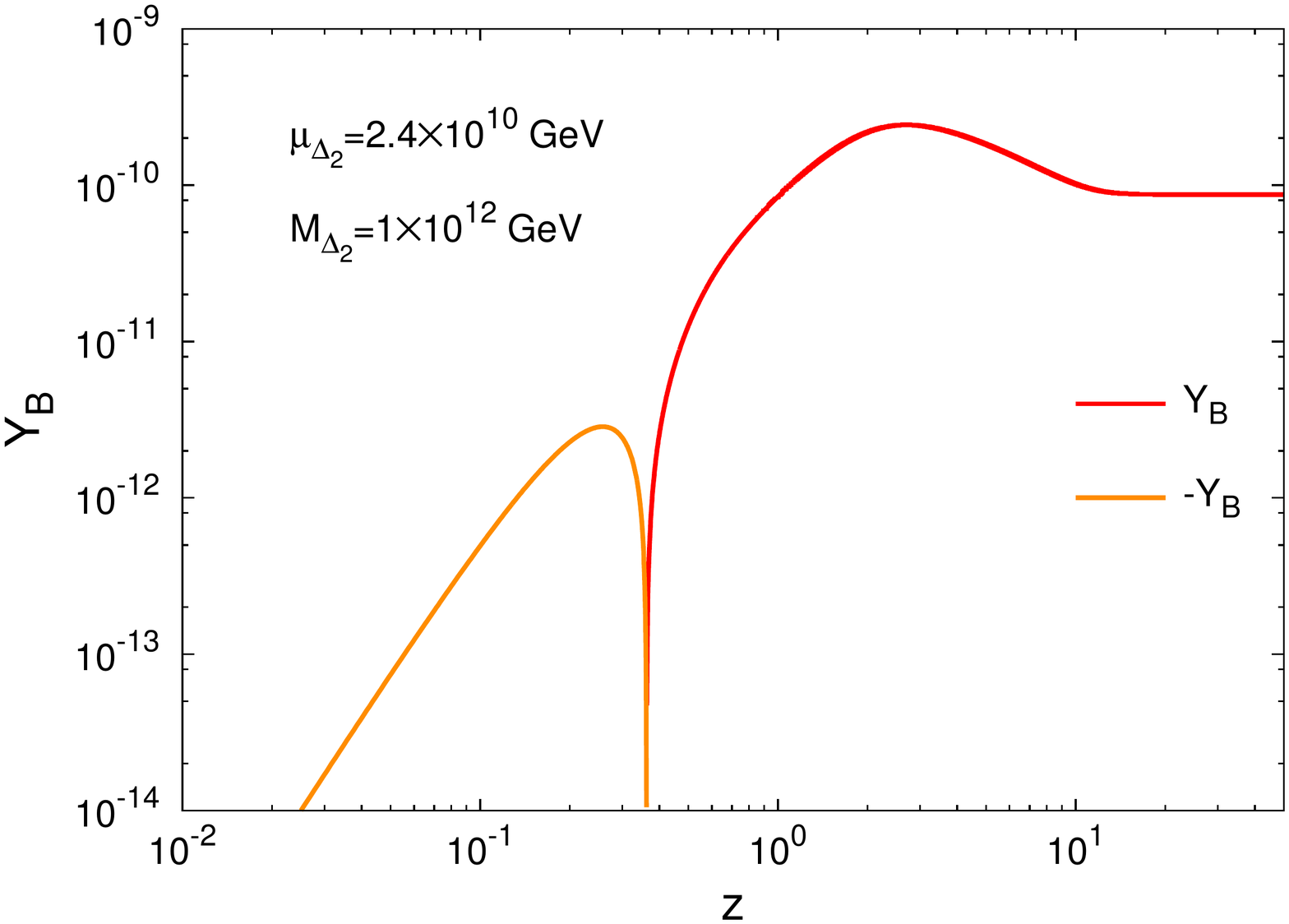}
\caption{Left panel: Evolution of different variables of the Boltzmann equation with $z$ for the fixed set 
($M_{\Delta_2}=10^{12},\mu_{\Delta_2}=2.4\times10^{10}$ GeV) for best fit to oscillation data with IO type neutrino masses.
The horizontal dashed line represents the experimental value of baryon asymmetry.
$Y_B$ and $Y_{\Delta_\Delta}$ are scaled by the modulus value of unflavoured CP asymmetry parameter denoted by $\varepsilon_\Delta$
in the plot;
Right panel: variation of $Y_B$ with $z$ for the  same fixed set of ($M_{\Delta_2},\mu_{\Delta_2}$).}
\label{yb_z_uf2}
\end{center}
\end{figure}

For the sake of completeness we have repeated the same analysis using IO for light neutrino masses i.e every other parameters
remains the same except for $(m_\nu)_{ij}$ we use eq.(\ref{eq:bfp_io}). The resulting plot for the final values of $Y_B$ with $\mu_{\Delta_2}$
is presented in Fig.\ref{yb_mu_uf2}.
It gives similar plot as that of the NO case, the only difference is that for a fixed $M_{\Delta_2}$ the value of $\mu_{\Delta_2}$ required
to produce same $Y_B$ is shifted slightly to a higher value. Again the evolution of different variables of the Boltzmann equation
with $z$ (for the fixed set ($M_{\Delta_2}=10^{12},\mu_{\Delta_2}=2.4\times10^{10}$ GeV)) is shown in Fig.\ref{yb_z_uf2}.
\\

In a simplistic approach if we neglect the triplet asymmetry term ($Y_{\Delta_\Delta}$) then we are left with only two
coupled differential equations involving $Y_\Sigma$ and $Y_{B-L}$. Again in case of very weak washout, the wash-out 
term can be neglected, i.e the 2nd Boltzmann equation contains only the source term. This assumption lead us to a set of
Boltzmann equations which are same as those presented in \cite{Ma-Us:1998}, i.e
\begin{eqnarray}
&& \dot{Y}_\Sigma=-\Big(\frac{Y_\Sigma}{Y_\Sigma^{eq}}-1 \Big )\gamma_D -2\Big[\Big(\frac{Y_\Sigma}{Y_\Sigma^{eq}}\Big )^2-1 \Big]\gamma_A \label{us_ma_uf1}\\ 
&& \dot{Y}_{\Delta_{B-L}}= - \Big( \frac{Y_\Sigma}{Y_\Sigma^{eq}} -1 \Big) \epsilon^{l}_\Delta ~.
 \label{us_ma_uf3}
\end{eqnarray}
Proceeding exactly in a similar manner we compute final $Y_B$ through solution of these two equations for two benchmark
values of $M_{\Delta_2}$ whereas $\mu_{\Delta_2}$ is varied over a wide range of values. The resulting plot of final $Y_B$
with $\mu_{\Delta_2}$ is presented below in Fig.\ref{yb_mu_without_wsh} for both the  NO and IO type neutrino mass orderings.
\begin{figure}[h!]
\begin{center}
\includegraphics[width=7.5cm,height=7cm,angle=0]{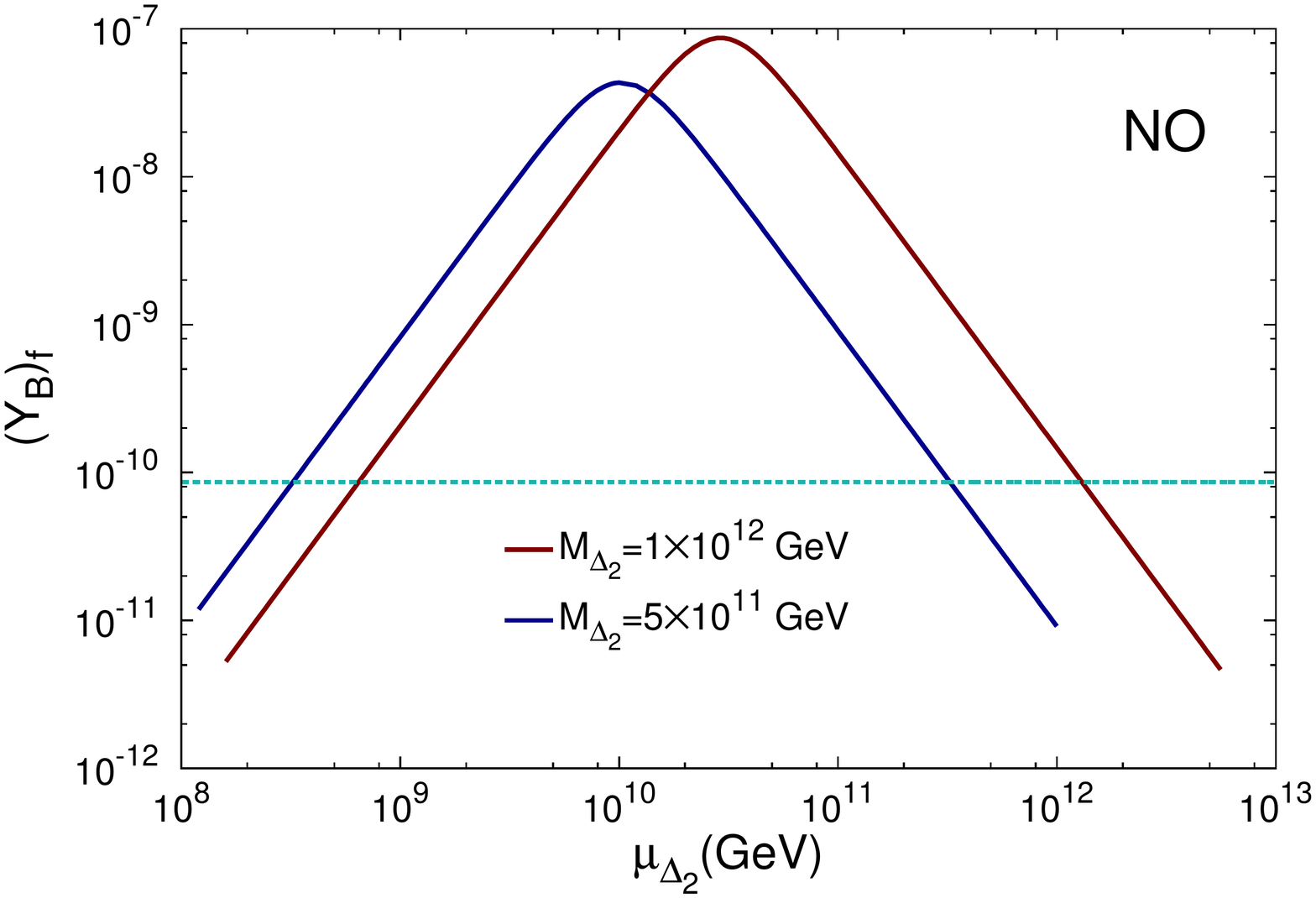}
\includegraphics[width=7.5cm,height=7cm,angle=0]{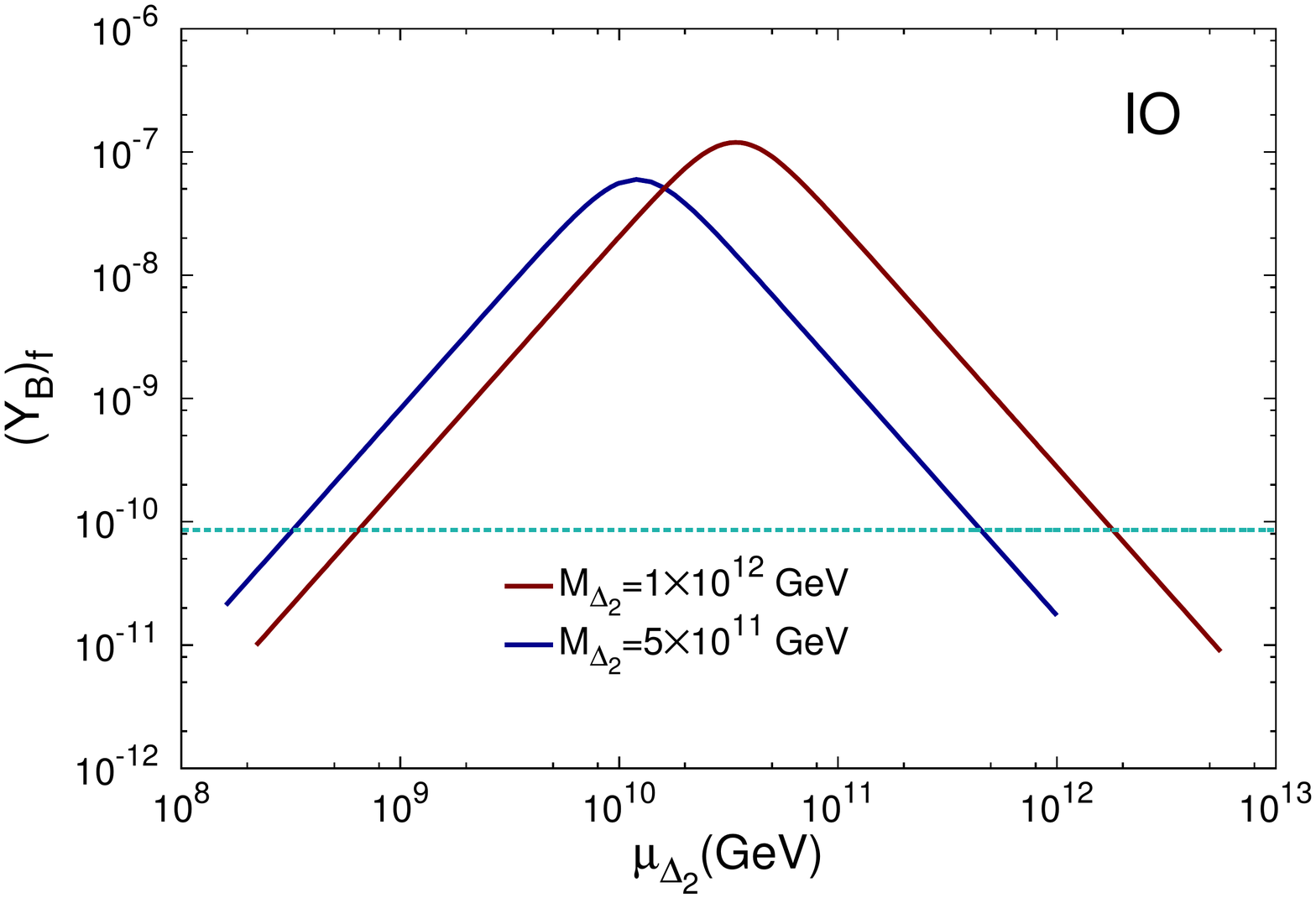}
\caption{Variation of  baryon asymmetry with trilinear coupling for two fixed benchmark values of lighter triplet mass 
neglecting triplet asymmetry and washout. The horizontal dashed line represents the experimental value of baryon asymmetry. The
left-panel (right-panel) represents solutions for NO (IO) type light neutrino mass hierarchy.}
\label{yb_mu_without_wsh}
\end{center}
\end{figure}
Due to the absence of the washout term the asymmetry produced in this case (for a fixed set of 
$(M_{\Delta_2},\mu_{\Delta_2})$) is much higher than the value given by solution of the full set of 
Boltzmann equations(Fig.\ref{yb_mu_uf1}, Fig.\ref{yb_mu_uf2}). The triplet asymmetry which has been 
omitted in this treatment has a non-trivial effect and also neglecting the washout term (without proper estimation
of the decay parameter $(K=\Gamma^{tot}_{\Delta_2}/H)$) may lead to overestimation of the asymmetry. We present
this analysis just for comparison. For all the future studies in this work we will use the full set of Boltzmann equations.\\

\begin{figure}[h!]
\begin{center}
\includegraphics[width=10cm,height=9cm,angle=0]{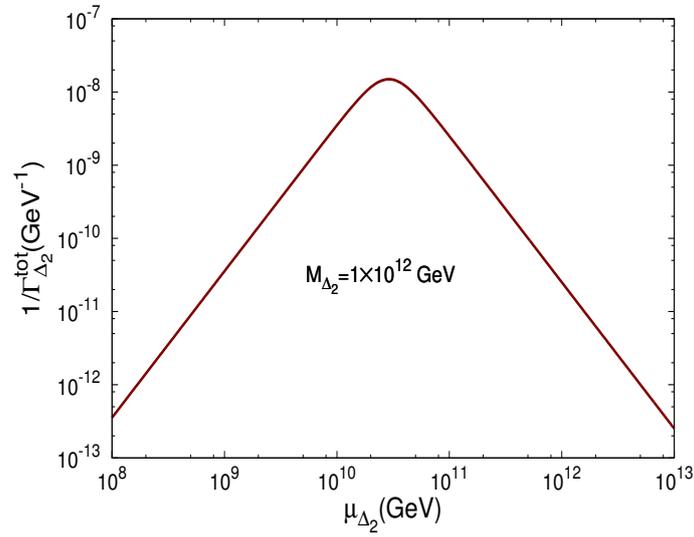}
\caption{Dependence of total decay width $\Gamma^{tot}_{\Delta_2}$ (of the lighter triplet) on the trilinear coupling
$\mu_{\Delta_2}$ for fixed a mass of the scalar triplet $\Delta_2$ taking NO for light neutrinos and best fit values of the oscillation data.}
\label{gamma_t}
\end{center}
\end{figure}

It is worthwhile to mention that pattern of the $(Y_B)_f$ vs $\mu_{\Delta_2}$ plot exactly follows the variation
of the CP asymmetry $(\epsilon^l_{\Delta})$ with $\mu_{\Delta_2}$.
In our analysis all the parameters except the trilinear coupling ($\mu_{\Delta_2}$) are fixed. In the expression of CP 
asymmetry parameter $\epsilon^l_{\Delta}$(eq.(\ref{unflav-asy1})), $\mu_{\Delta_2}$ dependence is contained only in the total triplet decay width $(\Gamma^{tot}_{\Delta_2})$, and $\epsilon^l_{\Delta} (|\mu_{\Delta_2}|) \sim 1/\Gamma^{tot}_{\Delta_2} (|\mu_{\Delta_2}|)$ 
where
\begin{equation}
  \Gamma^{tot}_{\Delta_2} (|\mu_{\Delta_2}|)= c_1 |\mu_{\Delta_2}|^2 +\frac{c_2}{|\mu_{\Delta_2}|^2},
\end{equation}
with $c_1=\frac{1}{8 \pi M_{\Delta_2}}$ and $c_2=\frac{M^5_{\Delta_2}Tr \left( m_\nu m^\dagger_\nu \right)}{8 \pi v^4}$.
In the plot (Fig.\ref{gamma_t}) below we depict the $\mu_{\Delta_2}$ dependence of $1/\Gamma^{tot}_{\Delta_2}$ which shows a peak near
\footnote{Its numerical value is $\mu_{\Delta_2}=2.9 \times10^{10}$ GeV assuming NO, taking best fit values of oscillation
data and triplet mass fixed at $M_{\Delta_2}=10^{12}$ GeV.}
$\mu_{\Delta_2}=\left( \frac{c_2}{c_1} \right)^{1/4} $. It is obvious that the CP asymmetry will exactly follow this pattern
and the final baryon asymmetry which is also directly proportional to CP asymmetry will also closely follow the same
type of $\mu_{\Delta_2}$ dependence.
\subsubsection{Identical mass ratio but random phase differences connecting $m^{(1)}_\nu$ and $m^{(2)}_\nu$}\label{uf_rand}
We denote the phase difference between two corresponding elements of $m^{(1)}_\nu$ and $m^{(2)}_\nu$ as
$\phi_{ij}=(\phi^{(1)}_{ij}-\phi^{(2)}_{ij})$. 
In the most general case,  we need a set of six independent phase parameters 
$\{ \phi_{11},\phi_{12},\phi_{13},\phi_{22},\phi_{23},\phi_{33} \}$ and modulus ratios $\{F_{11},F_{12},F_{13},F_{22},F_{23},F_{33}\}$ to connect 
$m^{(1)}_\nu$ and $m^{(2)}_\nu$ since they are both  complex symmetric Majorana type matrices. 
We denote the set of phases and ratios as a whole by $\Phi_k \equiv \{ \phi_{11},\phi_{12},\phi_{13},\phi_{22},\phi_{23},\phi_{33} \}_k$ and 
$\mathcal{F}_k \equiv  \{F_{11},F_{12},F_{13},F_{22},F_{23},F_{33}\}_k$,
where $k$ denotes one of the sets. Here we generate a large number sets $\Phi_{k=1,...N}$ for the phase differences where
each of the component phases is an absolute random number in the range $ -\pi \leq \phi_{ij} \leq \pi$. But 
for the sake of simplicity we have limited ourselves to the case of identical modulus ratio for all the elements ($i,j=1,3$), 
i.e $\mathcal{F}_{k=1,....N}=\{0.1,0.1,0.1,0.1,0.1,0.1\}$ for each and every sets marked as $k=1,....,N$. At first we estimate the unflavoured
CP asymmetry parameter $(\epsilon^l_\Delta)$ for the $N$ number of random sets and then choose those sets among them which
gives rise to negative value of the CP asymmetry parameter\footnote{For the unflavoured leptogenesis case there is 
relative negative sign in the formula connecting CP asymmetry($(\epsilon^l_\Delta)$) and the final baryon asymmetry
parameter($Y_B$). Therefore to get positive $Y_B$, the CP asymmetry($\epsilon^l_\Delta$) must be negative. }. Naturally,
imposition of this constraint ($\epsilon^l_\Delta<0$ or equivalently $Y_B>0$) reduces the number of allowed sets to $\sim N/2$.
To get the value of final baryon asymmetry parameter($Y_B$), the set of coupled Boltzmann equations, eq.(\ref{boltz_uf1})-eq.(\ref{boltz_uf3})
have to be solved approximately  $N/2$ times which is  time consuming or rather repetitive. Therefore, we pick a random set
\begin{equation}
\frac{\Phi_{k=k_{random}}}{\pi} \equiv \{ -0.3418,-0.0807,0.7850,0.9961,-0.4427,0.7244 \}~~  \label{rand}
\end{equation}
as a representative set among those $N/2$ and proceed further for the calculation of baryon asymmetry in NO and IO cases.
\begin{figure}[h!]
\begin{center}
\includegraphics[width=7.5cm,height=7cm,angle=0]{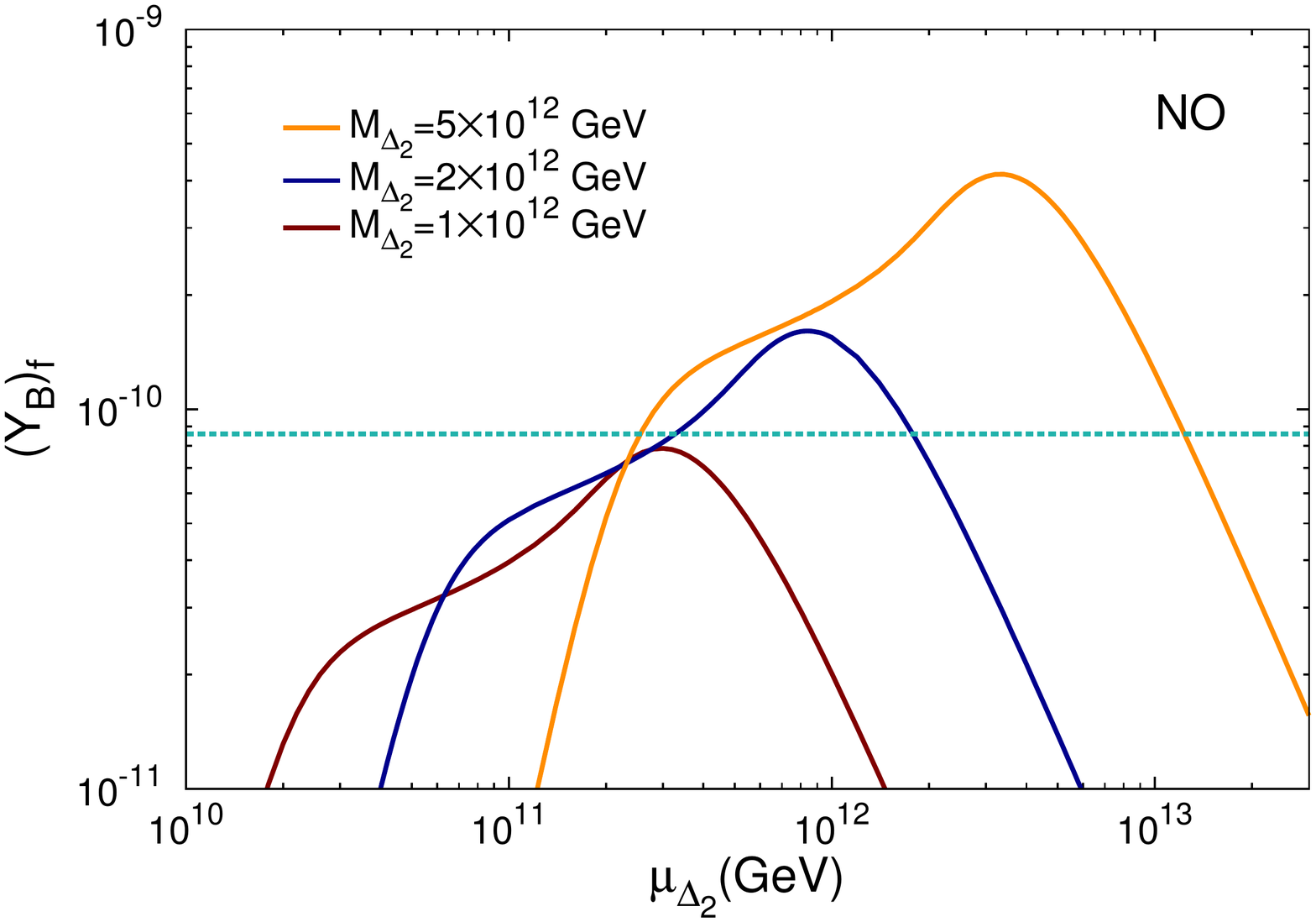}
\includegraphics[width=7.5cm,height=7cm,angle=0]{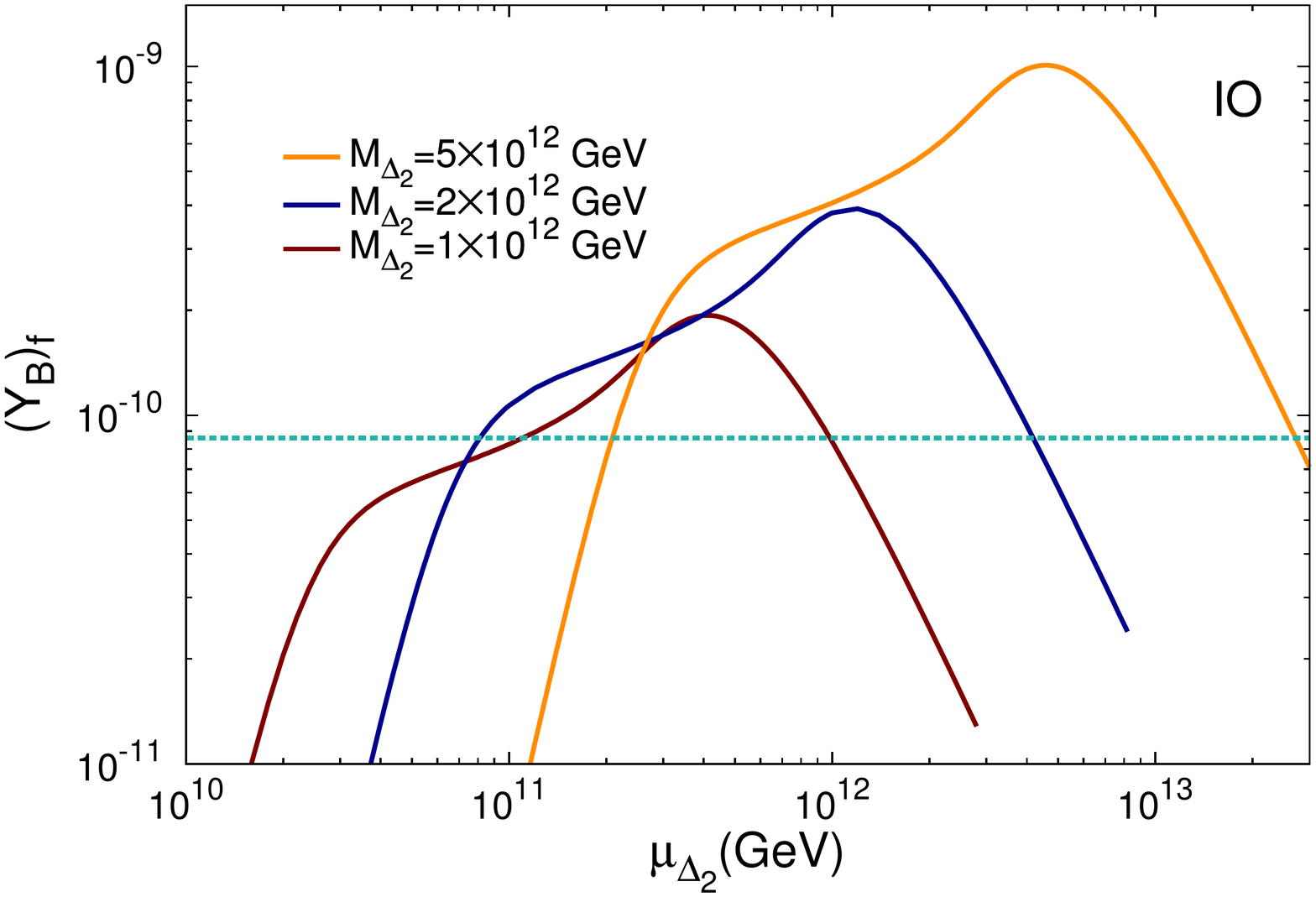}
\caption{Variation of  baryon asymmetry with trilinear coupling $\mu_{\Delta_2}$ for three fixed benchmark values of lighter triplet mass $M_{\Delta_2}$ derived  by the solution full set of Boltzmann equations. The horizontal dashed line represents the experimental value of baryon asymmetry.
The left-panel and the right-panel represent the result for NO and IO cases, respectively, consistent with the best fit to the oscillation data.}
\label{yb_mu_full_boltz1}
\end{center}
\end{figure}
In  Fig.\ref{yb_mu_full_boltz1}  we show the dependence of the final baryon asymmetry on the trilinear coupling $\mu_{\Delta_2}$ for three
benchmark values of the triplet mass $M_{\Delta_2}$ taking into account both the NO and IO types of light neutrino mass spectra consistent with the best fit to the oscillation data. It is
clear from  Fig.\ref{yb_mu_full_boltz1} (left-panel) that for this choice of phases given in eq.(\ref{rand})) in the NO case, even $M_{\Delta_2}=10^{12}$ GeV fails to generate 
adequate asymmetry within the experimental range for any value of $\mu_{\Delta_2}$.
But  the left-panel of the same Fig.\ref{yb_mu_full_boltz1} also shows that the
required value of  baryon asymmetry can be successfully generated for $M_{\Delta_2}=2\times 10^{12}$ GeV and higher values. 
Right-panel of  Fig.\ref{yb_mu_full_boltz1}  shows that for IO case $M_{\Delta_2}=10^{12}$ GeV is enough to generate baryon asymmetry 
within the experimental range.  

\begin{figure}[h!]
\begin{center}
\includegraphics[width=7.5cm,height=7cm,angle=0]{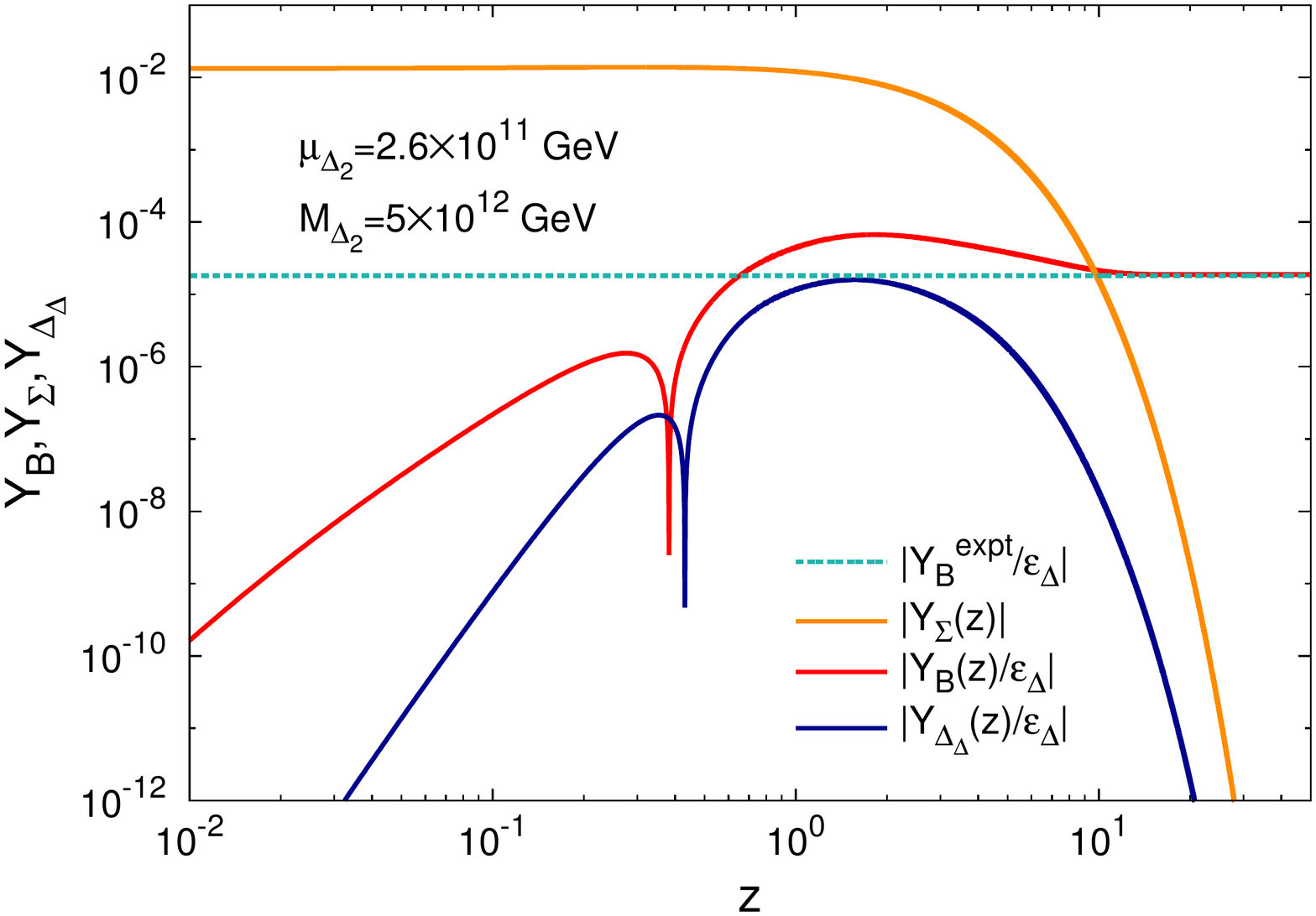}
\includegraphics[width=7.5cm,height=7cm,angle=0]{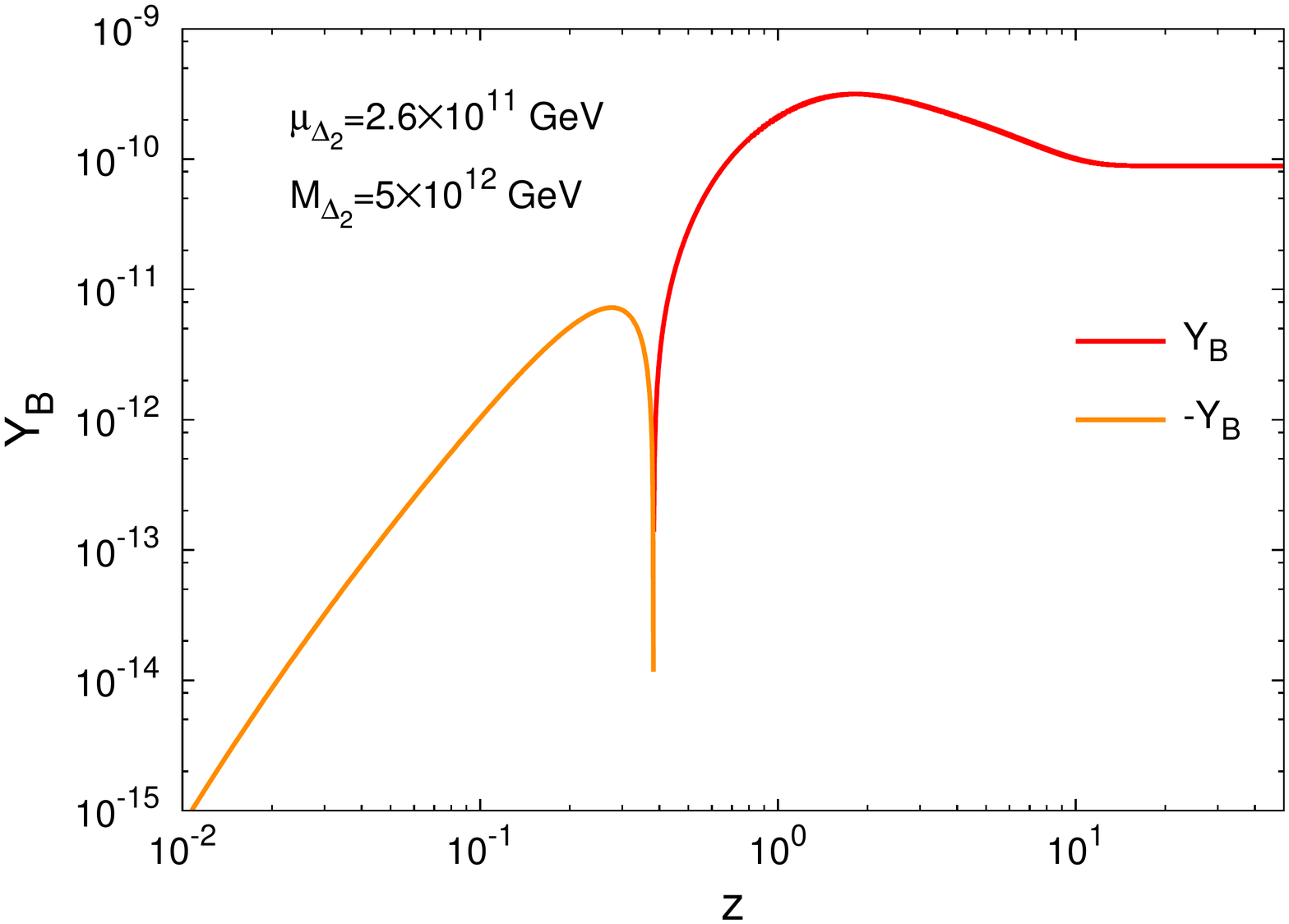}
\caption{Left-panel: Evolutions of different variables of the Boltzmann equation with $z$ for the fixed set 
($M_{\Delta_2}=5\times10^{12},\mu_{\Delta_2}=2.6\times10^{11}$ GeV) in NO case.
The horizontal dashed line represents the experimental value of baryon asymmetry.
Right panel: variation of $Y_B$ with $z$ for the same fixed set of $(M_{\Delta_2},\mu_{\Delta_2})$ as in the left-panel.}
\label{yb_z_uf3}
\end{center}
\end{figure}

\begin{figure}[h!]
\begin{center}
\includegraphics[width=7.5cm,height=7cm,angle=0]{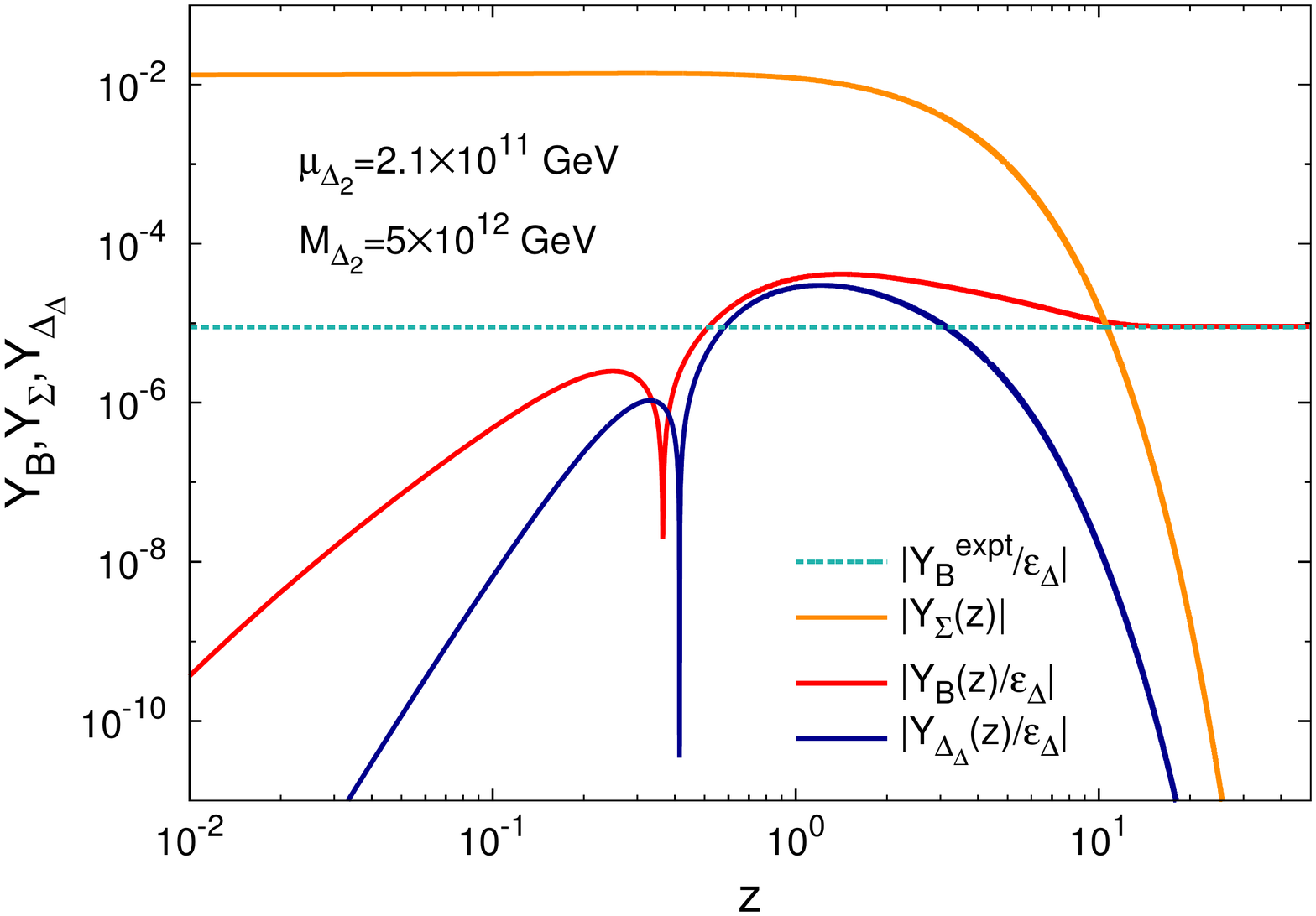}
\includegraphics[width=7.5cm,height=7cm,angle=0]{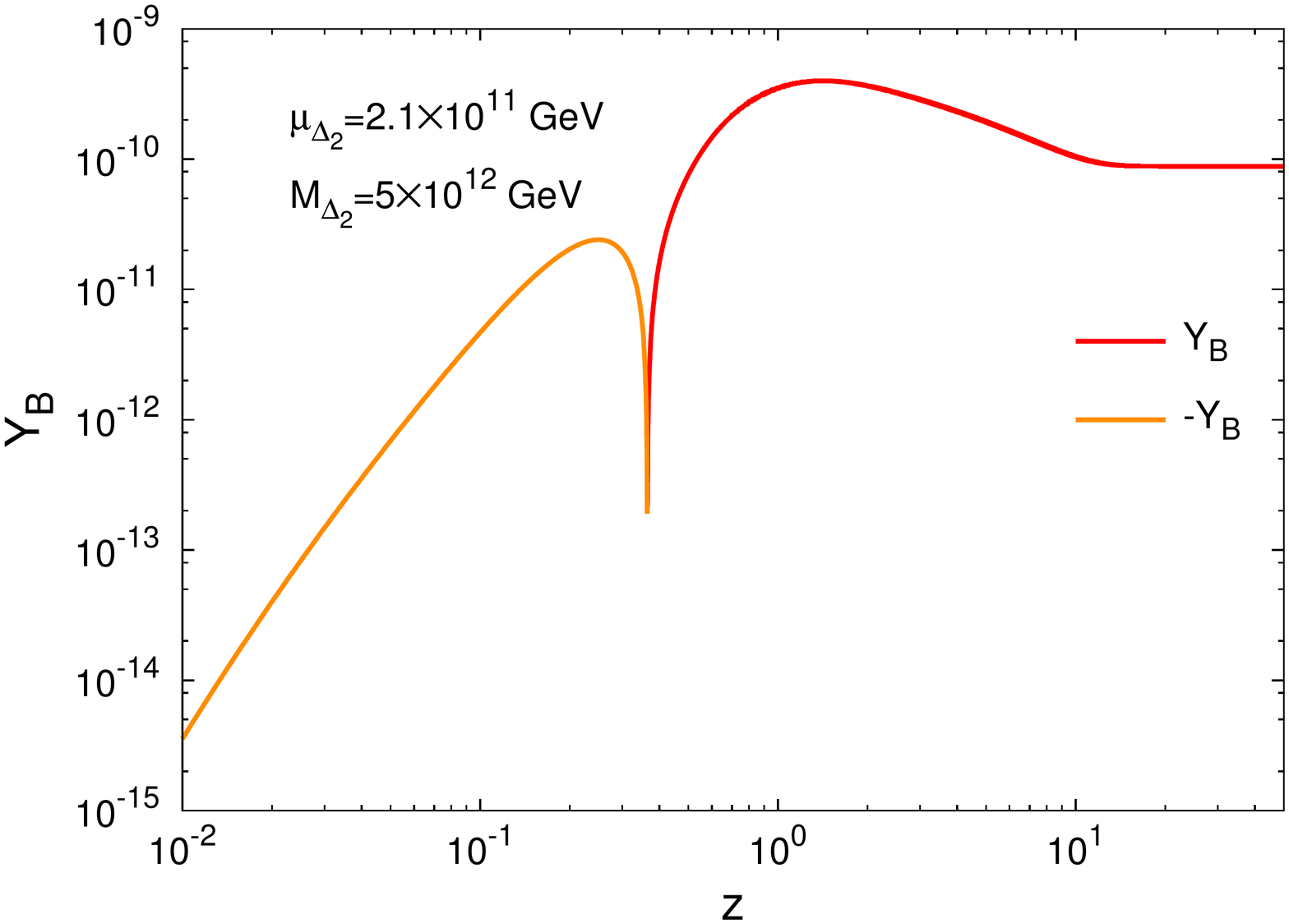}
\caption{Left-panel: Evolutions of different variables of the Boltzmann equation with $z$ for the fixed set 
($M_{\Delta_2}=5\times10^{12},\mu_{\Delta_2}=2.1\times10^{11}$ GeV) in IO case.
The horizontal dashed line represents the experimental value of baryon asymmetry.Right-panel: Variation of $Y_B$ with $z$ for the  same fixed set of ($M_{\Delta_2},\mu_{\Delta_2}$) as in the left-panel.}
\label{yb_z_uf4}
\end{center}
\end{figure}

Now we choose a specific set of triplet  mass and trilinear coupling 
($M_{\Delta_2}=5\times 10^{12},\mu_{\Delta_2}=2.6\times10^{11}$ GeV) for NO case, and
($M_{\Delta_2}=5\times 10^{12},\mu_{\Delta_2}=2.1\times10^{11}$ GeV) for IO case.
We show the evolution of relevant variables of Boltzmann equation as a function of  $z(=M_{\Delta_2}/T)$ in Fig.\ref{yb_z_uf3} and Fig.\ref{yb_z_uf4} for the NO and IO cases, respectively.

\subsection{Flavoured regime}
As already discussed above, the flavoured regime  can be approximately subdivided into two: (i) the two-flavoured or $\tau$-flavoured regime for which $10^9< \frac{M_{\Delta_2}}{{\rm GeV}} < 4\times 10^{11}$, and (ii) the three-flavoured or fully flavoured regime for which $M_{\Delta_2} <10^{9}$ GeV. For the sake of simplicity throughout the present work we confine our discussion to the \textquoteleft two-flavoured or $\tau$-flavoured \textquoteright~ regime only. 
Here we take the mass of the lighter triplet ($M_{\Delta_2}$) to be less than $4\times10^{11}$ but greater than $10^{9}$ GeV. 
Therefore,
according to the discussion presented in Sec.\ref{fl-decoh}, the decoherence of $\tau$ flavour has been achieved fully
whereas $e$ and $\mu$ still act as a coherent superposition which can be treated equivalently as a single flavour $a \equiv e+\mu$. Thus
 here we have two distinguishable flavours $a$ and $\tau$ \footnote{ Hence  the nomenclature of this regime  as 2-flavoured  or $\tau$-flavoured
regime is well known \cite{Sierra:2014tqa}.}. Accordingly we have two flavoured CP asymmetry parameters $\epsilon^\tau_\Delta$ and $\epsilon^a_\Delta$ 
where $\Delta \equiv \Delta_2$ and $\epsilon^a_\Delta=\epsilon^e_\Delta+\epsilon^\mu_\Delta$. They can be calculated
using the set of formulas given in eq.(\ref{flav_ep}), eq.(\ref{flav_ep1}), 
and eq.(\ref{flav_ep2}) of Sec.\ref{cpasy} for flavoured CP asymmetry.
Those flavoured CP asymmetry parameters are then used in the set of flavoured Boltzmann equations in 
eq.(\ref{boltz_big2}), eq.(\ref{boltz_big3}), and eq.(\ref{boltz_big4}) which have to be solved simultaneously upto a very high value of $z$ in order
to derive the final  freeze-in value of baryon asymmetry  $Y_B$. In the process of these computations we bear in mind that the lepton flavour
indices $(i,j,k)$ in those equations can take only two values $a$ and $\tau$. Therefore, the asymmetry vector $\vec{Y}_{\Delta}$ in this case consists of three entries given by $\left[ \vec{Y}_{\Delta} \right]^T \equiv \left( Y_{\Delta_\Delta}, Y_{B/3-{L_a}}, Y_{B/3-{L_\tau}}  \right)$. It can be understood that the dimensionality of the asymmetries coupling matrices $C^l_{ij}$, $C^\phi_k$ will be $2\times3$ and $1\times3$, respectively and their explicit numerical forms are given in Table.\ref{clcphi} of 
Appendix.\ref{sec:ap3}. The branching ratios 
$(B_{l_{ij}})$ of triplet decay to different lepton flavours can be regarded as a $2\times2$ matrix of the form
\begin{equation}
B_l =\left( \begin{array}{cc}
      B_{l_{aa}} & B_{l_{a\tau}} \\
      B_{l_{\tau a}} & B_{l_{\tau\tau}}
     \end{array} \right).
\end{equation}
The $22$ element is obvious $B_{l_{\tau\tau}}=\frac{M_{\Delta_2}}{8\pi \Gamma^{tot}_{\Delta_2}} |y^{(2)}_{33}|^2$, whereas the other 
three entries are given by
\begin{eqnarray}
&& B_{l_{aa}} = \frac{M_{\Delta_2}}{8\pi \Gamma^{tot}_{\Delta_2}} \sum_{i,j=1,2}|y^{(2)}_{ij}|^2 ,\\
&& B_{l_{a\tau}} = \frac{M_{\Delta_2}}{8\pi \Gamma^{tot}_{\Delta_2}} \sum_{i=1,2}|y^{(2)}_{i3}|^2 ,\\
&& B_{l_{\tau a}} = \frac{M_{\Delta_2}}{8\pi \Gamma^{tot}_{\Delta_2}} \sum_{j=1,2}|y^{(2)}_{3j}|^2~.
\end{eqnarray}
We are now in a position to solve the set of flavoured Boltzmann equations to find the value of the asymmetry parameters 
$Y_{B/3-{L_a}}(z\rightarrow z_f)$, $Y_{B/3-{L_\tau}}(z\rightarrow z_f)$ where $z_f$ is a large enough value of $z$ where asymmetry freezes, i.e it does not change 
furthermore  with decrease in temperature T. Then the final baryon asymmetry parameter $Y_B$ is evaluated by summing over 
$Y_{B/3-{L_a}}(z_f)$, $Y_{B/3-{L_\tau}}(z_f)$ followed by multiplication with sphaleronic factor and the $SU(2)$ factor 
as shown in eq.(\ref{yb}). Now the detailed numerical analysis has been subdivided in two categories depending upon the phase differences and modulus ratios in a manner similar to that of the unflavoured case.

\subsubsection{Identical ratio and phase differences connecting $m^{(1)}_\nu$ and $m^{(2)}_\nu$}
For numerical computations we use a fixed modulus ratio and phase difference as
$F_{ij}=0.1$  and $(\phi^{(1)}_{ij}-\phi^{(2)}_{ij})=-\pi/2$ for all $i,j$. Numerical values of the light neutrino
mass matrix elements $(m_\nu)_{ij}$ have been obtained by using the best fit values of the neutrino oscillation
data with NO type mass hierarchy of eq.(\ref{eq:bfp_no}).
The rest of the analysis has been carried out for two fixed benchmark values of the lighter triplet mass in the range $10^9{\rm GeV} < M_{\Delta_2} < 4\times 10^{11} {\rm GeV}$. The   trilinear LNV coupling  $\mu_{\Delta_2}$ is taken over a wider range of values
while keeping the ratio $\frac{\mu_{\Delta_2}}{M_{\Delta_2}}$ within the perturbative limit.
 After gathering  informations about all the required quantities, we first estimate the 
flavoured CP asymmetry parameters of eq.(\ref{flav_ep}), eq.(\ref{flav_ep1}) and eq.(\ref{flav_ep2}). In this context, it should be mentioned that the 
flavour violating (or the purely flavoured) part of the CP asymmetry parameter $\epsilon^{l_i(\not F)}_{\Delta}$ of eq.(\ref{flav_ep2}) 
vanishes identically since the phase differences $(\phi^{(1)}_{ij}-\phi^{(2)}_{ij})$ are assumed to be identical for any combination of
$i,j$. Therefore only the combined CP-asymmetry with (lepton number + flavour) violating part $\epsilon^{l_i(\not L,\not F)}_{\Delta}$ contributes to the 
asymmetry generation. \\

\begin{figure}[h!]
\begin{center}
\includegraphics[width=10cm,height=9cm,angle=0]{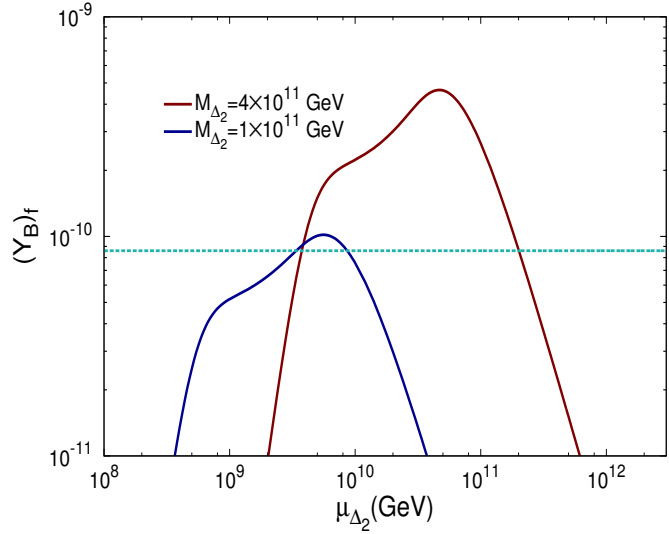}
\caption{Variation of  baryon asymmetry with trilinear coupling for two fixed benchmark values of the triplet mass
($M_{\Delta_2}=10^{11}$ GeV and $M_{\Delta_2}=4\times10^{11}$ GeV) in the NO case.
The horizontal dashed line represents the experimental value of baryon asymmetry.}
\label{yb_mu_2f1}
\end{center}
\end{figure}

\begin{figure}[h!]
\begin{center}
\includegraphics[width=7.5cm,height=7cm,angle=0]{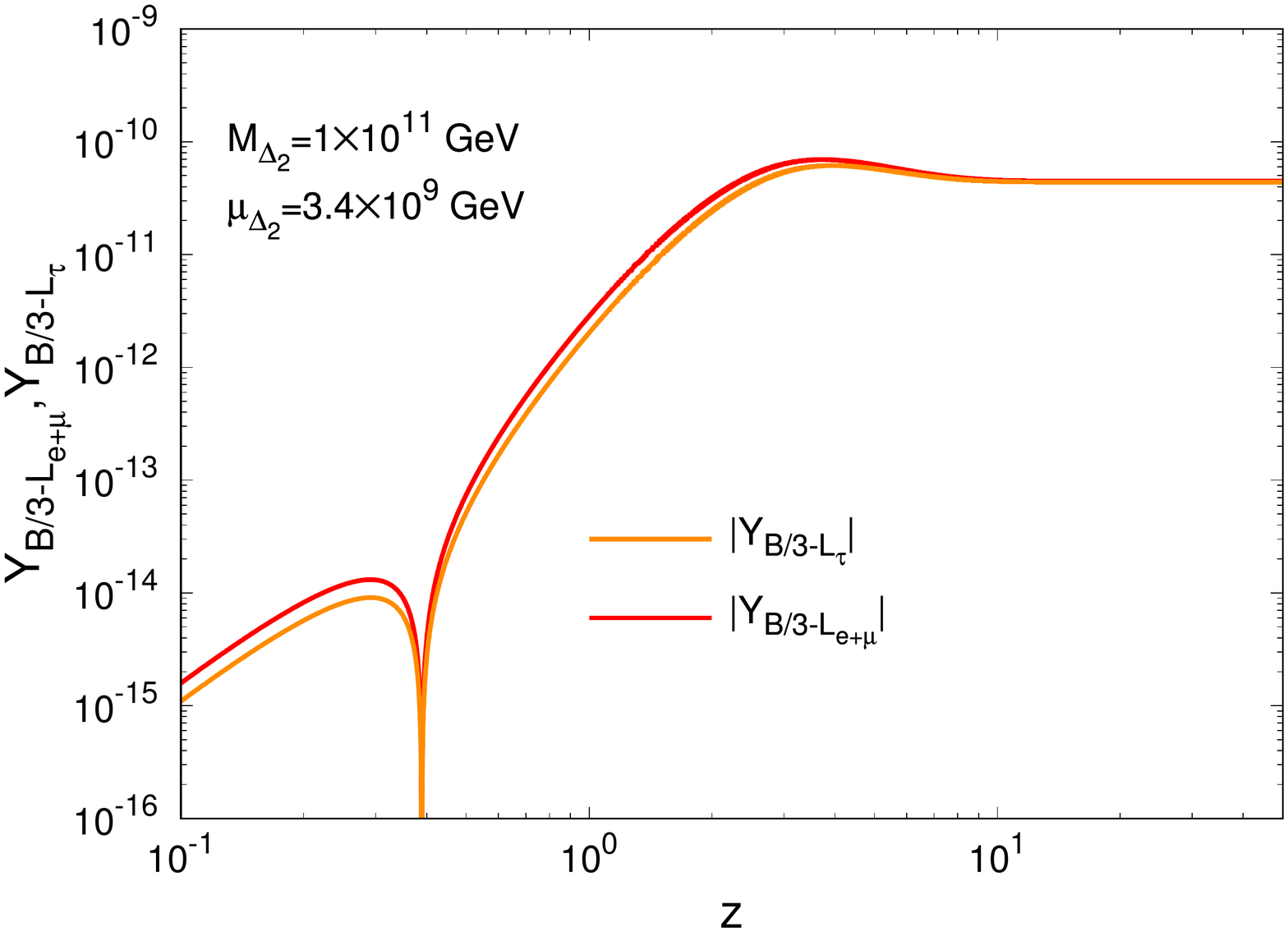}
\includegraphics[width=7.5cm,height=7cm,angle=0]{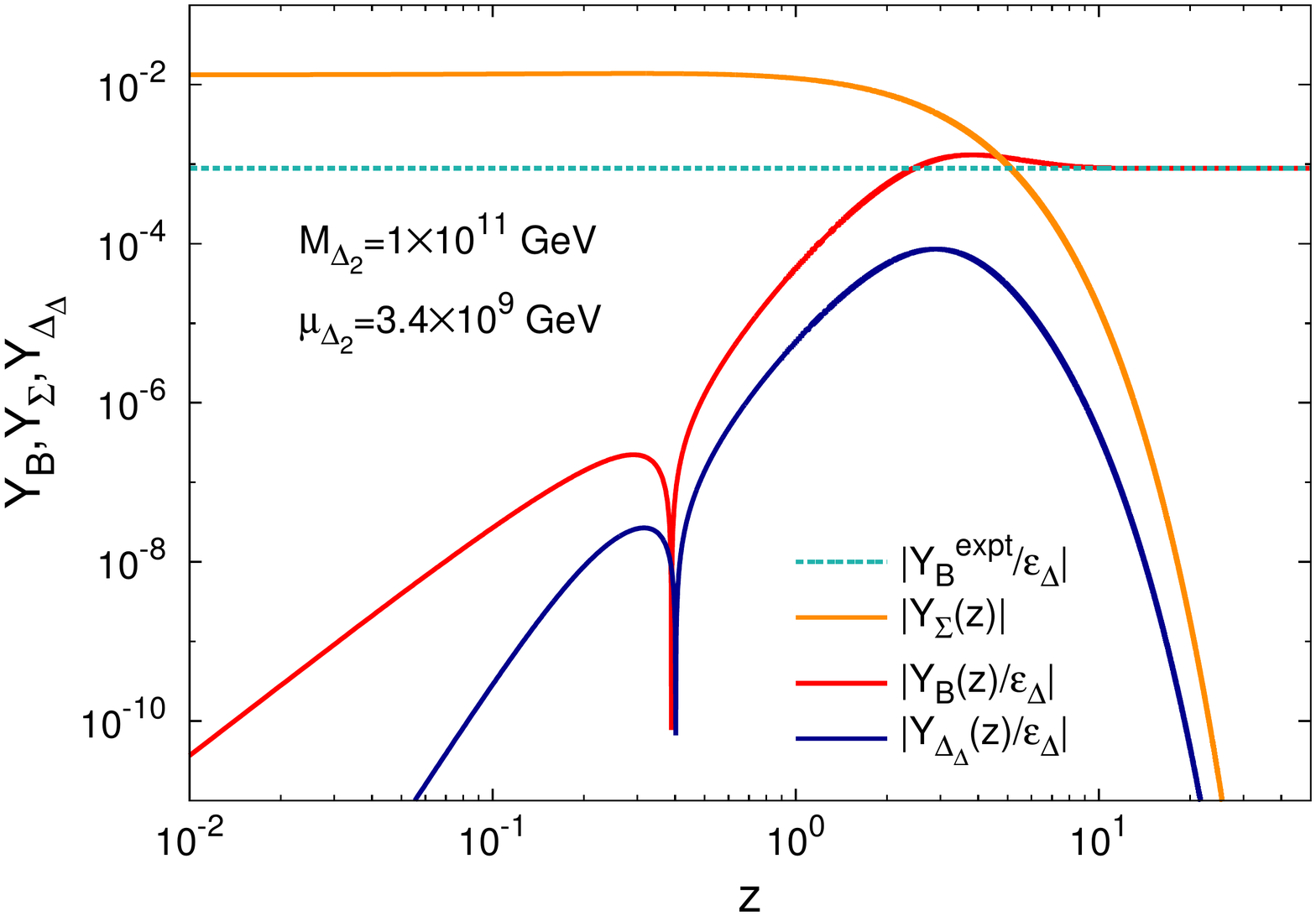}
\caption{Left-panel: Evolutions of different flavour asymmetry parameters with $z$ for the fixed set 
($M_{\Delta_2}=10^{11},\mu_{\Delta_2}=3.4\times10^{9}$ GeV).
Right-panel: Variation of $Y_\Sigma,Y_{\Delta_\Delta},Y_B$ with $z$ for the same fixed set of $(M_{\Delta_2},\mu_{\Delta_2})$.
The horizontal dashed line represents the experimental value of baryon asymmetry.
The whole analysis has been carried out using NO type light neutrino masses and the best fit.}
\label{yb_z_2f1}
\end{center}
\end{figure}

For NO type light neutrino mass hierarchy, the  ($Y_B$) parameter dependence on the trilinear coupling ($\mu_{\Delta_2}$) is shown in Fig.\ref{yb_mu_2f1} for two
fixed benchmark values of $M_{\Delta_2}=10^{11},4\times10^{11}$ GeV. For both the values of the triplet
mass, the curve intersects the horizontal line representing experimental baryon asymmetry at two places. It signifies that for each
fixed value of the triplet mass, enough asymmetry within the experimental range can be generated with two distinct values of trilinear
coupling $\mu_{\Delta_2}$, one before and the other after the peak of the curve.
Picking one such combination of triplet mass and trilinear coupling
($M_{\Delta_2}=10^{11},\mu_{\Delta_2}=3.4\times10^{9}$ GeV) we show the evolution of   different flavour asymmetry parameters with $z$ in the left-panel of Fig.\ref{yb_z_2f1} while the right-panel of the same figure depicts the variations of triplet density abundance
$(Y_\Sigma)$, triplet asymmetry $(Y_{\Delta_\Delta})$ and the baryon asymmetry parameter $(Y_B)$\footnote{In Fig.\ref{yb_z_2f1} we have scaled the variables $Y_{\Delta_\Delta},Y_B$ by sum of absolute value of the flavoured CP asymmetry parameters ($\varepsilon_\Delta=|\epsilon^{e+\mu}_\Delta|+|\epsilon^{\tau}_\Delta|$) to show all of them ($Y_\Sigma,Y_{\Delta_\Delta},Y_B$) in same figure.} which for a large value of $z$ freezes to the experimental value (shown by dashed line). 

\subsubsection{Identical ratio but random phase differences connecting $m^{(1)}_\nu$ and $m^{(2)}_\nu$}
\begin{figure}[h!]
\begin{center}
\includegraphics[width=7.5cm,height=7cm,angle=270]{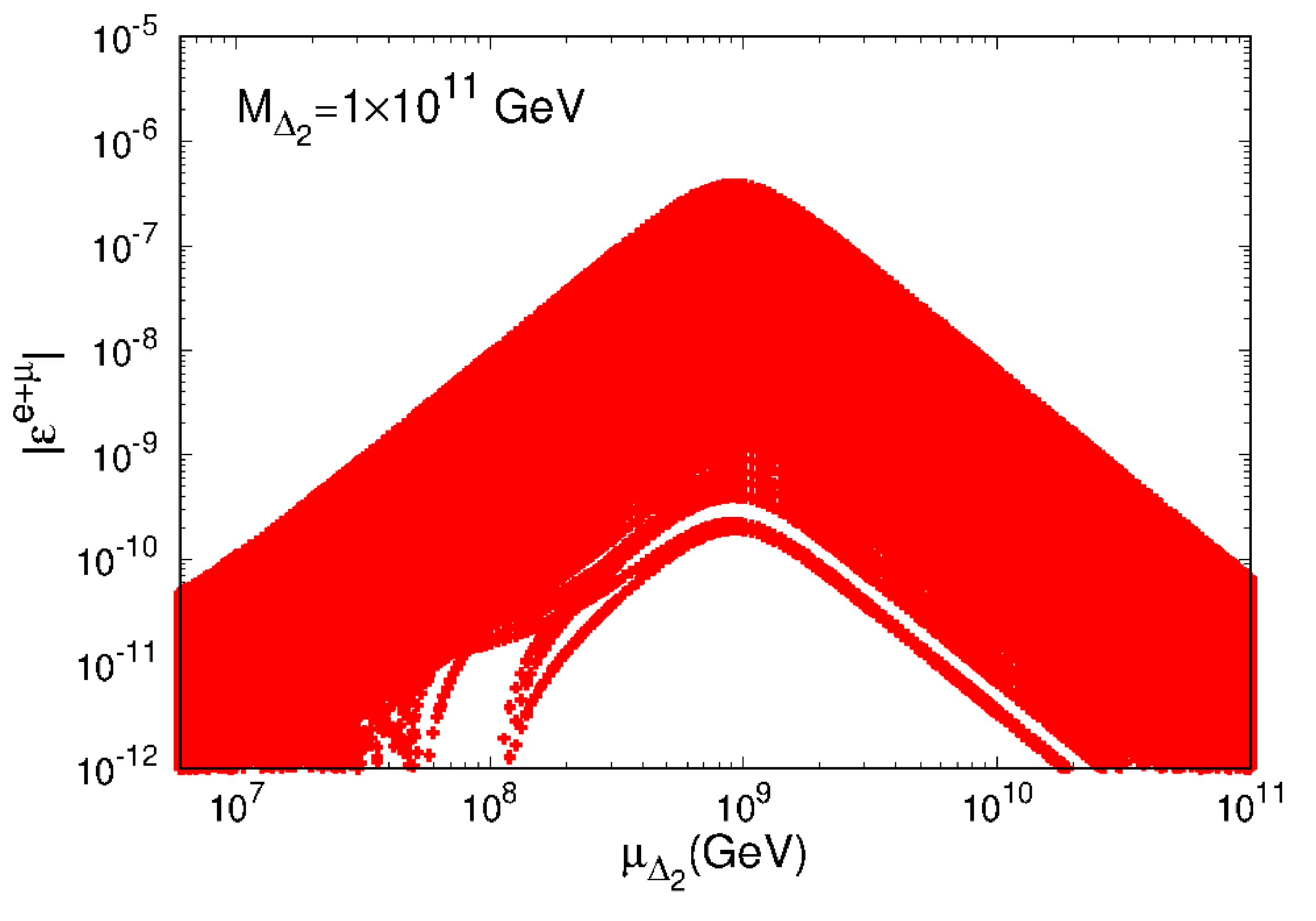}
\includegraphics[width=7.5cm,height=7cm,angle=270]{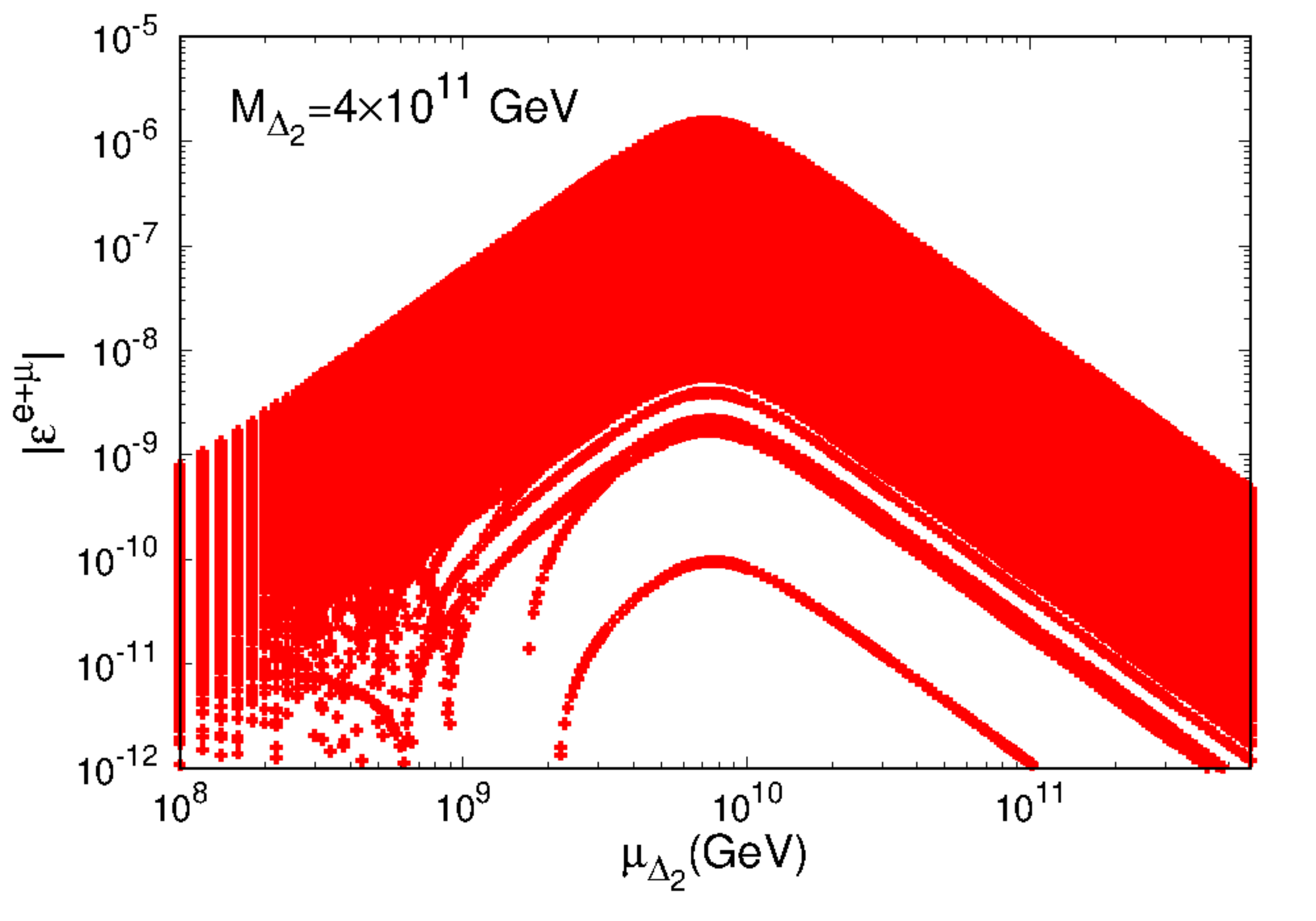}
\caption{CP asymmetry parameter corresponding to the flavour $a \equiv e+\mu$ for a wide range of values of $\mu_{\Delta_2}$. The thick
spread along the vertical axis signifies $N$ number of values of the asymmetry parameter corresponding to $N$ random sets of phase differences. Left- panel: $M_{\Delta_2}=10^{11}$ GeV (fixed); Right-panel: $M_{\Delta_2}=4\times10^{11}$ GeV (fixed).}
\label{cp_random}
\end{center}
\end{figure}

\begin{figure}[h!]
\begin{center}
\includegraphics[width=7.5cm,height=7cm,angle=0]{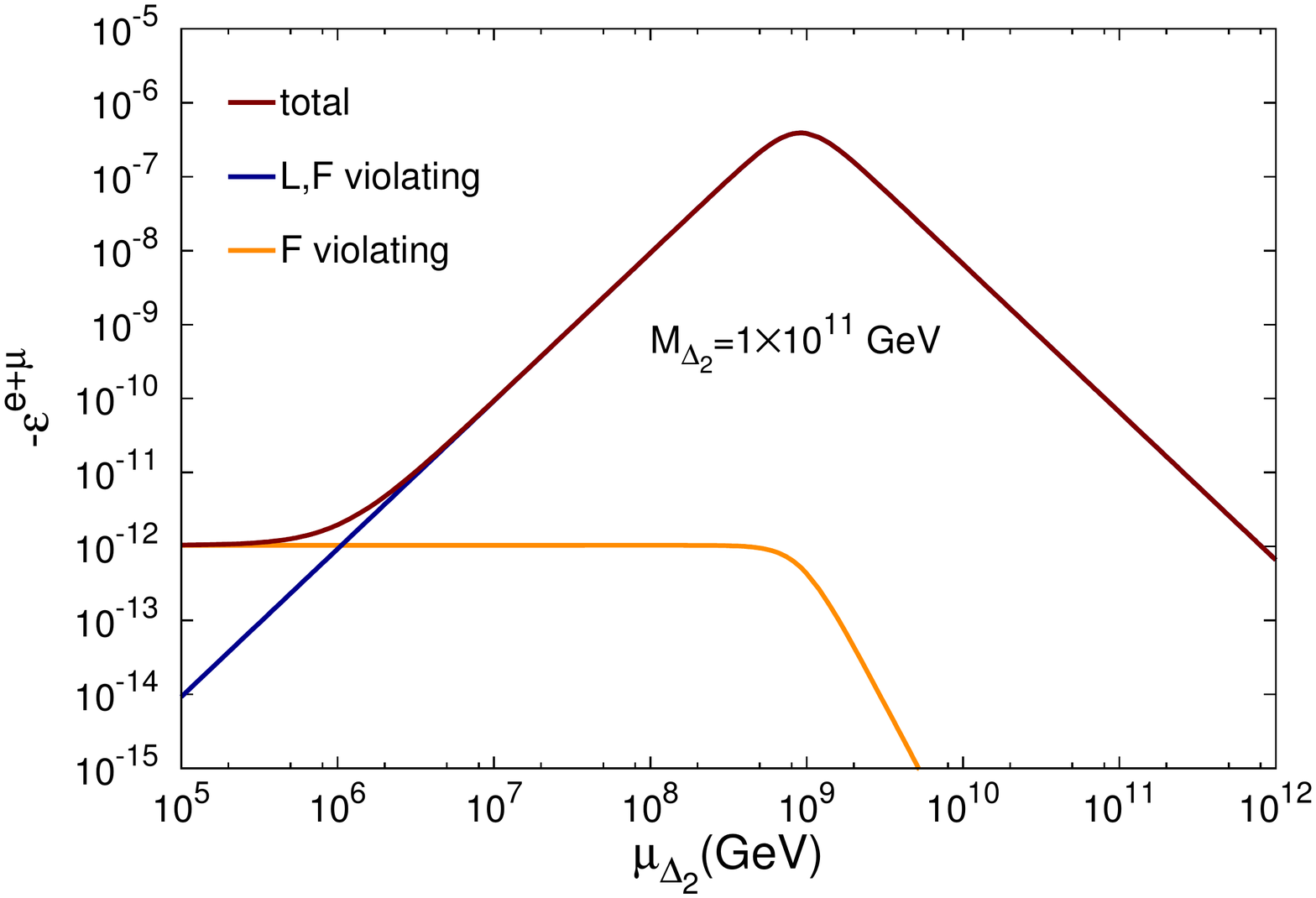}
\includegraphics[width=7.5cm,height=7cm,angle=0]{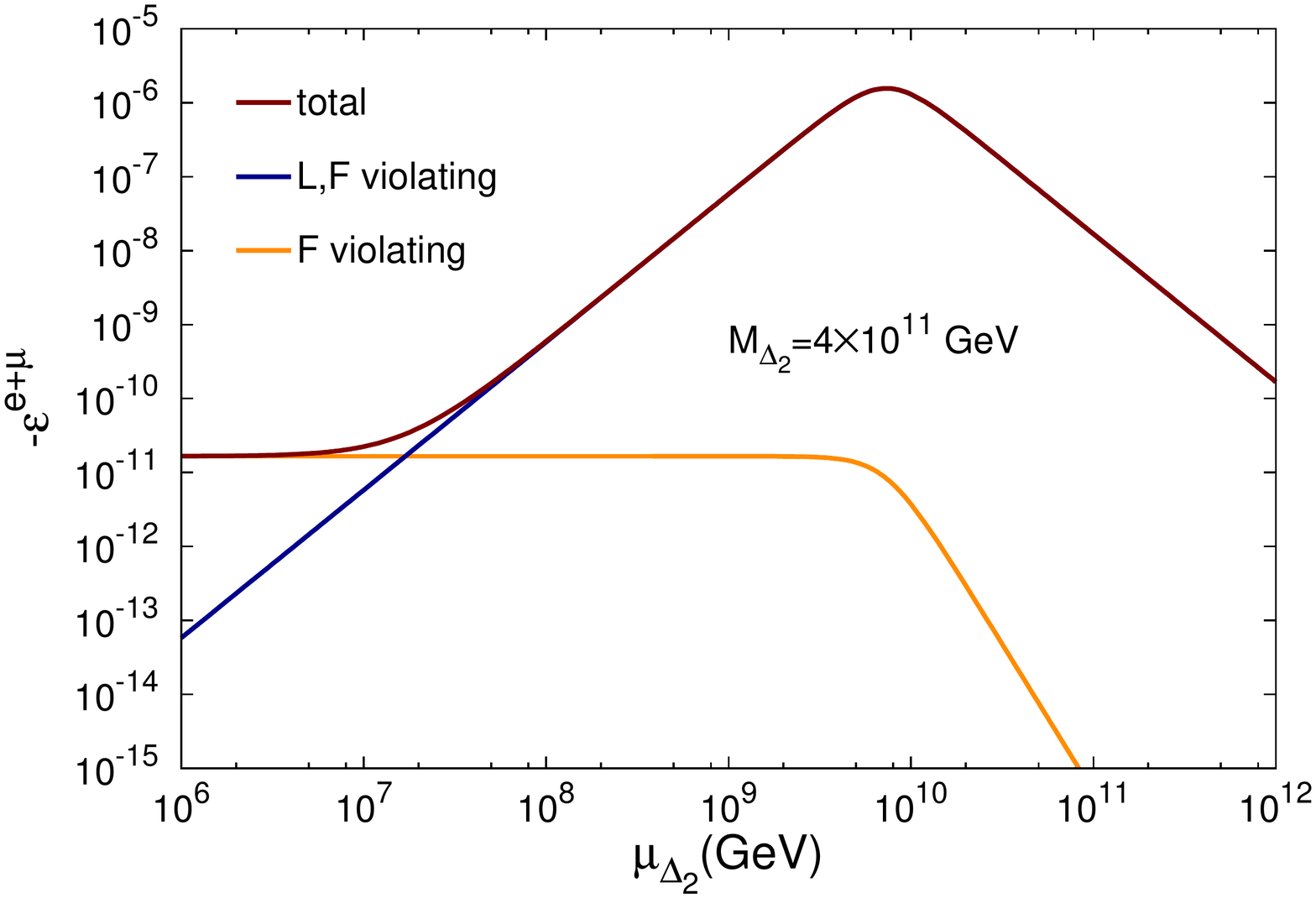}
\caption{ Relative magnitude of different components of the CP asymmetry parameter corresponding to the flavour $a \equiv e+\mu$ for a wide range of values of $\mu_{\Delta_2}$. Left panel: $M_{\Delta_2}=10^{11}$ GeV (fixed); 
Right panel: $M_{\Delta_2}=4\times10^{11}$ GeV (fixed). }
\label{cp_pfl}
\end{center}
\end{figure}

Following exactly the same procedure as  in Sec.\ref{uf_rand} we generate large number of sets of random phase differences
$\Phi_{k=1,...N} (N={\rm large~ integer})$ while taking identical values  for all ratios  $\mathcal{F}_{k=1,....N}=\{0.1,0.1,0.1,0.1,0.1,0.1\}$. Although theoretically we can find
the baryon asymmetry for all these $N$ number of sets, practically it is too time consuming. So we choose one specific set among those $N$ number of sets. Particularly, we  select that set which will produce maximum CP asymmetry. For this purpose we choose two fixed values of the lighter triplet mass $M_{\Delta_2}=10^{11},4\times10^{11}$ GeV, while permitting $\mu_{\Delta_2}$ over a wide range. For each combination of $(M_{\Delta_2},\mu_{\Delta_2})$ the flavoured CP asymmetries are 
computed taking into account all of the ($N$) random sets of phases \footnote{Here we have taken $N=10000$, i.e $10000$ random sets have been generated.}. The resulting plot is shown in Fig.\ref{cp_random} where the spread in values of CP asymmetry for a fixed value of $\mu_{\Delta_2}$ arises due to the $N$ random sets. Then we pick the top most value of  CP asymmetry from the plot and the set of phase differences as $\frac{\Phi_{k=k_{max}}}{\pi}= \{-0.711,-2.228,-1.798,-1.606,-1.809,-1.481 \}$.  Only this very set is used for all the future numerical computations. 
One interesting aspect of this scenario is that here we can have non-trivial value of the flavour violating (or purely flavoured)
CP asymmetry parameter $(\epsilon^{l_i(\not F)}_{\Delta})$ due to the unequal values of phase differences $(\phi^{(1)}_{ij}-\phi^{(2)}_{ij})$ ($i,j=1,3$) for different combinations of $i,j$. We examine whether the allowed range of parameters can meet the requirement of purely flavoured
leptogenesis (PFL) (i.e $(\epsilon^{l_i(\not F)}_{\Delta} \gg \epsilon^{l_i(\not L,\not F)}_{\Delta})$). 
In  Fig.\ref{cp_pfl} we show the relative magnitudes of the two components (only $F$ violating, ($L+F$ violating)) of the CP asymmetry parameter corresponding to the flavour $a \equiv e+\mu$ for a wide range of $\mu_{\Delta_2}$ for each value of  $M_{\Delta_2}$  kept fixed at $10^{11}$ GeV  or $4\times10^{11}$ GeV. 
It is clear from Fig.\ref{cp_pfl} that PFL condition is satisfied only over a short range of values of $\mu_{\Delta_2}$, i.e
$\mu_{\Delta_2} \lesssim 10^6$ GeV when $M_{\Delta_2}$ fixed at $10^{11}$ GeV and $\mu_{\Delta_2} \lesssim 10^7$ GeV for $M_{\Delta_2}=4\times10^{11}$ GeV. But the resulting values of the total CP asymmetry within these above mentioned range 
are too small ($\epsilon^{e+\mu}_\Delta \sim 10^{-12}-10^{-11}$) to produce enough baryon asymmetry. To get higher values of CP asymmetry we have to go to the higher values of $\mu_{\Delta_2}$ for which $\epsilon^{l_i(\not L,\not F)}_{\Delta}$ increases 
 but $\epsilon^{l_i(\not F)}_{\Delta}$ decreases with  $\mu_{\Delta_2}$ and, eventually, the (lepton number+flavour) violating asymmetry becomes much larger than the flavour violating one so that the total asymmetry merges with the $(\not L,\not F)$ component, i.e $\epsilon^{l_i}_\Delta \simeq \epsilon^{l_i(\not L,\not F)}_{\Delta}$. For successful leptogenesis (i.e to generate baryon asymmetry
at the experimental order through leptogenesis), the value of CP asymmetry required is around $(\sim 10^{-8}-10^{-6})$ depending
upon the mass of the lighter triplet. When the total CP-asymmetry lies around the range mentioned above, the flavour
violating component $\epsilon^{l_i(\not F)}_{\Delta}$ is negligibly small and total asymmetry can be considered to be
 constituted solely by the ( lepton number+flavour ) violating part ($\epsilon^{l_i(\not L,\not F)}_{\Delta}$)  which is clear from Fig.\ref{cp_pfl}. Therefore the discussions following Fig.\ref{cp_pfl} allow us to conclude that although we can generate adequate
baryon asymmetry ($Y_B$) through flavoured leptogenesis, condition of PFL can not be satisfied. In other words in the regime where PFL condition is satisfied, the resulting CP asymmetry comes out to be so small that it can not generate $Y_B$ within the experimental range.
\begin{figure}[h!]
\begin{center}
\includegraphics[width=10cm,height=9cm,angle=0]{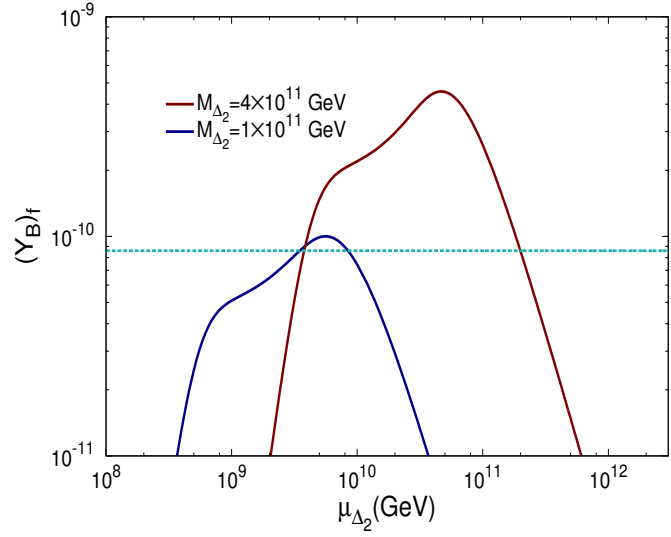}
\caption{Variation final value of baryon asymmetry with trilinear coupling for two fixed benchmark values of lighter triplet mass
($M_{\Delta_2}=10^{11}$ GeV and $M_{\Delta_2}=4\times10^{11}$ GeV).
The horizontal dashed line represents the experimental value of baryon asymmetry (the whole analysis has been carried out assuming NO 
for light neutrino masses).}
\label{yb_mu_2f2}
\end{center}
\end{figure}
\\

\begin{figure}[h!]
\begin{center}
\includegraphics[width=7.5cm,height=7cm,angle=0]{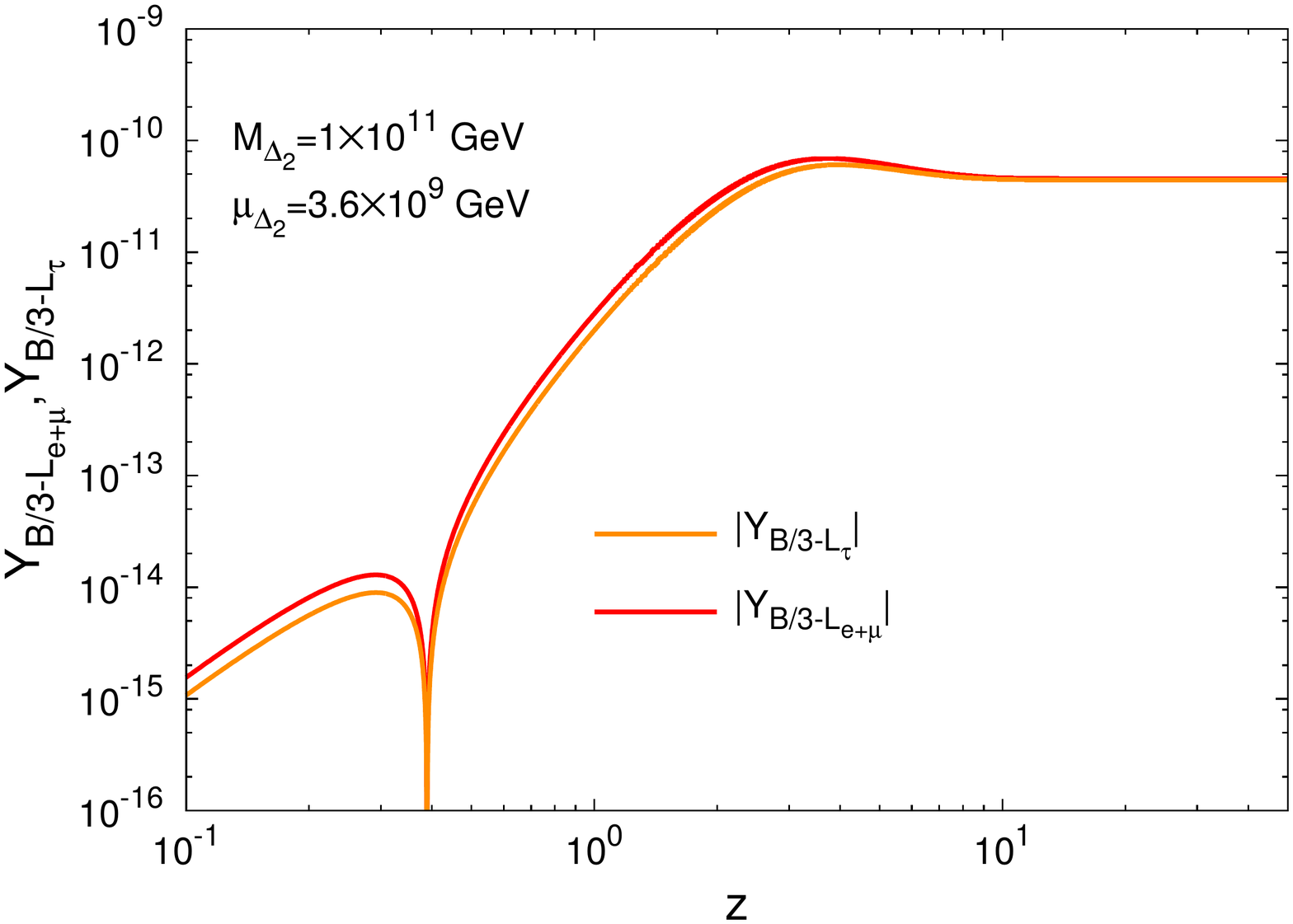}
\includegraphics[width=7.5cm,height=7cm,angle=0]{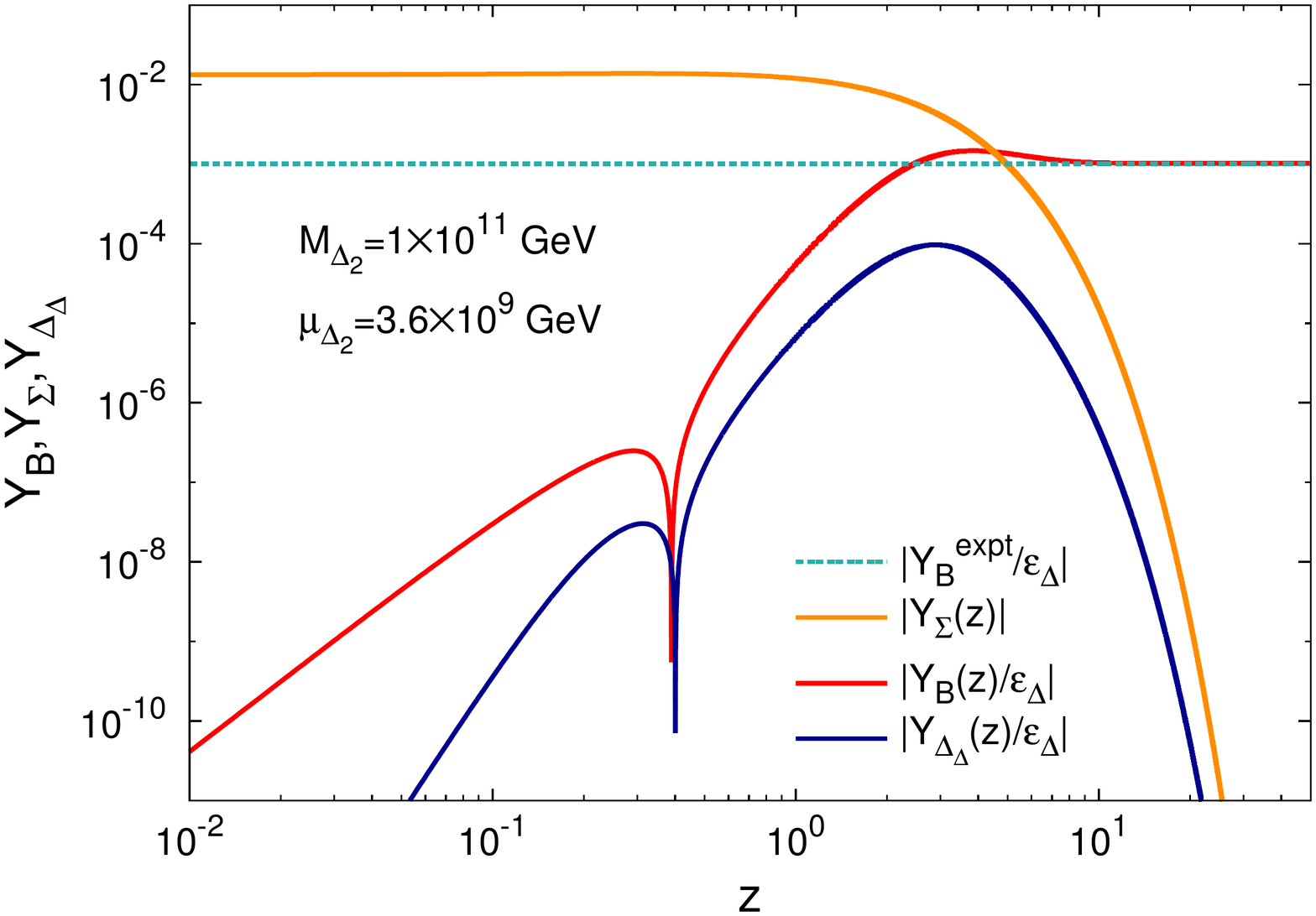}
\caption{Left-panel: Evolution of baryon asymmetry parameters with $z$ for  fixed values of
($M_{\Delta_2}=10^{11},\mu_{\Delta_2}=3.6\times10^{9}$ GeV) using best fit to neutrino oscillation data with NO type light neutrino masses.
Right panel: Variation of $Y_\Sigma,Y_{\Delta_\Delta},Y_B$ with $z$ for the same fixed set of $(M_{\Delta_2},\mu_{\Delta_2})$.
The horizontal dashed line represents the experimental value of baryon asymmetry.}
\label{yb_z_2f2}
\end{center}
\end{figure}

We now plot the final values of the baryon asymmetry parameter $({Y_B})_f$ as a function of the trilinear coupling $(\mu_{\Delta_2})$
in Fig.\ref{yb_mu_2f2} for two fixed benchmark values of the triplet mass $(M_{\Delta_2})$. Both the curves ($({Y_B})_f$ vs $(\mu_{\Delta_2})$) intersect the horizontal dashed line representing the experimental value of $Y_B$  at two 
places which in turn indicates that for each fixed value of $(M_{\Delta_2})$ there are two $(M_{\Delta_2},\mu_{\Delta_2})$ 
combinations which successfully generates the desired
value of baryon asymmetry in the experimental range. Out of the two intersecting points located on the horizontal line in the $({Y_B})_f$ vs $(\mu_{\Delta_2})$ curve, we choose the extreme left point corresponding to fixed $M_{\Delta_2}=10^{11}$ GeV curve which identifies this point with
($M_{\Delta_2}=10^{11},\mu_{\Delta_2}=3.6\times10^{9}$ GeV)). Correspondingly we show the variation of flavour asymmetry parameters
$(Y_{B/3-L_{e+\mu}},Y_{B/3-L_{\tau}})$ with $z$ in the left panel of Fig.\ref{yb_z_2f2} while the right panel of the same figure depicts the evolution of other variables  occurring in solutions of Boltzmann equations including  the baryon asymmetry. The left-panel and the right-panel of this figure clearly indicate that,
for this specific choice of parameters, the baryon asymmetry finally freezes-in to the desired constant value within the experimental range 
at a large enough value of $z$ (or at sufficiently low temperature).

\section{Minimal model extension for dark matter and vacuum stability}\label{sec:dmvac}
The inert scalar doublet model has radiative seesaw ansatz for neutrino masses and intrinsic capability for dark matter \cite{Ma:2006} which has been also shown to originate from SO(10) \cite{mkp:2011} with matter parity \cite{Kadastic:2009,Hambye:2010,Ma:2018}  as the stabilising discrete symmetry. More recently new
possible origin of scotogenic dark matter stability has been also suggested from softly broken global lepton number symmetry $U(1)_L$ \cite{Ma:2020}.
This inert doublet model \cite{Ma:2006} also  does not have vacuum instability problem in the associated scalar potential. But
the two heavy Higgs scalar triplet model \cite{Ma-Us:1998} (or the purely triplet seesaw model \cite{Sierra:2014tqa}), as such, does not possess
  dark matter through which it can explain
cosmological evidences including the observed  relic density 
$(\Omega_{\rm DM}h^2=0.1172-0.1224)$
\cite{DMexpt,Planck15,wmap}. The expected DM mass has been also
bounded from direct and indirect detection experiments
\cite{Akerib:2016,Aprile:2017,Aprile:2018,Cui:2017}. This issue has been also addressed in a number of ways in SM
extensions through a singlet scalar representing a weakly interacting
massive particle (WIMP)\cite{WIMP} as  DM candidate and the investigations have been also updated
more recently in
\cite{GAMBIT}. But most of the models discussed in \cite{GAMBIT} and earlier have not addressed  neutrino oscillation data, cosmological bound, and baryon asymmetry via leptogenesis. Also they have not addressed the issue on   
the vacuum stability of the associated scalar potential \cite{Espinosa,Lebedev}. Using corresponding renormalisation group evolutions (RGEs) discussed in the Appendix we find that  in the  two heavy Higgs
triplet model \cite{Ma-Us:1998} with $M_{\Delta_i}(i=1,2) \ge 10^{13}$ GeV, although the stability  has been  improved by predicting the Higgs quartic coupling $\lambda_{\phi}$ to be positive in an extended region
 with  
$|\phi| \ge  10^{13}$ GeV, the problem has not been completely resolved.  In particular, we note that in this model \cite{Ma-Us:1998} the standard Higgs quartic coupling $\lambda_{\phi}$ runs negative in the interval $|\phi|\simeq
10^{10}- 10^{13}$ GeV showing the persistence of vacuum instability of the scalar potential \cite{Espinosa,Lebedev}. Such instability also persists in the more recent investigation of the two-triplet model \cite{Sierra:2014tqa}.
In this section we discuss how the  heavy Higgs triplet model that
accounts for neutrino mass and baryon asymmetry as discussed above
can also be easily extended further to account for the phenomena of
WIMP DM while completing  vacuum stability through the
same scalar DM.
We add a real scalar singlet $\xi$ to the two Higgs triplet model \cite{Ma-Us:1998} and  assume
an additional $Z_2$ discrete symmetry under which $\xi$ and all SM fermions are odd. All other scalars including the SM Higgs $\phi$ and the two triplets are assumed to possess $Z_2=+1$. Thus the resulting Lagrangian after this real scalar extension has the symmetry $SU(2)_L\times U(1)_Y \times SU(3)_C \times Z_2
(\equiv G_{213}\times Z_2)$. The particle content and their charges in the minimally  extended model under this symmetry are shown in Table \ref{tab:smext}.

\begin{table}[!h]
\caption{Singlet scalar extensions of the two Higgs triplet model \cite{Ma-Us:1998} and its particle content with respective charges under $G_{213}\times Z_2$ symmetry. 
The  second and the third generation fermions not shown in this Table have identical transformation properties.} 
\label{tab:smext}
\begin{center}
\begin{tabular}{|c|c|c|}
\hline
{ Particle } & SM charges &{$Z_2$ charge}\\
\hline
$(\nu, e)^T_L$ & $(2,-1/2,1)$ & $-1$\\
$e_R$ & $(1,-1,1)$ & $-1$\\
$(u,d)_L^T$ & $(2,1/6,3)$ & $-1$\\
$u_R$ & $(1, 2/3,3)$ & $-1$\\
$d_R$ & $(1,-1/3,3)$ & $-1$\\
$\phi$ & $(2,1/2,1)$ & $+1$\\
$\Delta_1$ & $(3,-1,1)$ & $+1$\\
$\Delta_2$ & $(3,-1,1)$ & $+1$\\
$\xi$ & $(1,0,1)$ & $-1$\\
\hline
\end{tabular}
\end{center}
\end{table}

\subsection{Real scalar singlet dark matter }\label{sec:redm}     
At all lower mass scales $\mu \ll M_{\Delta_i}(i=1,2)$ noting that  the two heavy Higgs triplets in the Lagrangian of \cite{Ma-Us:1998} are expected to have decoupled leading, effectively, to the SM scalar potential
\par\noindent{\bf $\mu \ll M_{\Delta_i}(i=1,2)$:}\\
\beq
V_{SM}= -\mu_H^2\phi^{\dagger}{\phi}+\lambda_{\phi}(\phi^{\dagger}\phi)^2. \label{eq:smpot}
\eeq
It is well known that this SM potential alone  develops vacuum instability as the quartic coupling $\lambda_{\phi}$ runs 
negative at energy scales
 $\mu \ge 5\times 10^9$ GeV \cite{Espinosa,Lebedev}. In models with type-I see saw extensions of the SM, the negativity of
$\lambda_{\phi}$ is further enhanced due to RHN Yukawa interactions. This latter type of enhancement  due to
RHN is absent in the purely triplet leptogenesis model \cite{Ma-Us:1998}.
Using renormalisation group equations discussed in the Appendix we find that the SM Higgs quartic coupling remains positive for field values $|\phi|\le 5\times 10^9$ GeV and $|\phi|\ge 10^{13}$ GeV where the latter limit is due to $M_{\Delta_2}=10^{13}$ GeV in \cite{Ma-Us:1998}. Although such positive values 
of Higgs quartic coupling is a considerable improvement over purely SM running,
the model \cite{Ma-Us:1998} does not resolve the vacuum instability issue completely.
This is due to the fact that standard Higgs quartic coupling in the model \cite{Ma-Us:1998} acquires negative values in the region $|\phi|\simeq 5\times 10^9$ GeV to  $|\phi|\simeq 10^{13}$ GeV. Further details of discussion of this problem  has been made below in Sec.\ref{sec:vstab}.\\  
 
In order to resolve both the issues on DM and vacuum stability of the scalar potential, we make a
simple extension of the model \cite{Ma-Us:1998} by adding a real scalar singlet
$\xi$ whose mass we determine from DM relic density, direct detection experimental bounds and vacuum stability fits. For the stability of DM we impose a $Z_2$ discrete symmetry under
which $\xi$ and all SM fermions are odd, but all other scalars in the
extended model are even under $Z_2$ as shown in Table 
\ref{tab:smext}. The scalar potential is now modified in the presence of $\xi$ for mass scales $\mu < M_{\Delta_2}$
\begin{equation}
V_{\xi}=V_{SM} +\mu_\xi^2 \xi^2  +\lambda_\xi\xi^4 +
2 \lambda_{\phi \xi}  (\phi^\dagger \phi)\xi^2. \label{eq:vchi}
\end{equation}
In eq.(\ref{eq:vchi}) $\lambda_{\xi}=$  dark matter self coupling,
$\lambda_{\phi \xi}=$  Higgs portal coupling   and
 $\mu_{\xi}=$  mass of $\xi$. 
 The VEV of the standard Higgs doublet redefines the DM mass
 parameter        
\begin{eqnarray}
&& M_{DM}^2 = 2(\mu_\xi^2 +\lambda_{\phi \xi}^2 v^2), \nonumber\\
&& m_\phi^2 =2 \mu_H^2= 2 \lambda_\phi v^2. \label{eq:chim} 
\end{eqnarray}
For mass scales $\mu \ge M_{\Delta_2}$ the Higgs potential receives additional
contributions due to $\Delta_i (i=1,2)$ and its interactions with others \\
\par\noindent{$\mu \ge M_{\Delta_2}$:}\\
\begin{eqnarray}
V_{\xi \Delta}&=&V_{\xi}+\sum_{(i=1,2)}\left(M_{\Delta_i}^2{\rm
  Tr}(\Delta_i^{\dagger}\Delta_i)+\lambda_1^i[{\rm Tr}
  (\Delta_i^{\dagger}\Delta_i)]^2
+\lambda_2^i[{\rm Tr} (\Delta_i^{\dagger}\Delta_i)]^2-{\rm Tr}[(\Delta_i^{\dagger}\Delta_i)^2]\right)\,\,\nonumber\\
&+&\sum_{(i=1,2)}\left( \lambda_3^i(\phi^{\dagger}\phi){\rm Tr}(\Delta_i^{\dagger}\Delta_i)
+\lambda_4^i\phi^{\dagger}[(\Delta_i^{\dagger}\Delta_i)-(\Delta_i\Delta_i^{\dagger})]\phi+\left[\frac{\mu_i}{\sqrt
    2}\phi^Ti\tau_2\Delta_i^{\dagger}\phi+H.c.\right]\right)\nonumber\\
 &+&\sum_{(i=1,2)} \lambda_{\xi}^i{\rm Tr}(\Delta_i^{\dagger}\Delta_i)\xi^2  \label{eq:vchidel1}
\end{eqnarray}
where $V_{\xi}$ has been defined in eq.(\ref{eq:vchi}).\\

In order to  examine the allowed values of the Higgs portal
coupling $\lambda_{\phi \xi}$, we use two different kinds of
experimental results: (i) bounds on cosmological DM relic density
\cite{Planck15,wmap} $\Omega_{\rm DM}h^2=0.1172-0.1224$, (ii)bounds from  DM direct detection experiments such as LUX-2016\cite{Akerib:2016},
XENON1T\cite{Aprile:2017,Aprile:2018} and PANDA-X-II\cite{Cui:2017}.
Using our ansatz we  estimate the relic
densities for different combinations of  $m_\xi,\lambda_{\phi \chi}$. It is then easy to restrict the values of $m_\xi$ and $\lambda_{\phi \xi}$ using the bound on relic density mentioned above. In direct detection experiments it is assumed that WIMPs passing through earth scatter elastically from the target material of the detector.
The energy transfer to the detector nuclei  can be measured through various types of signals. All those direct detection experiments provide
DM mass vs DM-nucleon scattering cross section plot which clearly
separates the allowed regions below the predicted curve from the forbidden
regions above the curve.

\subsubsection{Estimation  of dark matter relic density}\label{sec:relic}
We assume the WIMP
DM particle $\xi$  to have  decoupled from the thermal bath at some early
epoch which has thus remained as a thermal relic. The following conventions are used at a certain stage of evolution of the Universe.
Denoting $\Gamma =$  particle decay rate and $H=$ Hubble parameter,  
 a particle species is
said to be  coupled  if  $\Gamma > H$. Similarly it is assumed to have
 decoupled if $\Gamma < H$. 
 The corresponding Boltzmann equation \cite{Kolb:1990,Bertone:2004} is solved for the estimation of the particle relic density
\begin{equation}
\frac{d n}{d t}+3Hn =-\langle \sigma v_0 \rangle (n^2 -n_{eq}^2) \label{boltz}
\end{equation}
Here $n=$  actual number density of $\xi$ at a certain
instant of time, $n_{eq}=$ its equilibrium number density,
 $v_0=$ velocity of $\xi$, and $\langle \sigma v_0 \rangle=$  thermally averaged
DM annihilation cross section. Approximate solution of Boltzmann equation gives the expression for the relic density \cite{Bertone:2004,Gondolo:1990dk} 
\begin{equation}
\Omega_{\rm DM} h^2= \frac{1.07\times10^9 x_F}{\sqrt{g_\ast} M_{ pl}\langle \sigma v_0 \rangle} \label{relic}
\end{equation}
where $x_F=m_\xi/T_F$,  $T_F=$ freeze-out temperature, $g_\ast=$  effective number of massless degrees of freedom and  $M_{pl}=1.22\times10^{19}$ GeV. This $x_F$ can be computed by iteratively solving the equation
\begin{equation}
x_F=\ln \left(\frac{m_\xi}{2\pi^3}\sqrt{\frac{45 M_{pl}^2}{8 g_\ast x_F}} \langle\sigma v_0 \rangle \right) .\label{xf}
\end{equation}
In  eq.(\ref{relic}) and eq.(\ref{xf}), the only particle physics
input is  the thermally averaged annihilation
cross section.
The total annihilation cross section is obtained by summing over all
the annihilation channels of the singlet DM which are $\xi\xi
\rightarrow { F}{\bar F},W^+W^-,ZZ,hh$
where the symbol $F$ represents all the associated fermions of SM. Using the expression of total annihilation 
cross section \cite{McDonald:1993ex,Guo:2010hq,Biswas:2011td} in eq.(\ref{xf}) at first we compute
$x_F$ which is then utilised in eq.(\ref{relic}) to yield the relic density. Two free parameters involved in this computation are mass of the 
 DM particle $m_\xi$ and the Higgs portal coupling $\lambda_{\phi \xi}$. The relic density has been estimated for a wide range of values of the 
DM matter mass ranging from few GeVs to few TeVs while the coupling $\lambda_{\phi \xi}$ is also varied simultaneously in the range $(10^{-4}-1)$.
The parameter space $(m_\xi,\lambda_{\phi\xi})$ is thus constrained by using
the bound on the relic density  reported by WMAP \cite{wmap} and
Planck \cite{Planck15}. In Fig.\ref{cons_para}
we show only those combinations of $\lambda_{\phi \xi}$  and $m_\xi$
which are capable of producing relic density in the experimentally
observed range.

\subsubsection{Dark matter mass bounds from direct detection experiments}\label{sec:bound}
\begin{figure}[h!]
\begin{center}
\includegraphics[scale=0.5,angle=270]{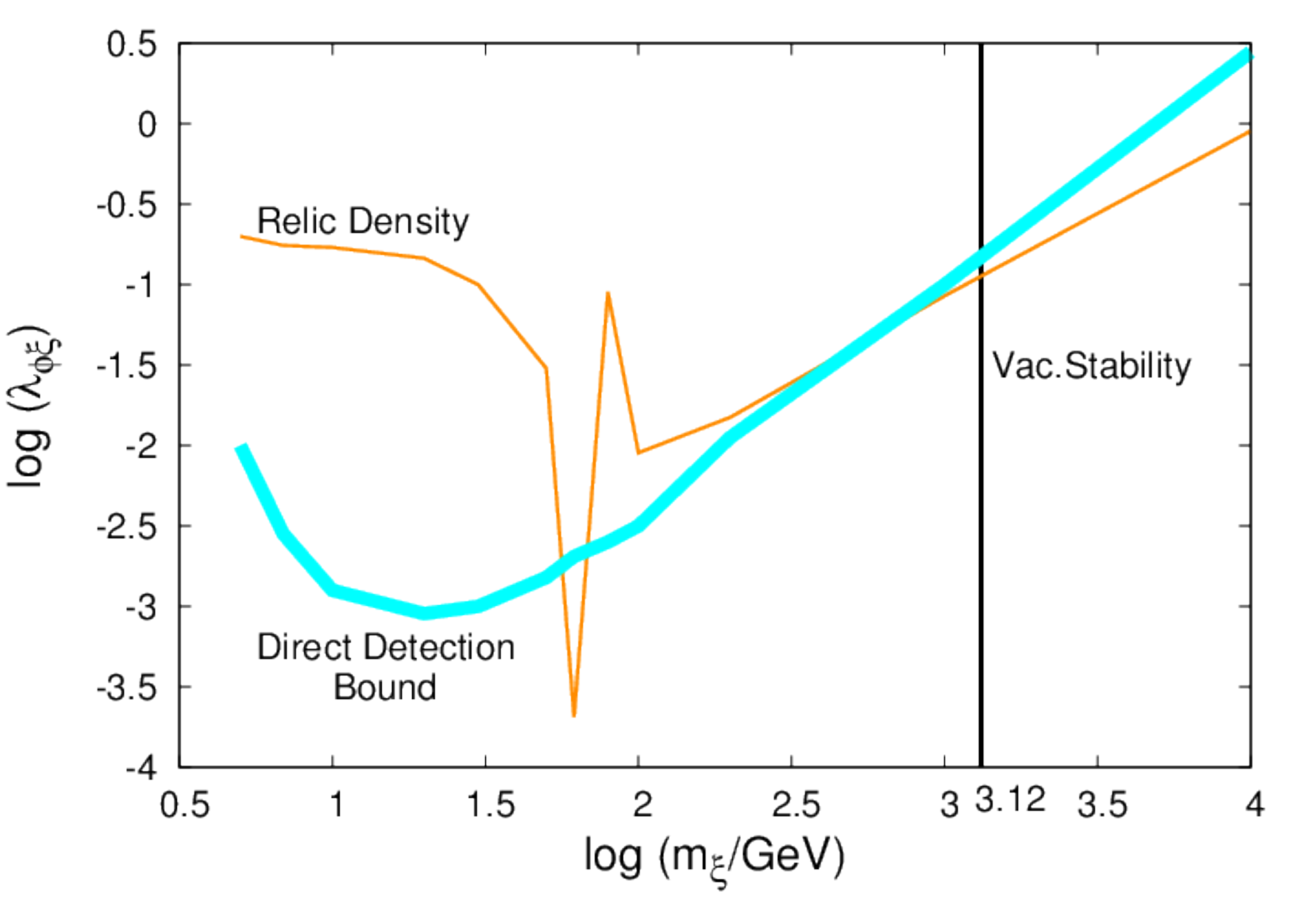}
\caption{ Determination of  dark matter mass from observed relic density,
direct detection experiments, and vacuum stability:
The yellow curve denotes the values of the parameters $(\lambda_{\phi\xi},m_\xi)$ allowed by the relic density bound $(\Omega_{\rm DM}h^2=0.1172-0.1224)$.
The cyan band represents overlapping exclusion plots from direct detection experiments of LUX-2016, XENON1T(2017)
and PANDA-XII(2017) for which any region below (above) the green band is  
allowed (forbidden). The vertical line at $\log (m_{\xi})=3.1$ ($m_{\xi}=1.3$ TeV) is due to limit set by vacuum stability  of the scalar potential as discussed
 in Sec.\ref{sec:vstab}. }
\label{cons_para}
\end{center}
\end{figure}

 We get exclusion plots of DM-nucleon scattering cross section and DM mass from different direct
detection experiments. The spin independent scattering cross section of singlet DM on nucleon is \cite{Cline:2013gha}
\begin{equation}
\sigma^{\rm SI}=\frac{f_n^2\lambda_{\phi \xi}^2 \mu_R^2 m_N^2}{4\pi m_\xi^2 m_h^4} ~~({\rm  cm}^2), \label{sigsi}
\end{equation}
where $m_h=$ mass of the SM Higgs ($\sim 125$ GeV), $m_N=$ nucleon mass $\sim 939$ MeV, $\mu_R=(m_\xi m_N)/(m_\xi+m_N)=$  reduced 
DM-nucleon mass and the factor $f_n \sim 0.3$. Using eq.(\ref{sigsi}) the exclusion plots in the $\sigma-m_\xi$ plane can be easily brought to $\lambda_{\phi\xi} - m_\xi$ plane. We superimpose the $\lambda_{\phi\xi}$ versus $m_\xi$ plots for different experiments on the plot of allowed parameter space constrained by relic density bound resulting in  Fig.\ref{cons_para}. Thus the Fig.\ref{cons_para} exhibits
 the parameter space $(\lambda_{\phi\xi}~vs.~m_\xi)$ constrained by both the relic density bound and the direct detection experiments. 
\paragraph{}
From Fig.\ref{cons_para} we note that  the points on the yellow curve lying below the green band are allowed by both relic density and direct detection experiments. This predicts lower values of DM mass  in the region $m_{\xi}\simeq 59-63$ GeV for the Higgs portal coupling  $\lambda_{\phi,\xi} \le 10^{-3}$ which is too small to be compatible with vacuum stability competion discussed below in Sec.\ref{sec:vstab}. All other values of DM masses  $ m_{\xi} \ge 750$ GeV are also allowed 
 by  relic density and direct detection experimental constraints.
But as discussed below this region will be further constrained by vacuum stability criteria.
 
\subsubsection{Resolution of vacuum instability}\label{sec:vstab}
\begin{figure}[h!]
\begin{center}
\includegraphics[scale=0.5,angle=0]{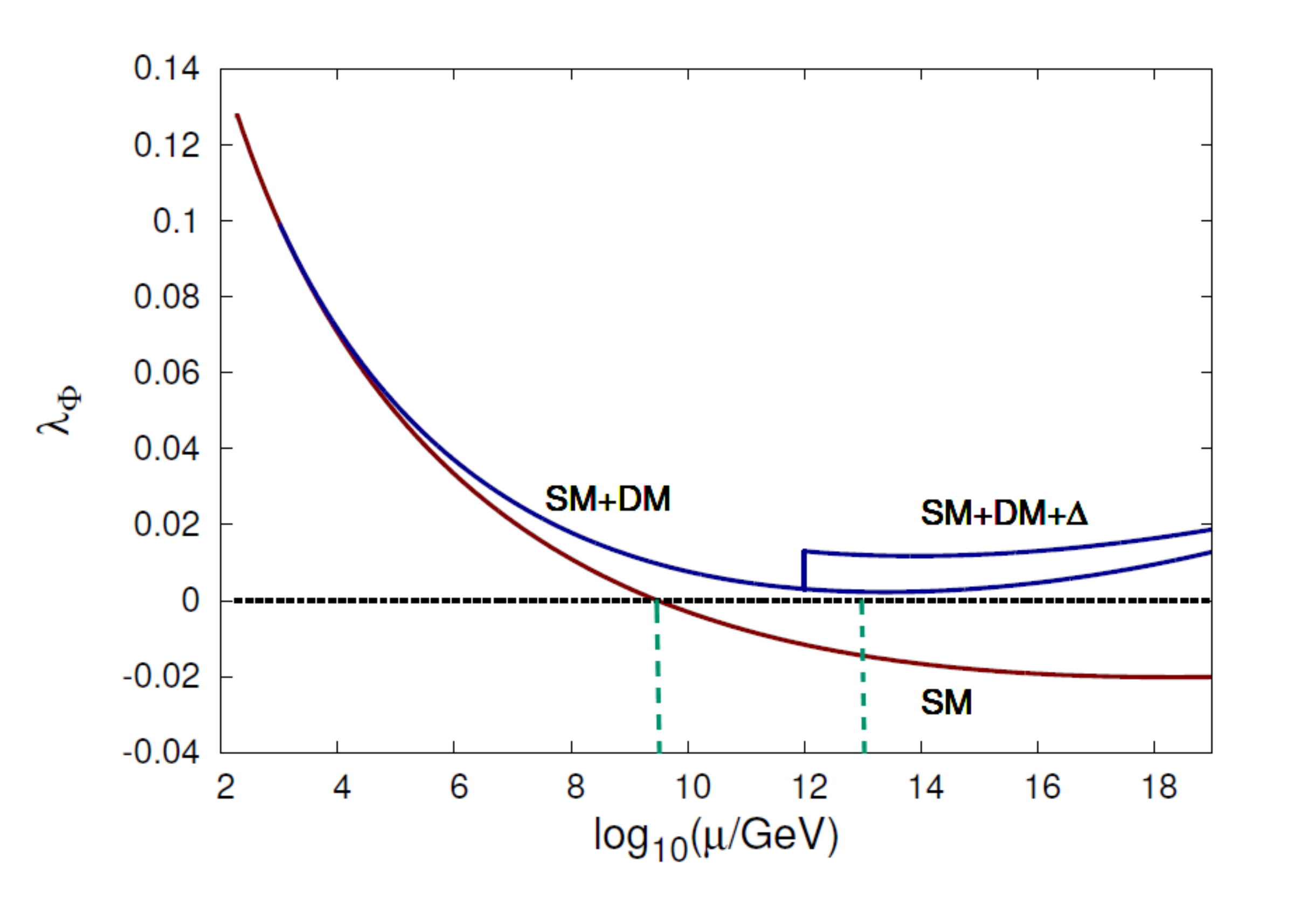}
\caption{Renormalization group evolution of Higgs quartic coupling ( denoted as
$\lambda_{\Phi}$) as a function of 
  scalar field value $\mu=|\phi|$ showing presence of vacuum 
  instability in the SM (lower red curve) for $\mu > 5\times 10^9$
  GeV. The vertical green dashed lines represent boundaries of the
  region within which 
  $\lambda_{\phi}$ runs negative for $5\times 10^9 {\rm \,\,GeV} < \mu
  <  M_{\Delta_2}\simeq 10^{13} {\rm \,\, GeV}$ in the model of
  \cite{Ma-Us:1998}. The middle blue curve marked as SM$+$DM represents evolution of
  $\lambda_{\phi}$ in the presence of real scalar DM $\xi$, excluding triplets, in the present model extension. Additional RG
  correction in the present model  due to triplet masses has been shown by the  uppermost curve marked as SM$+$DM$+$$\Delta$ where threshold enhancement due to $\Delta_2$ mass has been also included.}
\label{dm_stab}
\end{center}
\end{figure}

We have examined vacuum stability of different scalar potentials encountered in different regions of Higgs field value $\mu=|\phi|$ starting from $\mu=m_{top}-M_{Planck}$  through the
renormalisation group evolutions (RGEs) of the standard Higgs ($\phi$)
quartic coupling in the respective cases \cite{Chabab:2015,Haba:2016,Garg:2017,Kannike:2016} 
which have been given in Sec.\ref{sec:rge} of the Appendix.
At first using RGEs for Higgs quartic
coupling $\lambda_{\phi}$ and gauge and top quark Yukawa
couplings for the SM alone in the absence of DM $\xi$ or  heavy triplets, we have
plotted the quartic coupling against standard Higgs field values
$\mu=|\phi|=m_{top}-M_{Planck}$. As already noted \cite{Espinosa,Lebedev} $\lambda_{\phi}(\mu)$
runs negative for all field values $\mu \ge 5\times 10^9$ GeV  clearly
exhibiting vacuum instability of the SM Higgs potential. This has been
shown by the lower curve in Fig.\ref{dm_stab}. We next
examined the evolution of  $\lambda_{\phi}(\mu)$ in the two-heavy
Higgs triplet extension model \cite{Ma-Us:1998} using
$M_{\Delta_2} \simeq M_{\Delta_1}\simeq 10^{13}$ GeV but in the absence
of DM $\xi$. Besides being positive for $\mu < 5\times 10^9$ GeV, the Higgs
potential became definitely positive for field values $\mu \ge
10^{13}$ GeV with considerable improvement on the stability. However, the quartic coupling is found to be negative for field values in the range
 $\mu=5\times 10^{9}$ GeV  to $\mu=10^{13}$ GeV as
demarcated by the two vertical green dashed lines in
Fig.\ref{dm_stab}). We next included the effect of DM $\xi$ and the Higgs portal coupling $\lambda_{\phi\xi}$
 through the DM modified Higgs potential $V_{\xi}$ ignoring the presence of Higgs triplets in the model extension. The quartic coupling $\lambda_{\phi}$ was found to be
positive in the entire region of Higgs field values until the Planck
mass. This behaviour has been shown by the upper curve in
Fig.\ref{dm_stab} excluding the threshold like enhancement at
$\mu=10^{12}$ GeV. Finally the combined effects of DM $\xi$ and the
heavy Higgs triplets have been included on the Higgs quartic coupling running
where the effect of heavy Higgs triplets occurs only for $\mu
\ge M_{\Delta_2}$. In this region we have taken
$\lambda^{(2)}_{1}=\lambda^{(1)}_{2}\simeq 0.15$ and ignored the
effect of all other quartic couplings by setting their starting values
to be negligibly small. We have also retained small  threshold effect due
to $\Delta_2$ resulting in  $\Delta
\lambda_{\phi}=\frac{\mu_{\Delta_2}^2}{M_{\Delta_2}^2}$. Due to allowed heavier mass of
$\Delta_1$ its threshold effect has been treated to be negligible.

 Initial values of  the Higgs quartic coupling $\lambda_{\phi}$,
  DM self coupling $\lambda_{\xi}$,  DM Higgs portal 
coupling $\lambda_{\phi\xi}$, SM gauge couplings
$g_Y,g_{2L},g_{3c}$, and the top quark Yukawa coupling $h_t$ used for RG
evolution have been shown in Table.\ref{tab:cc} for $m_{\xi}=1.3$ TeV
and $m_{\xi}=2$ GeV. We find that at $m_{\xi}=1.3$ TeV, the one-loop evolution of
 of $\lambda_{\phi}$ reaches its minimum positive value around $|\phi|=10^{13}$ GeV. But if $m_{\xi} < 1.3$ TeV, 
then  $\lambda_{\phi}$ tends to run negative in the region
$10^{11}-10^{12}$ GeV even in the presence of heavy triplets which
have their 
masses $ > 10^{11}$ GeV in the present investigation. This
leads us to conclude that the vacuum stability predicts the real scalar DM mass to be $m_{\xi}\ge 1.3$ TeV. As the direct detection cross section rapidly decreases with increasing $m_{\xi}$ in this region, the predicted mass $m_{\xi}=1.3$ TeV  is expected to be more accessible to experiments compared to values $m_{\xi}\gg 1.3$ TeV, although the latter values are also allowed by the three constraints:relic density, direct detection, and vacuum stability.  

\begin{table}[h!]
\caption{Initial values of coupling constants at top quark mass
  $\mu=m_{top}=173.34$ GeV \cite{ATCM:2015,ATCM:2014} for  different values of the scalar singlet dark matter mass $m_{\xi}$. The input values of gauge couplings $g_i (i=Y,2L,3C)$ and top-quark
Yukawa coupling $h_t$ are due to PDG data \cite{PDG:2014} as explained in the Appendix. 
The predicted values of $\lambda_{\phi \xi}$ and $m_\xi$ are obtained from the
plot of constrained parameter space of Fig.\ref{cons_para}.
Values mentioned in 4th to 8th columns are common to all the dark matter masses.
}
\begin{center}
\begin{tabular}{|c|c|c|c|c|c|c|c|}  
\hline
$m_\chi$ ( TeV ) & $\lambda_{\phi \xi} $ &  $\lambda_{\xi} $ & $\lambda_{\phi}$ & $g_{1Y}$ & $g_{2L}$ &  $g_{3C}$ &  $h_{t}$ \\
& & & & & & & \\ \hline
$0.75$& $0.075$  & $0.190$ &  &  & &  &    \\
$1.3$ &  $0.118$ & $0.220$ &  &  & &  &    \\
$1.5$& $0.140$ &$0.165$  & $0.129$ & $0.35$ & $0.64$ & $1.16$ & $0.94$\\
$2$& $0.158$ & $0.100$ &  &  &  &  &
\\ \hline
\end{tabular} 
\end{center}
\label{tab:cc}
\end{table}
\subsubsection{Summary of dark matter mass prediction}\label{sec:sumDM}
We summarize below the results of theoretical and computational analyses on DM mass
carried out in this section.\\
\begin{itemize}
\item{Although the DM mass values in the narrow region $m_{\xi}=59-63$ GeV are permitted by both relic density \cite{wmap,exptomegaDM} and direct detection measurements, the corresponding Higgs portal coupling values $\lambda_{\phi,\xi}\simeq 1.7\times 10^{-4}-1.6\times 10^{-3}$ are too small to complete vacuum stability of the scalar potential.}\\
\item{All DM mass values $m_{\xi}\ge 750$ GeV easily satisfy both the relic density and the direct detection constraints. But for masses $0.75\,\, {\rm TeV}
 < \,\, m_{\xi}\,\,  <  1.3 \,\, {\rm TeV}$, the corresponding input values of $\lambda_{\phi,\xi}$ yield negative values for the RG evolution of $\lambda_{\phi}$ in the region $|\phi|\simeq 10^{10}-10^{11}$ GeV leading to vacuum instability of the scalar potential.}\\

\item{ Thus, we find that the present minimal extension of the two-triplet model  predicts real scalar singlet DM mass $ m_{\xi} \ge 1300$  GeV that satisfies all the three constraints: relic density, direct detection, and vacuum stability of the scalar potential. Out of these, the lowest limit $m_{\xi}=1.3$ TeV is expected to be comparatively more sensitive and accessible to direct detection experiments.}
\end{itemize}
\section{Radiative stability of Higgs mass and naturalness}\label{sec:nat}
In the present two-triplet model there are couplings of SM Higgs scalar ($\phi$) with triplets which are likely to  introduce large radiative correction $\delta m_{\phi} \propto M_{\Delta_1}\simeq 10^{13}$ GeV, the highest mass scale in the theory.
This would destabilise the SM Higgs mass prediction and electroweak gauge hierarchy, and calls for exploring naturalness criteria \cite{Vissani:1998,Casas:2004,Xing:2009,WRpsb:2017}, if any, to restrict such correction not to exceed the observed Higgs mass.
 A number of investigations have been carried out to constrain certain  non-SUSY seesaw model parameters to 
stabilise the Higgs mass near the electroweak scale \cite{Vissani:1998,Casas:2004,Xing:2009,WRpsb:2017}. In this section we discuss how  a naturalness constraint is available within this two-triplet model  without upsetting the model predictions for neutrino mass, leptogenesis, dark matter, and vacuum stability discussed in previous sections.

\begin{figure}[h!]
\begin{center}
\includegraphics[scale=0.4]{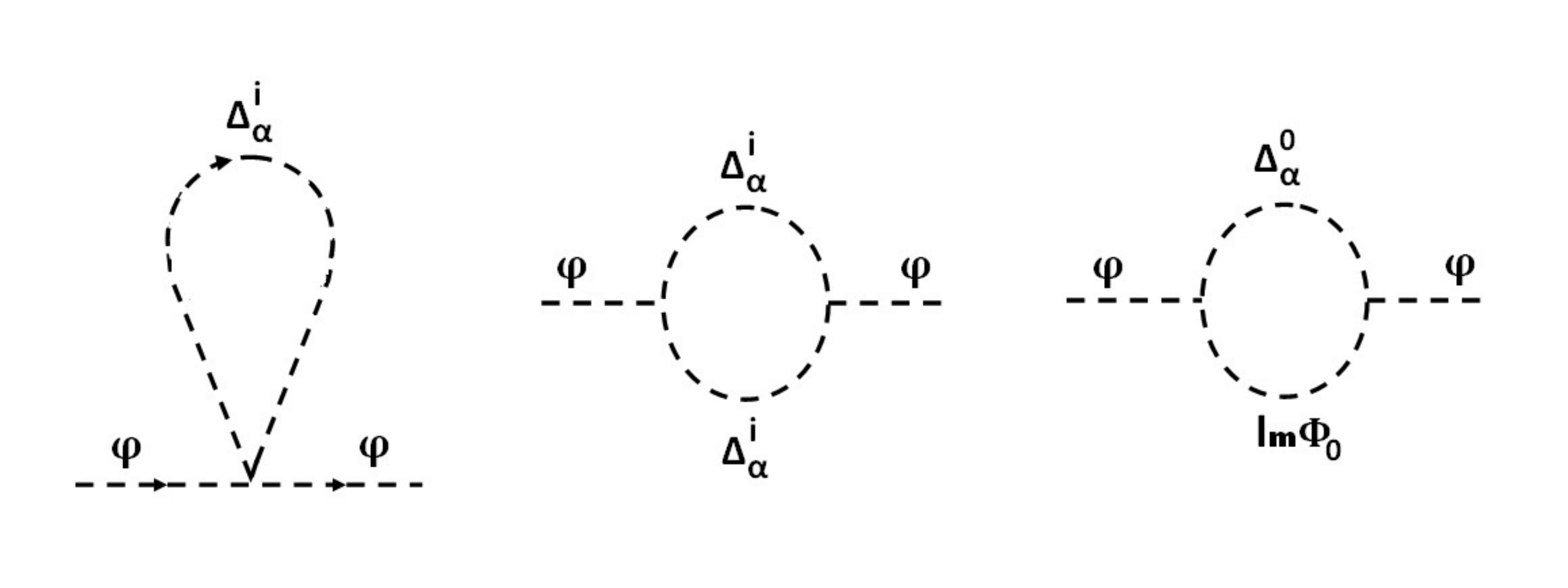}
\end{center}
\caption{ Feynman diagrams showing one-loop corrections to the standard Higgs mass  mediated  by charged and neutral components of two heavy scalar triplets  $\Delta_{\alpha}^i (\alpha=1,2; i=\pm\pm,\pm)$. Here ${\rm Im}\Phi_0$  is the  imaginary part of the neutral component of standard Higgs doublet $\Phi$.}
\label{fig:div}
\end{figure}

 The Feynman diagrams with loop-mediation by the components of the two heavy Higgs triplets  are shown in Fig.\ref{fig:div}.
Neglecting the contribution of the second diagram in Fig.\ref{fig:div} which is $\propto v^2$,  those due to the other two diagrams add up to  
\bea
\delta m_{\phi}^2&=&\frac{-3}{16\pi^2}\left( [\lambda_{3}^{(1)}+\frac{1}{2}|\lambda_6^{(1)}|^2] M_{\Delta_1}^2[1+\ln(\frac{\Lambda_R^2}{M_{\Delta_1}^2})]\right) \nonumber\\
&+&\frac{-3}{16\pi^2}\left( [\lambda_{3}^{(2)}+\frac{1}{2}|\lambda_6^{(2)}|^2] M_{\Delta_2}^2[1+\ln(\frac{\Lambda_R^2}{M_{\Delta_2}^2})] \right). \label{eq:div1}
\eea
 As explained earlier, the dimensionless couplings $\lambda^{(j)}_6=\frac {\mu_j}{M_{\Delta_j}}(j=1,2)$ are known from fitting the oscillation data and our leptogenesis ansatz in various cases leading to wider range of values for $\lambda_6^{(2)}$. The first (second) line in eq.(\ref{eq:div1}) represents the dominant radiative corrections to the Higgs mass due to the heavy triplet $\Delta_1(\Delta_2)$. Using the regularisation scale $\Lambda_R=M_{\Delta_1}$ which is the highest scalar mass in this model gives
\beq
\delta m_{\phi}^2=(\frac{-3}{16\pi^2})\left( [\lambda_{3}^{(1)}+\frac{1}{2}|\lambda_6^{(1)}|^2] M_{\Delta_1}^2+
 [\lambda_{3}^{(2)}+\frac{1}{2}|\lambda_6^{(2)}|^2] M_{\Delta_2}^2R_{L}\right),
 \label{eq:div2} 
\eeq
where 
\beq
R_L=1+\ln(\frac{M_{\Delta_1}^2}{M_{\Delta_2}^2}). \label{eq:RL}
\eeq 
Eq.(\ref{eq:div2}) leads to 
\beq
\frac{\delta m_{\phi}^2}{m_{\phi}^2}=0.019\left[\frac{M_{\Delta_1}}{100 {\rm GeV}}\right]^2
\left( \lambda_{3}^{(1)}+\frac{1}{2}|\lambda_6^{(1)}|^2 +[\lambda_3^{(2)}+\frac{1}{2}|\lambda_6^{(2)}|^2]R_{\Delta} R_{L}\right).\label{eq:div3}
\eeq
where
\beq
R_{\Delta}=\frac {M_{\Delta_2}^2}{M_{\Delta_1}^2}. \label{eq:div4}
\eeq
The naturalness criteria then suggests that RHS of eq.(\ref{eq:div3}) $\le 1$.
Thus the present model can be consistent with naturalness if the quantity inside the parenthesis in the RHS of this eq.(\ref{eq:div3}) is fine tuned to be zero and such cancellation should be ensured at least upto $2n_{\delta}-4$ places after the decimal point where $\frac{M_{\Delta_1}}{GeV}=10^{n_{\delta}}$. For  this purpose we note that although $0.1\le R_{\Delta} < 1$ and $R_L>1$, the product $0.1< R_{\Delta}R_L<1$ which make the cancellation possible within the perturbative limits of $\lambda_3^{(i)} (i=1,2)$.   
Contrary to various constraints available on $\phi\phi\Delta\Delta$ quartic couplings \cite{JWVFlam} in one-triplet extensions of the SM \cite{WRpsb:2017}, there does not seem to exist similar bounds in two-Higgs triplet model except for the universal 
 perturbativity  bound.

\begin{figure}[h!]
\begin{center}
\includegraphics[scale=0.5]{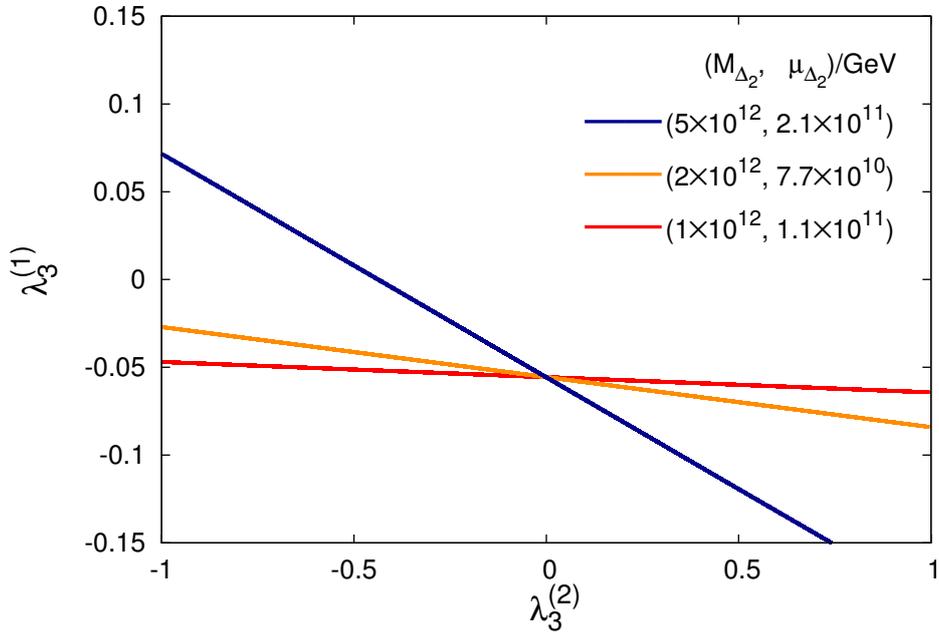}
\end{center}
\caption{ Allowed domains for fine-tuned naturalness solution in the two-heavy triplet seesaw and leptogenesis model consistent with neutrino mass, baryon asymmetry of the Universe, vacuum stability, and one-loop radiative 
stability of Higgs mass. }
\label{fig:NAT}
\end{figure}

With fixed values of
$(M_{\Delta_1}=3\times 10^{13},\mu_{\Delta_1}=10^{13})$ GeV but for three different sets of   $(M_{\Delta_2}=5\times 10^{12},\mu_{\Delta_2}= 2\times 10^{11})$  GeV, $(2\times 10^{12}, \mu_{\Delta_2}=7.7 \times 10^{10})$  GeV, and $(10^{12},
 \mu_{\Delta_2}= 1.1 \times 10^{11})$  GeV, allowed by neutrino mass and unflavoured leptogenesis, 
we make a parametric representaton of the naturalness criteria for those values of $(\lambda_3^{(1)} {\rm vs.} \lambda_3^{(2)})$ for which the RHS of eq.(\ref{eq:div3}) vanishes. These three  domains of naturalness solutions are presented by respective straight lines designated by the  corresponding  values of parameters
as shown  in Fig.\ref{fig:NAT}.
From Fig.\ref{fig:NAT} we find that the naturalness constraint that keeps RHS of eq.(\ref{eq:div3}) $\le 1$ is satisfied for all quartic couplings well within the perturbative limits.  
Some of the numerical values of $\lambda_3^{(i)}(i=1,2)$ satisfying naturalness constraint as shown in Fig.\ref{fig:NAT} have been already taken into account in the RGE of $\lambda_{\phi}(\mu)$ in our vacuum stability ansatz of Sec.\ref{sec:vstab}. Contributions of other larger values of these quartic couplings 
$\lambda_3^{(i)}(i=1,2)$ are found to keep the values of $\lambda_{\phi}(\mu)$ well below
 the perturbative limit even at $\mu \simeq M_{Planck}$.  
We have checked that similar plots displaying naturalness constraints on the quartic couplings are  possible for other two-flavoured leptogenesis solutions with $M_{\Delta_2}=1\times 10^{11}$ GeV and $M_{\Delta_2}=4\times 10^{11}$ GeV.
 Particularly, the small $\lambda_3^{(i)}(i=1,2)$ solutions represented by red and orange coloured straight lines in Fig.\ref{fig:NAT} might be relevant for quantum gravity \cite{Wetterich:2020} which predicts all quartic couplings to vanish
for $\mu > M_{Planck}$.

We thus conclude that the two triplet model can confront the Higgs mass naturalness problem via fine tuning of the model parameters without affecting neutrino mass, leptogenesis, dark matter and vacuum stability predictions.
\section{Summary and outlook}\label{sec:sum}
The original suggestion of purely triplet seesaw and leptogenesis \cite{Ma-Us:1998} addresses the interesting new possibility that both neutrino masses and  baryon asymmetry of the universe can be explained using only two heavy Higgs triplets in the absence of right-handed neutrinos. If neutrinos are quasi-degenerate with relatively larger mass scale, they can predict baryon asymmetry of the universe \cite{Ma-Us:1998} while manifesting in experimentally verifiable double beta decay. Noting that the recently determined cosmological bounds due to  Planck satellite data has severely restricted the sum of three neutrino masses to $ < 0.23$ eV (or even smaller bound $< 0.12$ eV), and the  recent neutrino oscillation data have revealed $\theta_{23}$ to be in the second octant with large Dirac CP-phase ($\simeq 214^{\circ}$), in this work we have examined this model predictions with hierarchical neutrino masses satisfying these cosmological bounds in concordance with the neutrino oscillation data.\\

 We have also attempted to explore the  model potential in addressing current issues on dark matter and vacuum stability of the scalar potential through a simple minimal extension of the model \cite{Ma-Us:1998}. We find that the original model can explain both the recent neutrino oscillation data while successfully predicting baryon asymmetry of the universe for both normal and inverted orderings where our best fit is consistent with the sum of the three neutrino masses to be nearly $25\%$($50\%$) of the Planck satellite  bound reported in  \cite{Planck15}(\cite{Sunny:2018}).\\

 We have further shown  the possibilities of both unflavoured and two-flavoured leptogenesis leading to experimental values of baryon asymmetry through detailed solutions to the respective set of CP-asymmetries and  Boltzmann equations where the 
 numerical results are consistent with generic choice of the two constituent light neutrino mass matrices $(m^{(1)}_\nu, 
 m^{(2)}_\nu)$. It has been shown (through proper graphical illustration) that adequate asymmetry as quoted by the experiments
 can be successfully generated even if elements of those two matrices are connected by completely random phases.\\ 

In addition we have also found that a simple
minimal extension of the two-Higgs triplet model \cite{Ma-Us:1998} successfully predicts a real scalar single dark matter in agreement with observed relic density and mass bounds set by direct and indirect detection experiments. Noting that in the scalar potential of the original model \cite{Ma-Us:1998}, the standard Higgs quartic coupling $\lambda_{\phi}$ runs negative in the region $5\times 10^9 {\rm \,\, GeV} \le |\phi|\le 10^{13}$ GeV, we have shown how the presence of this real scalar singlet DM also completes the vacuum stability. The lowest limit of the DM mass $m_{\xi}\simeq 750$ GeV that satisfies the existing data and constraints due to relic density and direct and indirect detection experiments,
is further pushed to  $1.3$ TeV under the constraint of  vacuum stability
completion.\\      

Noting that the two-triplet model, as such, is likely to destabilise the electroweak gauge hierarchy through large Higgs mass radiative correction, we have also found the corresponding naturalness constraint on the model parameters
that restricts this correction not to exceed the observed value of the Higgs mass itself. This naturalness constraint on model parameters does not affect our successful predictions of neutrino mass, leptogenesis, baryon asymmetry, dark matter and vacuum stability.\\
   
 In conclusion we note that the purely triplet seesaw model for neutrino mass  and leptogenesis \cite{Ma-Us:1998} is capable of successfully describing the most recent neutrino oscillation data 
including $\theta_{23}$ in the second octant and large Dirac CP-phase for both
normal and inverted ordering of neutrino masses in concordance with the existing cosmological bounds determined from Planck satellite measurement. Being non-supersymmetric 
the model has no gravitino problem and has a natural advantage of predicting cosmologically safe relic abundance of light elements. Supplemented by the underlying naturalness criteria the model is capable of ensuring Higgs mass radiative
stability. We further conclude that a simple minimal extension of this model \cite{Ma-Us:1998} successfully explains the direct and indirect evidences of dark matter; it also completes vacuum stability of the scalar potential.  Thus, a simple and minimal extension of the original model \cite{Ma-Us:1998} is capable of solving current puzzles confronting the SM: neutrino oscillation and baryon asymmetry of the universe within the cosmological bound but without gravitino problem, dark matter, vacuum stability, and Higgs mass radiative stability.\\

Although the dark matter stabilising $Z_2$ discrete symmetry has been assumed in the present model extension based upon the symmetry $SU(2)_L\times U(1)_Y\times SU(3)_C\times Z_2$ in the spirit of numerous other models including \cite{Ham-Ma:2006,GAMBIT,Ma:2006,Forengo:2006,Hambye:2008}, it would be interesting to explore its deeper gauge theoretic origin as in \cite{mkp:2011,Hambye:2010,Kadastic:2009,Krauss:1989,psb:2010} from  unified model perspectives \cite{cps:2019,Ma:2018,mkp:2020}. Because of dark-matter portal couplings with triplets, $\lambda_{\xi}^{(i)}(i=1,2)$, the real scalar DM mass prediction may be unstable against one loop radiative correction. But the model has a DM mass stability constraint similar to 
eq.(\ref{eq:div3}) that can restrict  the radiative correction close to or less than $m_{\xi}$ through fine-tuning of these couplings.     
   
\section{Acknowledgment}
M. K. P. acknowledges financial support through the research project
{\bf SB/S2/HEP-011/2013} awarded by the Department of Science and Technology,
Government of India.  M.C would like to acknowledge the 
financial support provided by SERB-DST, Govt. of India through the project {\bf EMR/2017/001434}. 
S.K.N. and R.S.  acknowledge the award of their
Ph. D. research scholarships  by Siksha 'O' Anusandhan, Deemed to be University.

\section{Appendix}

\subsection{Number density of  particle species}\label{sec:ap1}
The number densities for the massive as well as massless particles (assuming Maxwell Boltzmann distribution for both) are given by\cite{Sierra:2011ab,Sierra:2014tqa}
\begin{eqnarray}
&& n_\Sigma^{eq} (z)=n_\Delta^{eq} (z) +n_{N}^{eq} (z)^\dagger~, \\
&& n_\Delta^{eq} (z)=\frac{3 M_\Delta^3 K_2(z)}{2 \pi^2 z}, \\
&& n_{l,\phi}^{eq} (z)=\frac{2 M_\Delta^3}{\pi^2 z},
\end{eqnarray}
where $K_2(z)$ is the modified Bessel function of second kind. The expressions of entropy density and Hubble parameter are listed below.
\begin{eqnarray}
&& s(z)=\frac{4 g^\ast M_\Delta^3}{\pi^2 z^3},\\
&& H(z)=\sqrt{\frac{8 g^\ast}{\pi^2}}\frac{M_\Delta}{M_{\rm Planck} z^2},
\end{eqnarray}
with effective relativistic degrees of freedom $g^\ast=106.75$ and Planck mass $M_{\rm Planck}=1.22\times10^{19}$ GeV.
\subsection{Reaction densities} \label{sec:ap2}
Decay $(1\rightarrow 2)$ related reaction densities for lightest scalar triplet is given by\cite{Sierra:2011ab,Sierra:2014tqa}
\begin{eqnarray}
&& \gamma_D= \frac{K_1(z)}{K_2(z)} n_\Sigma^{eq} (z) \Gamma_\Delta^{tot} 
\end{eqnarray}
where $\Gamma_\Delta^{tot}$ is the total triplet decay width.
The generic expression of $(2\leftrightarrow2)$ scattering reaction densities is given by
\begin{equation}
 \gamma_s= \frac{M_\Delta^4}{64 \pi^4} \int_{x_{min}}^{\infty}  \sqrt{x} \frac{(z\sqrt{x}) \hat{\sigma_s}}{z} dx \label{gen_s}
\end{equation}
where $x=s^\prime/M_{\Delta}^2$ ($s^\prime$ is the centre of mass energy) and $\hat{\sigma_s}$ denotes reduced cross section.
For gauge induced process $x_{min}=4$ and Yukawa induced process it is $x_{min}=0$. The reduced cross sections for the gauge induced
processes is given by\cite{Sierra:2011ab,Sierra:2014tqa}
\begin{eqnarray}
 \hat{\sigma_A} & = &\frac{2}{72 \pi} \Big \{ (15 C_1-3 C_2) \omega + (5 C_2 -11 C_1 ) \omega^3 + 3 (\omega^2 -1)[2 C_1 + C_2 (\omega^2 -1) ] \ln \Big (\frac{1+\omega}{1- \omega} \Big) 
 \Big \} \nonumber \\ & + &
 \Big  (  \frac{50 g_{2L}^4 +41 {g_{1Y}}^4}{48 \pi} \Big ) \omega^{\frac{3}{2}}, \label{gauge_s}
\end{eqnarray}
where $\omega \equiv \omega (x) =\sqrt{1-4/x}$ and $C_1= 12 g_{2L}^4 +3 g_{1Y}^4 +12 g_{2L}^2 g_{1Y}^2,~C_2=6 g_{2L}^4+3 g_{1Y}^4 + 12 g_{2L}^2 g_{1Y}^2$. ($g_{2L}$ is the SM $SU(2)$ coupling and $g_{1Y}$ is the SM $U(1)_Y$ coupling.)
Reduced crosssections of $\Delta L=2$ scattering processes ($s$ channel and $t$ channel respectively) are given by
\begin{eqnarray}
&& \hat {\sigma}^{\phi \phi}_{l_i l_j} = 64 \pi B_\phi B_{l_{ij}} \delta^2 \Big[\frac{x}{(x-1)^2 +\delta^2}\Big],\\
&& \hat {\sigma}^{\phi l_j}_{\phi l_j} = 64 \pi B_\phi B_{l_{ij}} \delta^2 \frac{1}{x} \Big[ \ln (1+x) - \frac{x}{1+x} \Big ],
\end{eqnarray}
where $\delta=\Gamma^{tot}_\Delta/M_\Delta$. Similarly the reduced crosssections for lepton flavor violating processes ($s$ channel and $t$ channel) are represented as
\begin{eqnarray}
&& \hat {\sigma}^{l_n l_m }_{l_i l_j} = 64 \pi B_{l_{nm}} B_{l_{ij}} \delta^2 \Big[\frac{x}{(x-1)^2 +\delta^2}\Big],\\ 
&& \hat {\sigma}^{l_j l_m }_{l_i l_n} = 64 \pi B_{l_{nm}} B_{l_{ij}} \delta^2 \Big[ \frac{x+2}{x+1} -\ln(1+x) \Big] ~.\label{lfv2}
\end{eqnarray}
Reaction densities of different scattering processes $(\gamma_A,\gamma^{\phi\phi}_{l_i l_j},\gamma^{ l_n l_m}_{l_i l_j}~{\rm etc} )$ can be calculated using the expressions of reduced crosssections(eq.(\ref{gauge_s}) -eq.(\ref{lfv2})) in the generic 
formula (eq.(\ref{gen_s}))for the scattering reaction density. The Resonant intermediate state subtracted reaction densities are given by
\begin{eqnarray}
&&\gamma^{\prime \phi\phi}_{l_i l_j} = \gamma^{\phi\phi}_{l_i l_j} -B_{l_{ij}} B_\phi \gamma_D \\
&&\gamma^{\prime l_n l_m}_{l_i l_j} = \gamma^{ l_n l_m}_{l_i l_j} - B_{l_{ij}} B_{l_{nm}} \gamma_D ~.
\end{eqnarray}
\subsection{$C^l$ and $C^\phi$ matrices}\label{sec:ap3}
The lepton asymmetry and scalar doublet asymmetry are related to $(B/3 -L_i)$ and triplet asymmetry through the asymmetry coupling
matrices $C^l$ and $C^\phi$. These matrices are determined by solving a constrained set ( imposed by Global symmetry of the Lagrangian 
and chemical equilibrium relations) of equations involving chemical potentials.
Above a certain temperature ($\sim 10^{12}$ GeV) the lepton flavours act as a single entity ( a coherent superposition of 
three flavours ($e,\mu,\tau$)). The lepton flavour decoherence temperature (which signifies the temperature at which a specific 
lepton flavour loses its coherence and can be treated as a separate entity) is 
denoted by $T^{f_i}_{decoh}$, where $f_i$ stands for any specific flavour $(e,\mu,\tau)$.
So it is clear that above a temperature $T^\tau_{decoh}$ all three lepton flavours act indistinguishably and
the Boltzmann equation has to be solved for the quantity $(B-L)$. Again the regime
$T>T^\tau_{decoh}$ is subdivided into three windows. 
The structure of $C^l$ and $C^\phi$ matrices are different in those windows since the number of active chemical potentials and 
the governing constraint equations are different in each of these windows (This issue has been discussed extensively in Sec.3.1 
and Appendix B of Ref\cite{Sierra:2014tqa}). Similarly in the intermediate region between
$T^\tau_{decoh} ~ -~ T^\mu_{decoh}$ two lepton flavours $(a(\equiv e+\mu), \tau)$ are effectively active, whereas
below $T^\mu_{decoh}$ complete flavour decoherence is attained and all three lepton flavours are separately identifiable, 
thus the set of flavoured Boltzmann equations has to be solved in terms of $(B/3-L_e,B/3-L_\mu,B/3-L_\tau)$.  The asymmetry 
coupling matrices ( following Ref\cite{Sierra:2014tqa}) are shown in Table \ref{clcphi}.
\begin{table}[!h]
 \caption{$C^l$ and $C^\phi$ matrices in different temperature regimes}\label{clcphi}
 \begin{center}
  \begin{tabular}{|c|c|c|c|}
  \hline
   $T$ (GeV) & Flavours & $C^l$ & $C^\phi$\\
   \hline
    $\gtrsim 10^{15}$ & single & $\Big (0~\frac{1}{2} \Big )$ & $\Big (3~\frac{1}{2} \Big )$\\
    &  &  & \\
     $[10^{12},10^{15}]$ &  & $\Big (0~\frac{1}{2} \Big )$ & $\Big (2~\frac{1}{3} \Big )$ \\
     &  &  & \\
      $[T_{decoh}^\tau,10^{12}]$ &  & $\Big (0~\frac{3}{10} \Big )$ & $\Big (\frac{3}{4}~\frac{1}{8} \Big )$ \\
     &  &  & \\ 
    \hline
    $[10^9,T_{decoh}^\tau],$ &  two & $\left( \begin{array}{ccc}
              -\frac{6}{359} & \frac{307}{718} & -\frac{18}{359}\cr
              \frac{39}{359} & -\frac{21}{718}  & \frac{117}{359}\cr
             \end{array}\right)$ & $\Big (\frac{258}{359}~\frac{41}{359}~\frac{56}{359} \Big )$\\
     $[T_{decoh}^\mu,T_{decoh}^\tau]$ &  &  & \\
     &  &  & \\
     \hline
    $[10^5,T_{decoh}^\mu]$ &  three & $\left( \begin{array}{cccc}
              -\frac{6}{179} & \frac{151}{358} & -\frac{10}{179} &  -\frac{10}{179}\cr
              \frac{33}{358} & -\frac{25}{716}  & \frac{172}{537} & -\frac{7}{537}\cr
               \frac{33}{358} & -\frac{25}{716} & -\frac{7}{537} & \frac{172}{537}\cr
             \end{array}\right)$ & $\Big (\frac{123}{179}~\frac{37}{358}~\frac{26}{179} ~\frac{26}{179}\Big )$\\
       &  &  &  \\
       $\lesssim 10^{5}$ &  & $\left( \begin{array}{cccc}
              -\frac{9}{158} & \frac{221}{711} & -\frac{16}{711} &  -\frac{16}{711}\cr
              \frac{9}{158} & -\frac{16}{711}  & \frac{221}{711} & -\frac{16}{711}\cr
               \frac{9}{158} & -\frac{16}{711} & -\frac{16}{711} & \frac{221}{711}\cr
             \end{array}\right)$ &  $\Big (\frac{39}{79}~\frac{8}{79}~\frac{8}{79} ~\frac{8}{79}\Big )$\\ 
             & & & \\
             \hline      
  \end{tabular}
 \end{center}
\end{table}

\subsection{Renormalisation group equations for gauge and
  scalar couplings}\label{sec:rge}

We use the following electroweak precision data at $\mu\simeq m_{top}$
and the Higgs mass 
\cite{PDG:2014,ATCM:2015,ATCM:2014} as inputs in the in the bottom-up approach
\begin{eqnarray}
m_{top}&=& 173.34\pm 0.77\,{\rm GeV}\,\,\nonumber\\
\sin^2\theta_W&=&0.23129\pm 0.00005\,\, \nonumber\\
\alpha_S&=&0.1182\pm 0.0005 \,\, \nonumber\\
\frac{1}{\alpha}&=&127.9\pm0.02\,\,\nonumber\\
m_h&=&125.09\pm 0.237\,{\rm GeV}\, \label{eq:ewdata}
\end{eqnarray}
These values determine the initial boundary values $\lambda_{\phi}=0.129$, the SM gauge couplings $g_{1Y}=0.35, g_{2L}=0.64, g_{3C}=1.16$ and the top-quark Yukawa coupling $h_t=0.94$.
In addition we use the Higgs triplet masses, their trilinear
couplings, and scalar singlet DM mass as discussed in
Sec.\ref{sec:nufit},Sec.\ref{sec:bauest},Sec.\ref{sec:dmvac} and Sec.\ref{sec:redm}, 
\begin{eqnarray}
M_{\Delta_2}&=&10^{12}\,\,{\rm GeV}\,\,\nonumber\\
\mu_{\Delta_2}&=& 6\times 10^{10}\,\,{\rm GeV}\,\,\nonumber\\
m_{\xi}&=&1.3 \,\,{\rm TeV}\,\,\nonumber\\
M_{\Delta_1}&=&10^{13}\,\,{\rm GeV}\,\,\nonumber\\
\mu_{\Delta_1} &=& 10^{12}\,\,{\rm GeV}.\,\,\label{eq:para}
\end{eqnarray}
The RGEs for SM gauge couplings and top quark Yukawa coupling at two loop level are given by
\begin{eqnarray}
\frac{dh_t}{d\ln \mu} &= & {1 \over 16\pi^2}\left({9 \over 2}h_t^2-{17 \over 12}g_{1Y}^2
 -{9 \over 4}g_{2L}^2-8g_{3C}^2 \right)h_t \\ \nonumber
 & +  & {1 \over (16\pi^2)^2}[-{23 \over 4}g_{2L}^4-{3 \over 4}g_{2L}^2g_{1Y}^2+{1187 \over 216}g_{1Y}^4 + 9g_{2L}^2g_{3C}^2+{19 \over 9}g_{3C}^2g_{1Y}^2-108g_{3C}^4 
\\ \nonumber
 & +& \left({225 \over 16}g_{2L}^2+{131 \over 16}g_{1Y}^2+36g_{3C}^2 \right)h_t^2+6(-2h_{t}^4-2h_{t}^2\lambda_{\phi}+\lambda_{\phi}^2)],  \\ \nonumber
 {dg_{1Y} \over d \ln \mu} & = & {1 \over 16\pi^2}\left({41 \over 6}g_{1Y}^3\right)+{1 \over (16\pi^2)^2}\left({199 \over 18}g_{1Y}^2+{9 \over 2}g_{2L}^2+{44 \over 3}g_{3C}^2-
{17 \over 6}h_t^2\right)g_{1Y}^3 ,\\ \nonumber
{dg_{2L} \over d \ln \mu} & = & {1 \over 16\pi^2}\left(-{19\over 6}g_{2L}^3\right)+{1 \over (16\pi^2)^2}\left({3 \over 2}g_{1Y}^2+{35 \over 6}g_{2L}^2+12g_{3C}^2-
{3 \over 2}h_t^2\right)g_{2L}^3, \\ \nonumber
{dg_{3C} \over d \ln \mu} & = & {1 \over 16\pi^2}\left(-7g_{3C}^3\right)+{1 \over (16\pi^2)^2}\left({11 \over 6}g_{1Y}^2+{9 \over 2}g_{2L}^2-26g_{3C}^2-
2h_t^2\right)g_{3C}^3,\label{eq:smrge}
\end{eqnarray}
where $g_{2L},g_{1Y},g_{3C}$ are the gauge couplings of $SU(2)_L,U(1)_Y,SU(3)_C$, respectively, and $h_t$ is the top quark Yukawa coupling.
The RG equations for the scalar quartic couplings up to one loop level are 
\begin{eqnarray}
&&\frac{d \lambda_{\phi }}{d \ln \mu}= \frac{1}{16\pi^2}\left[ ( 12 h_t^2 -3 {g_{1Y}}^2 -9g_{2L}^2) \lambda_\phi -6 h_t^4 +
\frac{3}{8}\{ 2 g_{2L}^4 +({g_{1Y}}^2 +g_{2L}^2)^2 \} +24 \lambda_\phi^2 + 4 \lambda_{\phi \xi}^2 \right]~,\nonumber\\
&&\frac{d \lambda_{\phi \xi}}{d \ln \mu}= \frac{1}{16\pi^2}\left[\frac{1}{2}(12h_t^2 -3 {g_{1Y}}^2 -9 g_{2L}^2)\lambda_{\phi \xi}+
4\lambda_{\phi \xi} (3 \lambda_\phi +2 \lambda_\xi) + 8 \lambda_{\phi \xi}^2 \right]~, \nonumber\\
&&\frac{d \lambda_{ \xi}}{d \ln \mu}= \frac{1}{16\pi^2}\left[ 8 \lambda_{\phi \xi}^2 + 20 \lambda_\xi^2 \right].~\label{eq:rge}
\end{eqnarray}

For  mass scale $\mu \ge M_{\Delta_2}\simeq 10^{12}$ GeV, the
scalar potential is defined through eq.(\ref{eq:vchidel1}) of Sec.\ref{sec:redm}.

We define the respective beta functions through
\beq
16\pi^2{dC \over dt}=\beta_{C}~~(C=\lambda_{\phi},\lambda_{\phi
  \xi},\lambda_\xi,\lambda_1^i,\lambda_2^i,\lambda_3^i,\lambda_4^i, (i=1,2)).
\eeq
The beta functions for desired quartic couplings are
\begin{eqnarray}
\beta_{\lambda_\phi} &=&
	\lambda_\phi \left[ 12\lambda_\phi - \left( \frac{9}{5}g_{1Y}^2 + 9g_{2L}^2 \right) + 12h_t^2 \right]
	+ \frac{9}{4} \left( \frac{3}{25}g_{1Y}^4 + \frac{2}{5}g_{1Y}^2g_{2L}^2 + g_{2L}^4 \right) \nonumber \\
	&& +\sum_{(i=1,2)}\left( 6(\lambda_3^i)^2 + 4(\lambda_4^i)^2\right) - 12h_t^4,
\label{betalphi} 
\end{eqnarray}
For $i=1,2$ , the RGEs for respective quartic couplings are 
\begin{eqnarray}
 \beta_{\lambda_1^i} &=&
	\lambda_1^i \left[ 14\lambda_1^i + 4 \lambda_2^i
	- \left( \frac{36}{5} g_{1Y}^2 + 24g_{2L}^2 \right)
	+ 4 {\rm Tr} \left[T \right] \right]
	+ \frac{108}{25}g_{1Y}^4 + \frac{72}{5}g_{1Y}^2g_{2L}^2 + 18g_{2L}^4 \nonumber\\
	&& + 2 (\lambda_2^i)^2 + 4 (\lambda_3^i)^2 + 4(\lambda_4^i)^2
	- 8 {\rm Tr} \left[T^2 \right], \\
 \beta_{\lambda_2^i} &=&
	\lambda_2^i \left[ 12 \lambda_1^i + 3 \lambda_2^i 
	- \left( \frac{36}{5}g_{1Y}^2 + 24g_{2L}^2 \right)
	+ 4 {\rm Tr} \left[T  \right] \right]
	- \frac{144}{5}g_{1Y}^2g_{2L}^2 + 12g_{2L}^4 \nonumber\\
	&& - 8 (\lambda_4^i)^2 + 8 {\rm Tr} \left[T^2 \right],  \\
 \beta_{\lambda_3^i} &=&
	\lambda_3^i \left[ 6 \lambda_\phi + 8 \lambda_1^i + 2 \lambda_2^i + 4\lambda_3^i
	- \left( \frac{9}{2}g_{1Y}^2 + \frac{33}{2}g_{2L}^2 \right) 
	+ 6 h_t^2 + 2 {\rm Tr} \left[T \right] \right] \nonumber\\
	&& + \frac{27}{25}g_{1Y}^4 + 6g_{2L}^4
	+ 8 (\lambda_3^i)^2 - 4 {\rm Tr}\left[ T^2 \right], \\
 \beta_{\lambda_4^i} &=&
	\lambda_4^i \left[ 2 \lambda^i + 2\lambda_1^i - 2\lambda_2^i + 
8 \lambda_3^i
	- \left( \frac{9}{2}g_{1Y}^2 + \frac{33}{2}g_{2L}^2 \right)
	+ 6 h_t^2 + 2 {\rm Tr}\left[T \right] \right]
	- \frac{18}{5}g_{1Y}^2g_{2L}^2 \nonumber\\
	&& + 4 {\rm Tr}\left[T^2 \right],
\end{eqnarray}
 where $T$ is defined as $T={y^{(2)}}^\dagger y^{(2)}$ where $y^{(2)}\simeq m_{\nu}/VL2$. 
 and its beta function is expressed through the relation
\begin{eqnarray}
\beta_T =
	T \left[ 6\, T - 3 \left( \frac{3}{5} g_{1Y}^2 + 3 g_{2L}^2 \right)
	+ 2 {\rm Tr}[T] \right] . \label{eq:betaT}
\end{eqnarray}

We have examined  how vacuum stability of the scalar potential in this minimally extended model is ensured by the presence of the scalar singlet  DM even with
its lowest 
mass $m_{\xi}\simeq 1.3$ TeV and its associated Higgs portal coupling. We have estimated RG evolution of standard Higgs quartic coupling $\lambda_{\phi}$ in the presence of the DM as well as the heavy scalar triplets in the appropriate ranges of mass scales and Higgs field values. When the DM and the triplets are excluded we get the lowermost red curve \cite{Espinosa,Lebedev} of Fig.\ref{dm_stab} of Sec.\ref{sec:vstab} where $\lambda_{\phi}$ runs negative for all values of Higgs field 
$|\phi|> 5\times 10^9$ GeV showing unstable SM vacuum. When we exclude the scalar DM but include the two heavy triplets as in the original model  of \cite{Ma-Us:1998}, the negativity of the quartic coupling persists only in the interval $|\phi|=5\times 10^9-10^{13}$ GeV after which the quartic coupling has the ability to be positive due to the additional contribution of the triplets. Here a major compensation is caused by the
$\Delta_2$-threshold enhancement at $M_{\Delta_2}=10^{13}$ GeV not shown in Fig. \ref{dm_stab}. In Fig.\ref{dm_stab} the negative part of the red coloured curve bounded by vertical green dashed lines  is also predicted by the original model \cite{Ma-Us:1998} signifying vacuum instability in the model.  Excluding the triplets but including DM, the solution is given by the  upper blue curve of Fig.\ref{dm_stab} marked as SM$+$DM (excluding threshold enhancement). When effects of heavy triplets are also included along with DM
in the present model extension, the RG evolution for the quartic coupling develops threshold enhancement at $M_{\Delta_2}=10^{12}$ GeV (rather than $10^{13}$ GeV of \cite{Ma-Us:1998}) which has been predicted by matching the baryon asymmetry data in the present analysis. This threshold enhancement is $\Delta \lambda_{\phi}\simeq \mu_{\Delta_2}^2/M_{\Delta_2}^2 \simeq 0.005-0.01$. In addition we have also included the effects of small triplet portal couplings using $\lambda_3^{(2)}\simeq \lambda_4^{(2)} \simeq 0.1$.  The resulting corrections have been shown by the uppermost curve for $\mu > 10^{12}$ Gev in Fig.\ref{dm_stab} of Sec.\ref{sec:vstab}. This part of the curve has been marked as SM$+$DM$+$$\Delta$.


\begin{thebibliography}{99}
\bibitem{nudata} 
G.~L.~Fogli, E.~Lisi, A.~Marrone, D.~Montanino, A.~Palazzo and A.~M.~Rotunno,
Phys. Rev. D \textbf{86}, 013012 (2012)
[arXiv:1205.5254 [hep-ph]];
T.~Schwetz, M.~Tortola and J.~W.~F.~Valle,
New J. Phys. \textbf{13}, 063004 (2011)
[arXiv:1103.0734 [hep-ph]].
\bibitem{Forero:2014}
D.~V.~Forero, M.~Tortola and J.~W.~F.~Valle,
Phys. Rev. D \textbf{90}, no.9, 093006 (2014)
[arXiv:1405.7540 [hep-ph]];
M.~C.~Gonzalez-Garcia, M.~Maltoni and T.~Schwetz,
Nucl. Phys. B \textbf{908}, 199-217 (2016)
[arXiv:1512.06856 [hep-ph]].
\bibitem{Esteban:2018} 
I.~Esteban, M.~C.~Gonzalez-Garcia, A.~Hernandez-Cabezudo, M.~Maltoni and T.~Schwetz,
JHEP \textbf{01}, 106 (2019)
[arXiv:1811.05487 [hep-ph]].
\bibitem{BAUexpt} 
D.~N.~Spergel \textit{et al.} [WMAP],
Astrophys. J. Suppl. \textbf{148}, 175-194 (2003)
[arXiv:astro-ph/0302209 [astro-ph]];
E.~Komatsu \textit{et al.} [WMAP],
Astrophys. J. Suppl. \textbf{180}, 330-376 (2009)
[arXiv:0803.0547 [astro-ph]];
G.~Hinshaw \textit{et al.} [WMAP],
Astrophys. J. Suppl. \textbf{180}, 225-245 (2009)
[arXiv:0803.0732 [astro-ph]];
E.~Komatsu \textit{et al.} [WMAP],
Astrophys. J. Suppl. \textbf{192}, 18 (2011)
[arXiv:1001.4538 [astro-ph.CO]].
\bibitem{Planck15} 
P.~A.~R.~Ade \textit{et al.} [Planck],
Astron. Astrophys. \textbf{571}, A16 (2014)
[arXiv:1303.5076 [astro-ph.CO]];
P.~A.~R.~Ade \textit{et al.} [Planck],
Astron. Astrophys. \textbf{594}, A13 (2016)
[arXiv:1502.01589 [astro-ph.CO]].

\bibitem{DMexpt} F. Zwicky, Helv. Phys. Acta, {\bf 6} (1933) 110;
  D. N. Spergel {\em et al}, (WMAP Collaboration),
  Astrophys. J. Suppl. {\bf 170} (2007) 377;  J. Einasto, arXiv: 0901.0632[astro-ph,CO];
G.~R.~Blumenthal, S.~M.~Faber, J.~R.~Primack and M.~J.~Rees,
Nature \textbf{311}, 517-525 (1984);
 J. Angle {\em et. al}
 (XENON10 Collaboration), Phy. Rev. Lett {\bf 107}, 051301 (2011); ibid.
Phys. Rev. Lett {\bf 110}, 249901 (2013), arXiv: 1104.3088 [astro-ph,CO]; 
L.~E.~Strigari,
Phys. Rept. \textbf{531}, 1-88 (2013)
[arXiv:1211.7090 [astro-ph.CO]].
\bibitem{sphaleron} 
V.~A.~Kuzmin, V.~A.~Rubakov and M.~E.~Shaposhnikov,
Phys. Lett. B \textbf{191}, 171-173 (1987).
\bibitem{type-I} 
P.~Minkowski,
Phys. Lett. B \textbf{67}, 421-428 (1977);
 M. Gell-Mann, P. Ramond and R. Slansky,
  in {\it Supergravity}, edited by P. van Nieuwenhuizen and D. Freedman,
  (North-Holland, 1979), p.~315;
 S.L. Glashow, in Quarks and Leptons, Carg\`ese, eds. M. L\'evy et al.,
(Plenum, 1980, New-York), p. 707;
 T. Yanagida, in {\it Proceedings of the Workshop on the Unified Theory
  and the Baryon Number in the Universe}, edited by O. Sawada and
  A. Sugamoto (KEK Report No.~79-18, Tsukuba, 1979), p.~95;
R.~N.~Mohapatra and G.~Senjanovic,
Phys. Rev. Lett. \textbf{44}, 912 (1980)
\bibitem{Valle:1980} 
J.~Schechter and J.~W.~F.~Valle,
Phys. Rev. D \textbf{22}, 2227 (1980).
\bibitem{Fuku-Yana:1986} 
M.~Fukugita and T.~Yanagida,
Phys. Lett. B \textbf{174}, 45-47 (1986).

\bibitem{Nir:2008}
S.~Davidson, E.~Nardi and Y.~Nir,
Phys. Rept. \textbf{466}, 105-177 (2008)
[arXiv:0802.2962 [hep-ph]].
\bibitem{leptogenesis}   G.~Lazarides, Q.~Shafi and C.~Wetterich,  Nucl.\ Phys.\ B {\bf 181}, 287 (1981); 
 M.~A.~Luty, Phys.\ Rev.\ D {\bf 45}, 455 (1992); 
 A.~Acker, H.~Kikuchi, E.~Ma and U.~Sarkar, Phys.\ Rev.\ D {\bf 48}, 5006 (1993) [hep-ph/9305290];
 P.~J.~O'Donnell and U.~Sarkar,
  Phys.\ Rev.\ D {\bf 49}, 2118 (1994)
  [hep-ph/9307279].
  M.~Flanz, E.~A.~Paschos and U.~Sarkar,  Phys.\ Lett.\ B {\bf 345}, 248 (1995) [Erratum-ibid.\ B {\bf 382}, 447 (1996)] [hep-ph/9411366]; 
    M.~Flanz, E.~A.~Paschos, U.~Sarkar and J.~Weiss,  Phys. Lett.\ B {\bf 389}, 693 (1996) [hep-ph/9607310]; E. Ma, U. Sarkar, Phys. Rev. {\bf D 85} (2012) 075015, arXiv:1111.5350[hep-ph]; E. Ma, {\it Nucl. Phys. B Proc. Suppl.} {\bf 168} 
(2007) 347-349; T. Hambye, E. Ma, U. Sarkar, Nucl. Phys. {\bf B 590} (2000) 429,
 hep-ph/0006173; T. Hambye, E. Ma, U. Sarkar, Phys. Rev. {\bf D 62} (2000) 015010, hep-ph/9911422;  E. Ma, U. Sarkar, Phys. Lett. {\bf B 458} (1999) 73-78,
hep-ph/9812276; T. Hambye, E. Ma, U. Sarkar, Nucl. Phys. {\bf B 590} (2000) 429-452, hep-ph/0006173;  T. Hambye, E. Ma, M. Raidal, U. Sarkar, Phys. Lett. {\bf B 512} (2001) 373-378, hep-ph/0011197; 
A.~Pilaftsis,  Phys.\ Rev.\ D {\bf 56}, 5431 (1997)  [hep-ph/9707235]; 
  A.~Pilaftsis, Nucl.\ Phys.\ B {\bf 504}, 61 (1997)  [hep-ph/9702393]; 
 W.~Buchmuller and M.~Plumacher,  Phys.\ Lett.\ B {\bf 389}, 73 (1996)  [hep-ph/9608308];  
  W.~Buchmuller and M.~Plumacher, Phys.\ Lett.\ B {\bf 431}, 354 (1998) [hep-ph/9710460]; 
  W.~Buchmuller, P.~Di Bari and M.~Plumacher, Phys.\ Lett.\ B {\bf 547}, 128 (2002) [hep-ph/0209301]; 
  W.~Buchmuller, P.~Di Bari and M.~Plumacher, Nucl.\ Phys.\ B {\bf 643}, 367 (2002) [Erratum-ibid.\ B {\bf 793}, 362 (2008)] [hep-ph/0205349];  
  R.~Barbieri, P.~Creminelli, A.~Strummia and N.~Tetradis,  Nucl.\ Phys.\ B {\bf 575}, 61 (2000)  [hep-ph/9911315];
   K.~Hamaguchi, H.~Murayama and T.~Yanagida,  Phys.\ Rev.\ D {\bf 65}, 043512 (2002)  [hep-ph/0109030]; 
  T.~Hambye,  Nucl.\ Phys.\ B {\bf 633}, 171 (2002)  [hep-ph/0111089];
 J.~R.~Ellis and M.~Raidal,  Nucl.\ Phys.\ B {\bf 643}, 229 (2002)  [hep-ph/0206174];
  G.~C.~Branco, R.~Gonzalez Felipe, F.~R.~Joaquim, I.~Masina, M.~N.~Rebelo, C.~A.~Savoy, Phys.\ Rev.\ D {\bf 67},  (2003) 
073025,  [hep-ph/0211001];
  A.~Abada, S.~Davidson, A.~Ibarra, F.-X.~Josse-Michaux, M.~Losada, A.~Riotto
  JHEP {\bf 0609},  (2006) 010,
  [hep-ph/0605281].
  L.~Covi, E.~Roulet and F.~Vissani,
  Phys.\ Lett.\ B {\bf 384} (1996) 169,
  [hep-ph/9605319].
  Franco ~Buccella, Domenico ~Falcone, Chee ~Seng ~Fong, Enrico ~Nardi, Giulia.~Ricciardi,
  Phys.\ Rev.\ D {\bf 86} (2012) 035012,
  [arXiv:1203.0829 [hep-ph]].
  S.~Y.~Khlebnikov and M.~E.~Shaposhnikov,
  Nucl.\ Phys.\ B {\bf 308}, 885 (1988);
  J.~A.~Harvey and M.~S.~Turner,
  Phys.\ Rev.\ D {\bf 42}, 3344 (1990);
  F.~Iocco, G.~Mangano, G.~Miele, O.~Pisanti and P.~D.~Serpico,
  [arXiv:0809.0631 [astro-ph]];
  Enrico ~Bertuzzo, Pasquale ~Di Bari and Luca ~Marzola,
  Nucl.\ Phys.\ B {\bf 849}  (2011) 521,
  [arXiv:1007.1641 [hep-ph]];
  Chee ~Seng ~Fong, Enrico ~Nardi and Antonio ~Riotto,
  Adv.\ High Energy Phys.\  {\bf 2012}, 158303 (2012)
  [arXiv:1301.3062 [hep-ph];
     W.~Buchmuller, P.~Di Bari and M.~Plumacher,
  Annals Phys.\  {\bf 315} (2005) 305,
  [hep-ph/0401240];
J. C. Pati, Phys. Rev. {\bf D 68} (2003) 072002, hep-ph/0209160; J. C. Pati,
Int. J. Mod. Phys. {\bf A18} (2003) 4135-4156; H. An, Shao-Long Chen,
R. N. Mohapatra, Y. Zhang, JHEP {\bf 1003} (2010) 124,
arXiv:0911.4463; Pei-Hong Gu, Rabindra N. Mohapatra, 
Phys. Rev. {\bf D97} (2018) no.7, 075014, arXiv:1712.00420; C. Hagedorn, R. N. Mohapatra,
E. Molinaro, C. C. Nishi, S. T. Petcov, Int. J. Mod. Phys. {\bf A33}
(2018) no.05, 184206; S. K. Majee, M. K. Parida, A. Raychaudhuri,
Phys. Lett {\bf B 668}(2008) 299, arXiv:08073959[hep-ph]; S. K. Majee, M. K. Parida, A. Raychaudhuri,
U. Sarkar, Phys. Rev. {\bf D 75} (2007) 075003, hep-ph/0701179;   M. K. Parida, A. Raychaudhuri,
 Phys. Rev. {\bf D 82} (2010) 093017, arXiv:1007.5082[hep-ph];  I. K. Cooper, S. F. King, C. Luhn, Nucl. Phys. {\bf B859} (2012) 159-176, arXiv:1110.5676[hep-ph]; F. Bjorkeroth, F. J. de Anda, Ivo de Mederios Varzielas, S. F. King 
, JHEP {\bf 10} (2015) 104, arXiv:1505.05504[hep-ph]; P. Chen, G. -J. Ding, S.F. King, JHEP {\bf 03} (2016) 206, arXiv:1602.03873; M. Chianese, B. Fu, S. F. King, JCAP {\bf 03} (2020) 030, arXiv:1910.12916[hep-ph].

\bibitem{ASJWR} A. S. Joshipura, E. A. Paschos, W. Rodejohann, Nucl. Phys. {\bf B 611} (2001) 227-238, hep-ph/0104228;
A.~S.~Joshipura, E.~A.~Paschos and W.~Rodejohann,
JHEP \textbf{08}, 029 (2001)
[arXiv:hep-ph/0105175 [hep-ph]];
W. Rodejohann, Phys. Rev. {\bf D 70} (2004) 073010, hep-ph/0403236;
E.~K.~Akhmedov and W.~Rodejohann,
JHEP \textbf{06}, 106 (2008)
[arXiv:0803.2417 [hep-ph]];
Thomas Rink, Werner Rodejohann, Kai Schmitz, arXiv:2006.03021[hep-ph].

\bibitem{Adhikary:2014qba}
B.~Adhikary, M.~Chakraborty and A.~Ghosal,
Phys. Rev. D \textbf{93}, no.11, 113001 (2016)
[arXiv:1407.6173 [hep-ph]];
R.~Samanta, M.~Chakraborty, P.~Roy and A.~Ghosal,
JCAP \textbf{03}, 025 (2017)
[arXiv:1610.10081 [hep-ph]];
R.~Samanta and M.~Chakraborty,
JCAP \textbf{1902}, 003 (2019)
[arXiv:1802.04751 [hep-ph]];
M.~Chakraborty, R.~Krishnan and A.~Ghosal,
JHEP \textbf{09}, 025 (2020)
[arXiv:2003.00506 [hep-ph]].

\bibitem{Ma-Us:1998} 
  E.~Ma and U.~Sarkar,
  Phys.\ Rev.\ Lett.\  {\bf 80}, 5716 (1998)
  [hep-ph/9802445].
 
\bibitem{type-II}
 M.~Magg and C.~Wetterich, { Phys.\ Lett.\ {\bf B { 94}} (1980) 61};  
T. P. Cheng, L. F. Li, Phys. Rev. {\bf D 22} (1980) 2860; 
  G.~Lazarides, Q.~Shafi and C.~Wetterich, 
   Nucl.\ Phys.\ {\bf B { 181}} (1981) 287; 
R. N. Mohapatra, G. Senjanovic,Phys. Rev.  {\bf D 23}  (1981) 165; J. Schecter, J. W. F. Valle, Phys. Rev. {\bf D 25} (1982) 774.

\bibitem{Ham-Ma-Us:2001} 
  T.~Hambye, E.~Ma and U.~Sarkar,
  Nucl.\ Phys.\ B {\bf 602}, 23 (2001)
  [hep-ph/0011192].
\bibitem{Ma-Ss-Us:1999}
E. Ma, S. Sarkar and U. Sarkar, Phys. Lett {\bf B 458}, 73 (1999)  [hep-ph/9812276];
P.~H.~Gu, M.~Hirsch, U.~Sarkar and J.~W.~F.~Valle,
Phys. Rev. D \textbf{79}, 033010 (2009)
[arXiv:0811.0953 [hep-ph]].

\bibitem{Ham-gs:2003}
T.~Hambye and G.~Senjanovic,
Phys. Lett. B \textbf{582}, 73-81 (2004)
[arXiv:hep-ph/0307237 [hep-ph]].
\bibitem{Ham:2012} 
T.~Hambye,
New J. Phys. \textbf{14}, 125014 (2012)
[arXiv:1212.2888 [hep-ph]].

\bibitem{Ham-Ma:2006} 
T.~Hambye, K.~Kannike, E.~Ma and M.~Raidal,
Phys. Rev. D \textbf{75}, 095003 (2007)
[arXiv:hep-ph/0609228 [hep-ph]].
\bibitem{Ham-Strum:2006} 
T.~Hambye, M.~Raidal and A.~Strumia,
Phys. Lett. B \textbf{632}, 667-674 (2006)
[arXiv:hep-ph/0510008 [hep-ph]].
\bibitem{Ham:2005} 
T.~Hambye,
Nucl. Phys. B Proc. Suppl. \textbf{145}, 280-285 (2005).
\bibitem{Gu:2016} 
P.~H.~Gu, E.~Ma and U.~Sarkar,
Phys. Rev. D \textbf{94}, no.11, 111701 (2016)
[arXiv:1608.02118 [hep-ph]].
\bibitem{bbexpt} C. Alduino {\it
et al.}(CUORE Collaboration), Phys. Rev. Lett. {\bf 120} (2018)
132501; M. Agoshini {\it
et al.}(GERDA Collaboration), Phys. Rev. Lett. {\bf 120} (2018)
132503; C. E. Aalseth  {\it
et al.}(MAJORANA Collaboration), Phys. Rev. Lett. {\bf 120} (2018)
132502; A. Gando {\it
et al.}(KamLAND-Zen Collaboration), Phys. Rev. Lett. {\bf 117} (2016)
082503; J. B. Albert {\it
et al.}(EXO-200 Collaboration), Phys. Rev. Lett. {\bf 120} (2018)
072701; G. Anton {\it
et al.}(EXO-200 Collaboration), Phys. Rev. Lett. {\bf 123} (2019)
161802.   
\bibitem{Balaji:2000} K. R. S. Balaji, A. Dighe, R. N. Mohapatra, M. K. Parida,
 Phys. Rev. Lett. {\bf 84} (2000) 5034-5037,
  hep-ph/0001310; K. R. S. Balaji, A. S. Dighe, R. N. Mohapatra,
  M. K. Parida, Phys. Lett. {\bf B 481} (2000) 33-38, hep-ph/0002177; K. R. S. Balaji, R. N. Mohapatra,
  M. K. Parida, E. A. Paschos, Phys. Rev. {\bf D 63} (2001) 113002,
  hep-ph/0011263.


\bibitem{mpr:2004} R. N. Mohapatra, M. K. Parida, G. Rajasekaran,
  Phys. Rev. {\bf D 69} (2004) 053007, hep-ph/0301234; R. N. Mohapatra, M. K. Parida,
  G. Rajasekaran, Phys. Rev. {\bf D 71} (2005) 057301, hep-ph/0501275; R. N. Mohapatra,
  M. K. Parida, G. Rajasekaran, Phys. Rev. {\bf D 72} (2005) 013002, hep-ph/0504236; S. Agarwalla, M. K. Parida, R. N. Mohapatra, G. Rajasekaran, Phys. Rev. {\bf D 75} (2007) 033007, hep-ph/0611225;  W. G. Hollik, Phys. Rev. {\bf D 91} (2015) no.3, 033001, arXiv:1412.4585[hep-ph].
\bibitem{Sunny:2018} S. Vaqnozzi {\it et al}, Phys. Rev. {\bf D 94} (2016) 083522, arXiv:1605.04320[astro-ph.CO],  S. Vaqnozzi {\it et al}, Phys. Rev. {\bf 
D 96} (2017) 123503, arXiv:1701.08172[astro-ph.CO], S. Vaqnozzi {\it et al}, 
 Phys. Rev. {\bf D 98} (2018) 123526, arXiv:1802.08694[astro-ph.CO]. 

 \bibitem{Ohlsson:2019} 
M.~Ghosh, T.~Ohlsson and S.~Rosauro-Alcaraz,
JHEP \textbf{03}, 026 (2020)
[arXiv:1912.10010 [hep-ph]];
S.~Choubey and T.~Ohlsson,
Phys. Lett. B \textbf{739}, 357-364 (2014)
[arXiv:1410.0410 [hep-ph]].

\bibitem{KATRIN} 
M.~Aker \textit{et al.} [KATRIN],
Phys. Rev. Lett. \textbf{123}, no.22, 221802 (2019)
[arXiv:1909.06048 [hep-ex]].
\bibitem{Sierra:2014tqa} 
  D.~Aristizabal Sierra, M.~Dhen and T.~Hambye,
  JCAP {\bf 1408}, 003 (2014)
  [arXiv:1401.4347 [hep-ph]].
\bibitem{Sierra:2011ab} D.~Aristizabal Sierra, F. Bazzocchi, I. de
  Mederios Verzilas,  Nucl. Phys. {\bf B858} (2012) 196, arXiv:1112.1843[hep-ph]. D.~Aristizabal Sierra, L. A. Munoz, E. Nardi, Phys. Rev. {\bf D 80}
 (2009) 016007; D.~Aristizabal Sierra, M. Losado, E. Nardi, JCAP {\bf
   0912} (2009) 015, arXiv:0905.0662.  

\bibitem{cps:2019}
M.~Chakraborty, M.~K.~Parida and B.~Sahoo,
JCAP \textbf{01}, 049 (2020)
[arXiv:1906.05601 [hep-ph]].


   



\bibitem{DI:2002} 
S.~Davidson and A.~Ibarra,
Phys. Lett. B \textbf{535}, 25-32 (2002)
[arXiv:hep-ph/0202239 [hep-ph]].

\bibitem{Khlopov} M. Yu. Khlopov, A. D. Linde, Phys. Lett.
{\bf B 138}
 (1984) 265; J. Ellis, J. Kim, and D.
V. Nanopoulos, Phys. Lett. {\bf B 145} (1984) 181; 
J. Ellis, G. B. Gelmini, J. L. Lopez, D.
V. Nanopoulos, and S. Sarkar, Nucl. Phys.
{\bf B 373} (1992) 399; M. Kawasaki and T. Moroi, Progr. Theor. Phys.
{\bf 93} (1995) 879; V. S. Rychkov, A. Strumia, Phys. Rev.
{\bf D 75} (2007) 075001.
\bibitem{Drees:1996} 
M.~Drees,
[arXiv:hep-ph/9611409 [hep-ph]].
\bibitem{Vissani:1998} 
F.~Vissani,
Phys. Rev. D \textbf{57}, 7027-7030 (1998)
[arXiv:hep-ph/9709409 [hep-ph]].
\bibitem{Casas:2004}  J. A. Casas, J. R. Espinosa, and I. Hidalgo, JHEP 11, 057 (2004), arXiv:hep-ph/0410298 [hep-ph];
 M. Farina, D. Pappadopulo, and A. Strummia, JHEP 08, 022 (2013), arXiv:1303.7244 [hep-ph]; J. D. Clarke, R. Foot, and R. R. Volkas, Phys. Rev. D91, 073009 (2015), arXiv:1502.01352 [hep-ph]; M. Fabbrichesi and A. Urbano, Phys. Rev. D92, 015028 (2015), arXiv:1504.05403 [hep-ph]; J. D. Clarke, R. Foot, and R. R. Volkas, Phys. Rev. D92, 033006 (2015), arXiv:1505.05744 [hep-ph].


\bibitem{Akerib:2016} 
D.~S.~Akerib {\it et al.} [LUX Collaboration],
  Phys.\ Rev.\ Lett.\  {\bf 118}, no. 2, 021303 (2017)
  [arXiv:1608.07648 [astro-ph.CO]].


\bibitem{Aprile:2017} 
E.~Aprile {\it et al.} [XENON Collaboration],
  Phys.\ Rev.\ Lett.\  {\bf 119}, no. 18, 181301 (2017)
  [arXiv:1705.06655 [astro-ph.CO]].

\bibitem{Aprile:2018} 
  E.~Aprile {\it et al.} [XENON Collaboration],
  Phys.\ Rev.\ Lett.\  {\bf 121}, no. 11, 111302 (2018)
  [arXiv:1805.12562 [astro-ph.CO]].
  
  

\bibitem{Cui:2017} 
 X.~Cui {\it et al.} [PandaX-II Collaboration],
  Phys.\ Rev.\ Lett.\  {\bf 119}, no. 18, 181302 (2017)
  [arXiv:1708.06917 [astro-ph.CO]].


\bibitem{wmap} 
WMAP Collaboration, D. N. Spergel et al,  
Astrophys. J. Suppl. 170 (2007)377 [astro-ph/0603449] [INSPIRE]; 
  D.~Larson, J.~Dunkley, G.~Hinshaw, E.~Komatsu, M.~R.~Nolta, C.~L.~Bennett, B.~Gold and M.~Halpern {\it et al.},
  Astrophys.\ J.\ Suppl.\  {\bf 192}, 16 (2011),
  arXiv:1001.4635 [astro-ph.CO].\\
 
\bibitem{Espinosa} 
J.~Elias-Miro, J.~R.~Espinosa, G.~F.~Giudice, H.~M.~Lee and A.~Strumia,
JHEP \textbf{06}, 031 (2012)
[arXiv:1203.0237 [hep-ph]].
\bibitem{Lebedev} 
O.~Lebedev,
Eur. Phys. J. C \textbf{72}, 2058 (2012)
[arXiv:1203.0156 [hep-ph]].

\bibitem{WIMP} 
B.~W.~Lee and S.~Weinberg,
Phys. Rev. Lett. \textbf{39}, 165-168 (1977);
K.~Griest and M.~Kamionkowski,
Phys. Rev. Lett. \textbf{64}, 615 (1990)
\bibitem{GAMBIT} 
P.~Athron \textit{et al.} [GAMBIT],
Eur. Phys. J. C \textbf{77}, no.8, 568 (2017)
[arXiv:1705.07931 [hep-ph]].



\bibitem{Beringer:2012}
J.~Beringer {\it et al.} [Particle Data Group],
  Phys.\ Rev.\ D {\bf 86}, 010001 (2012);  C. Patrignani {\it et al.} (Particle Data Group), Chin. Phys. {\bf C40} (2016) no.10, 100001.
01;
  
\bibitem{Sahoo:2018wdt}
B.~Sahoo, M.~Chakraborty and M.~K.~Parida,
Adv. High Energy Phys. \textbf{2018} (2018), 4078657
[arXiv:1804.01803 [hep-ph]].
  


\bibitem{Ma:2006} 
E.~Ma,
Phys. Rev. D \textbf{73}, 077301 (2006)
[arXiv:hep-ph/0601225 [hep-ph]].
\bibitem{mkp:2011} 
M.~K.~Parida,
Phys. Lett. B \textbf{704}, 206-210 (2011)
[arXiv:1106.4137 [hep-ph]].
\bibitem{Hambye:2010} 
T.~Hambye,
PoS \textbf{IDM2010}, 098 (2011)
[arXiv:1012.4587 [hep-ph]];
M.~Frigerio and T.~Hambye,
Phys. Rev. D \textbf{81}, 075002 (2010)
[arXiv:0912.1545 [hep-ph]].
\bibitem{Kadastic:2009}
M.~Kadastik, K.~Kannike and M.~Raidal,
Phys. Rev. D \textbf{80}, 085020 (2009)
[arXiv:0907.1894 [hep-ph]];
M.~Kadastik, K.~Kannike and M.~Raidal,
Phys. Rev. D \textbf{81}, 015002 (2010)
[arXiv:0903.2475 [hep-ph]].
\bibitem{Ma:2018} 
E.~Ma,
Phys. Rev. D \textbf{98}, no.9, 091701 (2018)
[arXiv:1809.03974 [hep-ph]].
\bibitem{Ma:2020} 
E.~Ma,
[arXiv:2004.12949 [hep-ph]].
\bibitem{Kolb:1990} E.~W.~Kolb and M.~S.~Turner,
  Front.\ Phys.\  {\bf 69}, 1 (1990).
\bibitem{Bertone:2004} G.~Bertone, D.~Hooper and J.~Silk, 
  Phys.\ Rept.\  {\bf 405}, 279 (2005),
  hep-ph/0404175.
\bibitem{Gondolo:1990dk} 
P.~Gondolo and G.~Gelmini,
  Nucl.\ Phys.\ B {\bf 360}, 145 (1991).
\bibitem{McDonald:1993ex} 
J.~McDonald,
  Phys.\ Rev.\ D {\bf 50}, 3637 (1994),
  hep-ph/0702143.



\bibitem{Guo:2010hq}  
 W.~L.~Guo and Y.~L.~Wu,
  JHEP {\bf 1010}, 083 (2010),
  arXiv:1006.2518 [hep-ph].
\bibitem{Biswas:2011td} 
A.~Biswas and D.~Majumdar,
  Pramana {\bf 80}, 539 (2013),
  arXiv:1102.3024 [hep-ph].
\bibitem{Cline:2013gha} 
 J.~M.~Cline, K.~Kainulainen, P.~Scott and C.~Weniger,
  Phys.\ Rev.\ D {\bf 88}, 055025 (2013)
  Erratum: [Phys.\ Rev.\ D {\bf 92}, no. 3, 039906 (2015)]
  [arXiv:1306.4710 [hep-ph]].
\bibitem{Chabab:2015} 
M.~Chabab, M.~C.~Peyranere and L.~Rahili,
Phys. Rev. D \textbf{93}, no.11, 115021 (2016)
[arXiv:1512.07280 [hep-ph]].
\bibitem{Haba:2016} 
N.~Haba, H.~Ishida, N.~Okada and Y.~Yamaguchi,
Eur. Phys. J. C \textbf{76}, no.6, 333 (2016)
[arXiv:1601.05217 [hep-ph]];
R.~Padhan, D.~Das, M.~Mitra and A.~Kumar Nayak,
Phys. Rev. D \textbf{101}, no.7, 075050 (2020)
[arXiv:1909.10495 [hep-ph]].
\bibitem{Garg:2017} I. Garg, S. Goswami, Vishnudath K. N., N. Khan, Phys. Rev. D 96, 055020
(2017),arXiv:1706.08851 [hep-ph];
 M. Gonderinger, Y. Li, H. Patel, M. J. Ramsey-Musolf, JHEP 1001, 053 (2010)
 C.S. Chen, Y. Tang, JHEP 1204, 019 (2012) [arXiv:1202.5717 [hep-ph]].
 N. Khan, S. Rakshit, Phys. Rev. D 90, no. 11, 113008 (2014) [arXiv:1407.6015 [hep-ph]].
\bibitem{Kannike:2016} 
K.~Kannike,
Eur. Phys. J. C \textbf{76}, no.6, 324 (2016)
[arXiv:1603.02680 [hep-ph]];
K.~Kannike,
Eur. Phys. J. C \textbf{72}, 2093 (2012)
[arXiv:1205.3781 [hep-ph]].
\bibitem{ATCM:2015} 
G.~Aad \textit{et al.} [ATLAS and CMS],
Phys. Rev. Lett. \textbf{114}, 191803 (2015)
[arXiv:1503.07589 [hep-ex]].
\bibitem{ATCM:2014}
 [ATLAS, CDF, CMS and D0],
[arXiv:1403.4427 [hep-ex]].
\bibitem{PDG:2014} 
K.~A.~Olive \textit{et al.} [Particle Data Group],
Chin. Phys. C \textbf{38}, 090001 (2014);
S.~Bethke,
Nucl. Phys. B Proc. Suppl. \textbf{234}, 229-234 (2013)
[arXiv:1210.0325 [hep-ex]].
\bibitem{exptomegaDM} 
P.~A.~R.~Ade \textit{et al.} [Planck],
Astron. Astrophys. \textbf{571}, A16 (2014)
[arXiv:1303.5076 [astro-ph.CO]].
\bibitem{Xing:2009} 
Z.~z.~Xing,
Prog. Theor. Phys. Suppl. \textbf{180}, 112-127 (2009)
[arXiv:0905.3903 [hep-ph]].

\bibitem{WRpsb:2017} 
P.~S.~B.~Dev, C.~M.~Vila and W.~Rodejohann,
Nucl. Phys. B \textbf{921}, 436-453 (2017)
[arXiv:1703.00828 [hep-ph]].
\bibitem{JWVFlam} C. Bonilla, R. M. Fonseca, J. W. F. Valle, Phys. Rev. {\bf D92} (2015) 075028, arXiv:1508.02323 [hep-ph]; A. Arhrib, R. Benbrik, M. Chabab, G. Moultaka, M. C. Peyranere, L. Rahili, and J. Ramadan, Phys.
Rev. D84, 095005 (2011), arXiv:1105.1925 [hep-ph]; D. Das and A. Santamaria, Phys. Rev. D94, 015015 (2016), arXiv:1604.08099 [hep-ph].
\bibitem{Wetterich:2020} Guillem Domie${\hat e}$nech, Mark Godsell, Christof Wetterich, [arXiv:2008.04310[hep-ph]].

\bibitem{Forengo:2006} M. Cirelli, N. Forengo, A. Strummia, Nucl. Phys. {\bf B753} (2006) 178. 
\bibitem{Hambye:2008} T. Hambye, M. H. Tytgat, Phys. Lett. {\bf B659} (2008) 651-655, [arXiv:0707.0633[hep-ph]]; R. Barbieri, L. J. Hall, V. Rychkov, Phys. Rev. {\bf D 74} (2006) 015007, hep-ph/0603188.
\bibitem{Krauss:1989} L. M. Krauss, F. Wilczek, Phys. Rev. Lett. {\bf 62} (1989) 1221.
\bibitem{psb:2010}  M. K. Parida, P. K. Sahu, K. Bora, Phys. Rev. {\bf D 83} (2011) 093004,
[arXiv:1011.4577[hep-ph]].

\bibitem{mkp:2020} M. K. Parida, R. Samantaray, Eur.Phys. J. ST (2020)(accepted), [arXiv:2002.06869 [hep-ph]].

\end{thebibliography}
\end{document}